\documentclass[twocolumn,fleqn]{aastex631}

\usepackage{amsmath}
\usepackage{amsbsy}
\usepackage{lipsum}

\usepackage{subfigure}

\usepackage{natbib}

\shorttitle{Jet stability}
\shortauthors{Wu Fan et al.}

\graphicspath{{./}{figures/}}

\makeatletter
\let\longtable*\empty
\makeatother

\begin{document}

\title{Laplace-Fourier linear stability analysis of non-relativistic magnetized rotational jets: mode identification by Hamiltonian analysis and the occurrence of mode transitions}
\author{Wu Fan}
\author{Yamada Shoichi}
\affiliation{Department of Applied and Pure Physics, Waseda University 3-4-1 Okubo, Shinjuku City, Tokyo 169-8555, Japan}

\begin{abstract}
In this paper, we conduct a linear stability analysis of magnetized and/or rotating jets propagating in ambient matter that is also magnetized and/or rotating, having in mind the application to the jet penetrating the core/envelope of a massive star. We solve the linearized magneto-hydrodynamic (MHD) equations in the non-relativistic regime by Laplace transform in time and Fourier transform in space. In this formulation all unstable modes with the same translational and azimuthal wave numbers can be obtained simultaneously by searching for pole singularities in the complex plane. In order to determine unambiguously their driving mechanisms, we evaluate the second-order perturbation of the MHD Hamiltonian for individual eigenfunctions derived at these singular points. We identify in our non-rotating models the Kelvin-Helmholtz instability (KHI) as one of the shear-driven modes and the current-driven instability such as the kink instability (KKI). In rotational models we also find the magnetorotational instability (MRI) as another shear-driven mode. In some cases, we find that a mode changes its character continuously from KKI to KHI (and vice versa) or from MRI to KHI as the jet velocity is increased.
\end{abstract}

\keywords{MHD; High energy astrophysics; Jets}

\section{Introduction} 

Neutron stars are magnetized in general (\citealt{Konar_2017}; \citealt{Igoshev_2021}). Strong magnetic fields, combined with rotation, are supposed to produce an energetic subgroup of core-collapse supernovae (CCSNe) referred to as the hypernovae (\citealt{Janka_2012}). They are thought to help neutrinos revive a stalled shock wave even for the ordinary CCSNe (\citealt{Kuroda_2020}; \citealt{Matsumoto_2022}; \citealt{Varma_2023}). One of the features commonly observed in the simulations of magnetic supernovae is jet formation (\citealt{Mösta_2014}; \citealt{Obergaulinger_2017}; \citealt{Bugli_2021}). The magnetic field is existent already in massive progenitor stars (\citealt{Donati_2009}; \citealt{Keszthelyi_2023}) and is amplified by compression and also by winding if the progenitor is rotating during the collapse. After core bounce, the magnetorotational instability (MRI) is believed to amplify it further in the rotational case (\citealt{Akiyama_2003}; \citealt{Rembiasz_2016}). The magnetic field taps the rotational energy of the supernova core, which can amount to more than ten times the canonical explosion energy of CCSNe, to produce the hypernovae. The jet is thought to play an important role also in gamma-ray bursts (\citealt{MacFadyen_1999}; \citealt{Obergaulinger_2020}; \citealt{Kumar_2015} for a recent review).

Astrophysical jets are prone to various instabilities. The most common and most extensively studied one is the Kelvin-Helmholtz instability (KHI), which feeds on shear motions in fluid and, as such, is expected to occur rather commonly in the jet pushing through a surrounding matter. The research of KHI in the jet has a long history. It was conducted initially to understand some radio sources (\citealt{Turland_1976}; \citealt{Blandford_1976}; \citealt{Ferrari_1978}). The linear stability analysis based on the normal mode was performed for idealized settings such as a uniform jet with a discontinuous edge at the boundary with the ambient matter: for example, \citet{Bodo_1989} and \citet{Bodo_1996} studied both axisymmetric and non-axisymmetric KHIs of uniformly rotating {non-relativistic} jets. \citet{Hardee_1997} considered cooling jets and showed that the non-adiabaticity can strongly affect the KHI growth in the linear phase. \citet{Perucho_2004} focused on relativistic planar jets and reproduced numerically the linear growth of KHI. \citet{Sobacchi_2017} studied the surface mode in the relativistic regime. \citet{Berlok_2019} performed not only linear analysis but also numerical simulations for a smooth velocity profile.

The study of the instability in magnetized plasmas was initiated for nuclear fusion experiments (\citealt{Bateman_1978}). Later the research was directed to the current-driven instability (CDI) in astrophysical jets. CDI is associated with toroidal magnetic fields. The kink instability (KKI), for example, induces radial displacements of the jet barycenter and may lead to a disruption of the jet. \citet{Appl_1992} derived the dispersion relation for CDI in a current-carrying jet by solving the linearized ideal MHD equation under the force-free condition. It was later extended to cold jets with trans-magnetosonic velocities (\citealt{Appl_1996}) and the dependence on the pitch angle {of helical magnetic} fields was discussed (\citealt{Appl_2000}).

The astrophysical jets, especially those from AGNs, were observed to be strongly collimated, which seemed to contradict the possible existence of KKI. \citet{Spruit_1997} suggested the suppression of KKI by the longitudinal component of the magnetic field. \citet{Bonanno_2011} claimed that the jet is unstable to high -$m$ modes, with $m$ being the azimuthal wave number, even in the strong longitudinal field. \citet{Begelman_1998}, taking into account the gradient of the toroidal magnetic field that is in balance with the plasma pressure, obtained the criterion for CDI in the relativistic jet. Later the discussion was expanded to non-relativistic jets with a longitudinal field (\citealt{Das_2019}). \citet{Lyubarskii_1999} derived the growth rate of KKI for rotating relativistic jets under the force-free condition. \citet{Nalewajko_2012}, on the other hand, derived the dispersion relation for magnetized plasma jets with a smooth and nonuniform radial profile of jet velocity. \citet{Narayan_2009} conducted linear analysis on CDI in relativistic force-free jets. Jets with the so-called Lundquist force-free field were investigated both in the Newtonian (\citealt{Kim_2015}) and relativistic regimes by \citet{Kim_2017}; \citet{Kim_2018}. \citet{Sobacchi_2017} extended the parameter range for the relativistic force-free jets. KHI was also investigated for the magnetized jet. \citet{Hardee_2007}, for example, paid attention to the relativistic MHD jets with a spine-sheath structure and derived the dispersion relation of KHI. \citet{Bodo_2013} performed a systematic linear analysis on KHI and CDI in the magnetized non-rotating relativistic cold jets.

Astrophysical jets are expected to be rotating in general, since they are often associated with accretion disks. Rotation tends to stabilize instabilities like KKI via the Coriolis effect as confirmed both in linear analysis (\citealt{Carey_2009}) and numerical simulations (\citealt{Nakamura_2004}). \citet{Gourgouliatos_2018} generalized the Rayleigh criterion for the axisymmetric hydrodynamic jet in special relativity. It was extended later to the magnetized jet in the Newtonian (\citealt{Komissarov_2019}) and relativistic (\citealt{Matsumoto_2021}) regimes. They both showed that the magnetic field tends to suppress the Rayleigh instability in the rotating jet. The combination of rotation and magnetic field can drive MRI (\citealt{Balbus_1991}; \citealt{Balbus_1998}). \citet{Kim_2000} took into account the gravity in a non-relativistic cold MHD flow and derived the dispersion relations for buoyancy modes such as the Parker instability and MRI. \citet{Pessah_2005} studied the effect of strong toroidal fields in linear analysis and found that MRI is suppressed at low wave numbers. Bodo et al. also conducted linear analysis on various modes including MRI for cold jets both in the Newtonian (\citealt{Bodo_2016}) and relativistic (\citealt{Bodo_2019}) regimes.

Numerical simulations are an indispensable tool to study non-linear evolutions of the instabilities. \citet{Mizuno_2009} performed 3D MHD simulations in the co-moving frame of the jet and observed a total disruption of the jet in the nonlinear phase of KKI that followed its linear evolution. \citet{O'Neill_2012} performed an ensemble of MHD jet simulations in force-free, pressure- and rotation-supported equilibrium configurations and found that the initial force-balance has a considerable impact on the jet morphology. \citet{Bromberg_2016} studied magnetic-pressure {dominant relativistic MHD} jets with different morphologies such as those with/without a head. \citet{Porth_2015} investigated the dependence of the {jet} stability on {the} ambient pressure and showed that the external pressure {obeying} a particular power law suppresses instabilities of all kinds. \citet{Singh_2016} explored KKI in rotating jets with different density profiles and suggested that the magnetic reconnection {plays} an important role in the emission process. \citet{Bromberg_2019} carried out 3D MHD jet simulations with high values of magnetization and demonstrated {the importance of dissipations via} the reconnection and turbulence after KKI is saturated. \citet{Mattia_2023} suggested that the shock wave driven by KHI strongly affects the dynamics of non-relativistic jets and discussed possible radiative features. \citet{BodoG_2021}, \citet{Nikhil_2021} studied X-ray emissions from a KKI- or KHI-dominant jet with simulations meant for AGNs, respectively. \citet{Bodo_2022}, on the other hand, investigated the dissipations in current sheets generated by the saturation of CDI in relativistic jets. The discussion was extended to pressure-driven instability by considering nonvanishing pressure gradients (\citealt{Musso_2024}). The possible growth of KKI and its subsequent disruption of a narrow jet in the magnetorotational CCSNe was first reported in their 3D MHD simulations by \citet{Mösta_2014}. Similar observations were made later by \citet{Bugli_2021}; \citet{Kuroda_2020}.

There are indeed many researches on the stability of astrophysical jets. More or less, the authors of these studies had some target objects in mind, and the unperturbed models they employed reflected them. In this paper we also conduct a linear analysis of nonrelativistic jets, having in mind the application to the jet propagating in the stellar core/envelope such as those in core-collapse supernovae. It has been discussed that the MRI in the supernova core can amplify the magnetic field (\citealt{Akiyama_2003}; \citealt{Obergaulinger_2009}), and the resultant strong magnetic field will extract the rotational energy to produce a jet. As we mentioned at the very beginning of this introduction, the magnetic field in combination with rotation is believed to play a crucial role in the generation of hypernovae. We hence consider in this paper rotating magnetized jets surrounded by ambient matter that is also magnetized and rotating in general, unlike other previous studies. The jet boundary is assumed either to be a discontinuous surface or to be a smooth transition layer. 

The methodology is also different. Unlike the previous studies, we employ the Laplace-Fourier analysis: the linearized ideal MHD equations are Laplace-transformed in time and Fourier-transformed in $z$ (the jet direction) and $\phi$ (the azimuthal direction). The resultant ordinary differential equation in $r$ (the radial direction transverse to the jet) is solved numerically; we look for the singularities in the complex plane of $s$ (the variable conjugate to time in the Laplace transformation) that correspond to unstable modes. The growth rate is read off from the pole position and the eigenfunction is derived from the homogeneous solution there (see Section 2.2.1 for the definition). We are thus able to treat various unstable modes such as KHI, CDI and MRI simultaneously on the same footing if they exist. However, their identities are sometimes not obvious. In fact the continuity argument often fails because the mode identity can change continuously with a parameter, such as the jet velocity, that specifies the background configuration. We hence evaluate the second variation of the Hamiltonian for each mode to determine the mode identity unambiguously. This is a direct extension of the conventional stability argument for stationary states to steady motions. The extension of the Laplace-Fourier linear analysis employed in this paper to the relativistic jet is straightforward and currently underway, with the application to gamma-ray bursts and AGNs in mind.

The paper is organized as follows. The method is described in detail in Section 2. The jet models are also presented there. The numerical results are given in Section 3. Section 4 is devoted to some discussions. The paper is concluded in Section 5.





\section{Methods}
\subsection{Basic equations}
In this paper, we systematically investigate the linear stability of cylindrical jets with axisymmetry. The jet is assumed to be rotating and/or magnetized and propagating through an ambient matter, which we also assume is rotating and magnetized in general. Ignoring dissipative/radiative processes, we employ the ideal MHD equations:
\begin{gather}
\frac{\partial \rho}{\partial t} + \nabla \cdot (\rho \boldsymbol{v})  = 0 ,
\\
\frac{\partial \rho \boldsymbol{v}}{\partial t} + \nabla \cdot (\rho \boldsymbol{v v}) = -\nabla P + \nabla \cdot (\boldsymbol{BB}) ,
\\
\frac{\partial \boldsymbol{B}}{\partial t} = \nabla \times (\boldsymbol{v} \times \boldsymbol{B}) ,
\end{gather}
to describe both unperturbed static configurations and perturbations to them. In these equations,  $\rho$, $\boldsymbol{v}$ are the matter density and velocity, respectively; $\boldsymbol{B}$ is the magnetic field and ${P}
=p+{\boldsymbol{B}^2}/{2}$ denotes the total pressure with $p$ being the matter pressure; the magnetic permeability $\mu_0$ is assumed to be unity as usual. The energy equation is replaced by the adiabaticity condition. The unperturbed states are time-independent solutions to these equations. 

The time evolution of the perturbation is described by the following linearized equations:
\begin{gather}
\frac{\partial\rho_1}{\partial t}+(\boldsymbol{v}_0\cdot \nabla)\rho_1 +(\boldsymbol{v}_1\cdot \nabla)\rho_0+ \rho_0\nabla\cdot \boldsymbol{v}_1 =0 ,
\\
\begin{split}
\rho_0 \left(\frac{\partial}{\partial t}+\boldsymbol{v}_0\cdot\nabla\right)\boldsymbol{v}_1 + \rho_1(\boldsymbol{v}_0\cdot\nabla)\boldsymbol{v}_0+ \rho_0(\boldsymbol{v}_1\cdot\nabla)\boldsymbol{v}_1 \\
= -\nabla P_1 +(\boldsymbol{B}_1\cdot \nabla)\boldsymbol{B}_0 +(\boldsymbol{B}_0\cdot \nabla)\boldsymbol{B}_1 ,
\end{split}
\\
\begin{split}
\frac{\partial \boldsymbol{B}_1}{\partial t}+ & (\boldsymbol{v}_0\cdot \nabla )\boldsymbol{B}_1 +(\boldsymbol{v}_1\cdot \nabla)\boldsymbol{B}_0 \\
& = (\boldsymbol{B}_0\cdot \nabla)\boldsymbol{v}_1 -(\nabla\cdot v_1)\boldsymbol{B}_0 +(\boldsymbol{B}_1\cdot \nabla)\boldsymbol{v}_0 ,
\end{split}
\\
\frac{1}{\rho_0}(\rho_1+(\boldsymbol{\xi} \cdot \nabla)\rho_0)=\frac{1}{\Gamma p_0}(p_1+(\boldsymbol{\xi} \cdot \nabla)p_0 ).
\end{gather}
In the above equations, the quantities with subscript~1 are perturbations whereas those with subscript 0 are unperturbed quantities. Equation (7) implies that we consider adiabatic perturbations alone; $\Gamma$ is the adiabatic constant and $\boldsymbol{\xi}$ denotes the Lagrangian displacement defined in Eq. (69). The unperturbed states are rotating and magnetized in general: the unperturbed velocity and magnetic field are expressed as $\boldsymbol{v}_0 $ $=(0, \Omega_0 r, V_0)$ and $\boldsymbol{B}_0=(0, B_{\phi 0}, B_{z0})$, respectively. All unperturbed quantities are functions of $r$ alone and are summarized later in Section 2.5. Here and in the following, we employ the cylindrical coordinates $(r, \phi, z)$. 

We introduce the Laplace transform in $t$ and the Fourier transform in $\phi$ and $z$:
\begin{equation}
\begin{split}
& \hat{q}_1(r, s, k, m)  = \mathcal{T}[q_1] = \\
& \frac{1}{2\pi} \int_{-\pi}^{\pi}\! d\phi \int_{-\infty}^{\infty}\! dz \int_{0}^{\infty}\!dt\ q(r, \phi, z, t)\  e^{-i(m \phi + kz)- st} .
\end{split}
\end{equation}
Above and hereafter the hat mark on a variable indicates that it is a Laplace-Fourier-transformed quantity. In the above equation, $m$ and $k$ are the wave numbers in the {$\phi$-} and {$z$-}directions, respectively; $s$ is a complex variable conjugate to time $t$. The equations for the Laplace-Fourier-transformed quantities are given componentwise as follows:
\begin{gather}
\begin{split}
\rho(i \sigma \hat{U}_1 - & U_* - 2 \Omega_0 \hat{\Omega}_1 r) - \hat{\rho}_1 {\Omega_0}^2 r \\
& = - \frac {d\hat{P}_1}{dr} + iF \hat{{B_r}_1} -     \frac {2}{r} {B_\phi}_0 \hat{B}_{\phi 1} ,
\end{split}
\\
\begin{split}
\rho_0 \Big( i \sigma \hat{\Omega}_1 r & - \Omega_* r + \Big( 2\Omega_0 + r \frac{d\Omega_0}{dr} \Big) \hat{U}_1 \Big)\\
& = -i \frac {m}{r} \hat{P}_1 + i F \hat{B}_{\phi 1} + \frac{1}{r} \frac{d}{dr} (r B_{\phi 0}) \hat{B}_{r 1} ,
\end{split}
\\
\begin{split}
\rho_0 \Big( i \sigma \hat{V}_1 & - V_* + \frac {dV_0}{dr}\hat{U}_1 \Big) \\
& = -i k \hat{P}_1 + i F \hat{B}_{z1} +  \frac{d {B_z}_0}{dr} \hat{B}_{r1} ,
\end{split}
\\
i \sigma \hat{B}_{r1} - {B_r}_*= i F \hat{U}_1 , 
\\
\begin{split}
i \sigma \hat{B}_{\phi 1} - {B_\phi}_* = i F & \hat{\Omega}_1 r + \left( \frac { {B_\phi}_0 }{r} - \frac {d{B_\phi}_0}{dr    }\right) \hat{U}_1 \\
& + r \frac {d\Omega_0}{dr} \hat{B}_{r1} - {B_\phi}_0 \mathcal{T}[\nabla \cdot  \mathbf{v}_1] ,
\end{split}
\\
\begin{split}
i \sigma \hat{B}_{z1} - {B_z}_* = i F & \hat{V}_1 - \frac {d{B_z}_0}{dr} \hat{U}_1 \\
& + r \frac {dV_0}{dr} \hat{B}_{r1} - {B_z}_0 \mathcal{T}[\nabla \cdot  \mathbf{v}_1] ,
\end{split}
\\
\frac{\Gamma p_0}{\rho_0}\bigg(\hat{\rho_1}+\bigg( \hat{\xi_r} + \frac{\xi_*}{i\sigma} \bigg)\frac{\partial \rho_0}{\partial r}\bigg)=\hat{p_1}+\bigg(\hat{\xi_r}+ \frac{\xi_*}{i\sigma} \bigg)\frac{\partial p_0}{\partial r} ,
\end{gather}
with the following notations
\begin{gather}
\sigma = -s +m\Omega_0 + kV_0 , 
\\
F=\frac{m}{r} B_{\phi 0}+kB_{z0} .
\end{gather}
In the above equations, $\hat{U}_1$ and $\hat{V}_1$ are the $r-$ and $z-$components of the perturbed velocity, respectively. Quantities with $*$ are the initial perturbations, which can be taken into account explicitly in the Laplace-transform analysis. 
The above equations can be reduced to two equations for $\hat{Y}_1$ and $\hat{P}_1$:
\begin{gather}
AS\frac{d\hat{Y}_1}{dr} - \frac{r}{i\sigma} I^Y_* = C_1 \hat{Y}_1 - r C_2 \hat{P}_1 ,
\\
AS\frac{d\hat{P}_1}{dr} + \frac{1}{r\sigma}I^P_* = \frac{1}{r} C_3 \hat{Y}_1 - C_1 \hat{P}_1 ,
\end{gather}
where we define
\begin{equation}
\hat{Y}_1 = \frac{r\hat{U}_1}{i \sigma} ,
\end{equation}
which can be rewritten by the Lagrangian radial displacement $\xi_r=\hat{U}_1/i\sigma$ as $\hat{Y}_1=r\xi_r$, and the perturbed total pressure
\begin{equation}
\hat{P}_1 = \hat{p} + B_{\phi 0}\hat{B}_{\phi 1}+B_{z 0}\hat{B}_{z 1}.
\end{equation}
Other functions in Eqs. (18) and (19) are summarized below:
\begin{gather}
C_1 = \rho_0 \sigma^2 \frac{Q}{r} - 2mS \frac{T}{r^2} ,
\\
C_2 = (\rho_0 \sigma^2)^2 - \left(\frac{m^2}{r^2} + k^2 \right) S ,
\\
C_3 = AS \Bigg( A+r \frac{d}{dr} \Bigg( \frac { {{B_\phi}_0}^2 - \rho_0 {\Omega_0}^2 r^2}{r^2} \Bigg)\Bigg) - 4S \frac{T^2}{r^2} + \frac{Q^2}{r^2} ,   
\\
\begin{split}
I^Y_* & =  F^2 A \frac{\Gamma p_0}{\rho_0}\rho_* +  F^2A\bigg( \frac{\Gamma p_0}{\rho_0} \frac{\partial \rho_0}{\partial r} - \frac{\partial p_0}{\partial r} \bigg)\xi_{r*} \\
& + \sigma \rho_0 ( \sigma^2 \rho_0 F r {B_\phi}_0 - m S ) \Omega_* \\
& +\sigma \rho_0 ( \sigma^2 \rho_0 F {B_z}_0 - k S ) V_* + i \Big( S \Big( \Big( \frac{m}{r^2} \frac{d}{dr} \Big( r {B_\phi}_0 \Big) \\
& + k \frac{d {B_z}_0}{dr} \Big) + \frac{F}{\sigma} \frac{d\sigma}{dr} \Big)  - \sigma^2 \rho_0 \Big( \sigma \rho_0 \Big( {B_\phi}_0 r \frac{d\Omega_0}{dr} \\
& + {B_z}_0 \frac{dV_0}{dr} \Big)  + F \Big( \frac{1}{2} \frac{d}{dr} ({B_0}^2) + \frac{ {{B_\phi}_0}^2 }{r} \Big) \Big) \Big) {B_r}_*  \\
& -  F S \Big( \frac{m}{r} {B_\phi}_* + k {B_z}_* \Big) \\
& + \sigma^4 {\rho_0}^2 \big( {B_\phi}_0 {B_\phi}_* + {B_z}_0 {B_z}_* \big) ,
\end{split}
\\
\begin{split}
I^P_* & = \bigg( i r^2 A S {\Omega_0}^2 + i A Q \frac{\Gamma p_0}{\rho_0} \bigg) \rho_* +  i A Q\bigg( \frac{\Gamma p_0}{\rho_0} \frac{\partial \rho_0}{\partial r} \\ 
&- \frac{\partial p_0}{\partial r} \bigg)\xi_{r*} - r A S \sigma \rho_0 U_* + i \sigma \rho_0 r ( - 2 S T \\
&+  F Q {B_\phi}_0 ) \Omega_* + i \sigma F Q \rho_0 {B_z}_0 V_* 
+ \Big( \rho_0 \alpha r \frac{d\Omega_0}{dr} \\ 
&+ \sigma Q \rho_0 {B_z}_0 \frac{dV_0}{dr}- 2 ST \frac{1}{r} \frac{d}{dr}(r {B_\phi}_0) + F Q \frac{ {{B_\phi}_0}^2 }{r} \\
& -rAFS + \frac{1}{2}\sigma QF\frac{\partial}{\partial r}(B_0^2)\Big) {B_r}_* - i \sigma \rho_0 \alpha {B_\phi}_* \\
& + i \sigma^2 Q \rho_0 {B_z}_0 {B_z}_* ,
\end{split}
\\
A=\rho_0 \sigma^2 - F^2 ,  
\\
S=B_0^2 \rho_0 \sigma + \Gamma p A ,
\end{gather}
\begin{gather}
Q = (2 {B_{\phi}}^2 - \rho_0 {\Omega_0}^2 r^2 ) A + 2 B_{\phi}FT ,
\\
T = F {B_\phi}_0 - \sigma \rho_0 \Omega_0 r ,
\\
\alpha = \sigma Q {B_\phi}_0 + 2 S ( F r \Omega_0 - \sigma {B_\phi}_0 ) . 
\end{gather}

These equations are discretized in $r$ and solved numerically with the 4th-order Runge-Kutta method. The concrete numerical setting will be given later in Section 2.5. The jet and the ambient matter are treated separately and their solutions are matched later at the boundary between the two regions as will be explained in Section 2.2.3. The boundary conditions are also given on the jet axis as well as at a large $r$, where the asymptotic solution is available. 

\subsection{Boundary conditions}
\subsubsection{The inner boundary condition at $r=0$}
We decompose the solution of Eqs. (18) and (19) into the solution to the homogeneous part with $I_{*}^{Y} = I_{*}^{P} = 0$ and the solution to the original inhomogeneous equations. We denote the former as $\hat{Y}_{1, \rm{H}}$ and $\hat{P}_{1, \rm{H}}$, whereas we write the latter as $\hat{Y}_{1, \rm{I}}$ and $\hat{P}_{1, \rm{I}}$. We treat their boundary conditions separately. For the latter contribution, we simply impose $\hat{Y}_{1,\rm{I}} = \hat{P}_{1,\rm{I}} = 0$. 

For the former contributions, on the other hand, we consider the close vicinity of the jet axis. We expand the background quantities in $r$ and retain only the lowest-order terms: under axisymmetry, we have $B_\phi \propto r$; we also find $F, S, A \propto r^0$, $T \propto r^1$, and  $Q \propto r^2$. Substituting these expansions into Eqs. (18) and (19) with Eqs.~(22)~-~(31) and setting $I_{*}^{Y} = I_{*}^{P}=0$, we obtain for $m\not= 0$
\begin{gather}
r\frac{d\hat{Y}_1}{dr} = -\left(2m\frac{T/r}{A}\right)\hat{Y}_1+{\frac{m^2}{A}}\hat{P}_1 ,
\\
r\frac{d\hat{P}_1}{dr} = \left(A-4\frac{T^2/r^2}{A}\right)\hat{Y}_1+{2m\frac{T/r}{A}}\hat{P}_1.
\end{gather}
{The} solution is derived as $\hat{Y}_1 = r^mY_{m}$ and  $\hat{P}_1 = r^mP_{m}$ with the following relation {between two coefficients:}
\begin{equation}
    \frac{Y_{m}}{P_{m}} = \frac{m}{2T/r+A} .
\end{equation}
We employ this {asymptotic solution} as the inner boundary condition for {$m \ne 0$}.

For $m=0$, on the other hand, {Eqs. (18) and (19)} are reduced to
\begin{gather}
AS\frac{d\hat{Y}_1}{dr} = -r{\rho_0^2\sigma^4-k^2S}\hat{P}_1 ,
\\
AS\frac{d\hat{P}_1}{dr} = 0.
\end{gather}
{The solution may be written as} $\hat{Y}_1 = r^2Y_{m0}$ and $\hat{P}_1 = r^0P_{m0}$, {in which} the following relation {holds}:
\begin{equation}
\frac{Y_{m0}}{P_{m0}} = \frac{1}{2}(-\rho_0^2 \sigma^4 + k^2 S).
\end{equation}
{We adopt this solution} as the inner boundary condition for $m=0$.

\subsubsection{The outer boundary conditions at $r \rightarrow \infty$} 
We proceed in a similar way for the outer boundary condition. We assume $\rho \propto r^0$ and $V=0$ for the ambient matter in the static background. Depending on the asymptotic behavior of $B_{\phi}$, $B_{z}$, $\Omega$ and $p$ at $r \rightarrow \infty$, we employ three different asymptotic solutions that go to zero at $r\to \infty$.
\\
\\
\textit{Case A}: $B_\phi \propto 0, r^{-1}, r^{-2}$; $B_z \propto 0, r^{0}, r^{-1}, r^{-2}$; $\Omega \propto 0, r^{-3/2}, r^{-2}$; $p \propto r^{0}$.
\\
We write the $r$ dependence as $B_\phi=r^nb_\phi$, $B_z=r^nb_z$, $\Omega=r^n\omega$ and $p=r^n p_a$. Then Eqs. (18) {and} (19) are rewritten as 
\begin{gather}
\frac{d\hat{Y}_1}{dr} = \frac{1}{r^2}C_{A1}\hat{Y}_1-rC_{A2} \hat{P}_1 ,
\\
\frac{d\hat{P}_1}{dr} = \frac{1}{r}C_{A3}\hat{Y}_1- \frac{1}{r^2}C_{A1}\hat{P}_1.
\end{gather}
Here we keep only the leading-order terms: $C_1/AS\sim C_{A1}$, $C_2/AS\sim C_{A2}$ and $C_3/AS\sim C_{A3}$ with $C_{A1}$, $C_{A2}$ and $C_{A3}$ being constant. Further neglecting $1/r^2$ terms, we obtain the asymptotic solutions at $r\to \infty$ as
\begin{gather}
\hat{Y}_1=C\exp \left(\beta r \right)r^{\alpha} ,
\\
\hat{P}_1=\frac{C}{rC_{A2}}\exp\left(\beta r\right)(\beta r^{\alpha}) ,
\end{gather}
with
\begin{gather}
\beta={- \sqrt{-C_{A2}C_{A3}}},
\\
\alpha=\frac{1}{2},
\end{gather}
and a common factor $C$. Note that the models with the vanishing $p$ and a constant $B_z$ also employ this asymptotic solution.
\\
\\
\textit{Case B}:  $B_\phi \propto 0, r^{-1}, r^{-2}$; $B_z\propto r^{-1}, r^{-2}$; $\Omega\propto r^{-3/2}$; $p\to 0$.
\\
As the same as Case 1, by keeping the leading order we have $C_1/AS \sim C_{B1}/r$, $C_2/AS\sim r^2C_{B2}$ and $C_3/AS\sim C_{B3}/r$. Then the following asymptotic solutions are obtained:
\begin{gather}
\hat{Y}_1=C\exp\left(\beta r^\frac{3}{2} \right)r^{\alpha} ,
\\
\hat{P}_1=\frac{C}{rC_{B2}}\exp\left(\beta r^\frac{3}{2}\right)(\beta r^{\alpha}) ,
\end{gather}
with
\begin{gather}
\beta=- \frac{2}{3}\sqrt{-C_{B2}C_{B3}},
\\
\alpha=\frac{3}{4}+\frac{1}{2}C_{B1}.
\end{gather}
\\
\\
\textit{Case C}:  $B_\phi \propto 0, r^{-1}, r^{-2}$; $B_z\propto r^{-1}, r^{-2}$; $\Omega\propto r^{-2}$; $p\to 0$.
\\
The leading order terms in this case are $C_1/AS \sim C_{C1}/r$, $C_2/AS\sim r^3C_{C2}$ and $C_3/AS\sim C_{C3}/r$. We obtain the asymptotic solution as
\begin{gather}
\hat{Y}_1=C\exp\left(\beta r^2 \right)r^{\alpha} ,
\\
\hat{P}_1=\frac{C}{r^3C_{C2}}\exp\left(\beta r^2\right)(\beta r^{\alpha}) ,
\end{gather}
with
\begin{gather}
\beta=- \sqrt{C_{C2}C_{C3}},
\\
\alpha=C_{C1}-3.
\end{gather}

\subsubsection{The matching condition at the jet boundary}
The total pressure $\hat{P}_1$ and the Lagrangian displacement $\xi_r={\hat{U}_1}/{i\sigma}$ (and hence $\hat{Y_1}$ also) should be continuous across the jet boundary. Since the solutions to Eqs. (18) and (19) are expressed as a sum of the homogeneous and inhomogeneous contributions and the latter contribution is assumed to be non-vanishing only inside the jet, we can cast the matching condition in the following form at the jet boundary:
\begin{gather}
C_{\mathrm{am}}\hat{Y}_{1,\mathrm{am,H}}= C_{\mathrm{jet}}\hat{Y}_{1,\mathrm{jet,H}}+\hat{Y}_{1,\mathrm{jet,I}},
\\
C_{\mathrm{am}}\hat{P}_{1,\mathrm{am,H}}= C_{\mathrm{jet}}\hat{P}_{1,\mathrm{jet,H}}+\hat{P}_{1,\mathrm{jet,I}},
\end{gather}
in which the subscripts ``am'' and ``jet'' refer to the ambient and jet quantities, respectively; the subscripts $\rm{H}$ and $\rm{I}$ denote, respectively, the homogeneous and inhomogeneous solutions; $C_{\rm am}$ and $C_{\rm jet}$ are constants to be determined. 

Equations (52) and (53), linear in $C_{\rm am}$ and $C_{\rm jet}$, can be solved uniquely if
\begin{equation}
\begin{vmatrix}
\hat{Y}_{1,\mathrm{am,H}} & -\hat{Y}_{1,\mathrm{jet,H}} \\
\hat{P}_{1,\mathrm{am,H}} & -\hat{P}_{1,\mathrm{jet,H}}
\end{vmatrix}
\ne 0.
\end{equation}
The values of $s$ that violate this condition correspond to unstable modes. We will consider their eigenfunctions in the next subsection. Note that Eqs. (52) and (53) depend on the initial perturbations through the inhomogeneous solution. They are not important for the asymptotic behavior at $t \rightarrow \infty$ of the unstable modes. We hence choose the simplest initial perturbation in the following: either $B_{\phi *}$ or $V_{*}$ is non-vanishing in the jet along. The choice does not have any influence on the properties of unstable modes.

\subsection{Eigenfunctions}
In the Laplace-transform analysis, we obtain the time evolution from an arbitrary initial condition in principle. In this paper, however, we are interested only in the asymptotic behavior of instabilities in the linear regime. It is characterized by the growth rate and the eigenfunction just as in the ordinary Fourier-transform analysis of normal modes. In the Laplace-transform analysis, the growth rates are obtained from the positions of poles in the complex $s$-plane, whereas the eigenfunctions are given as the homogeneous solutions at the poles. 

Once the eigenfunctions of $\hat{P}_1$ and $\hat{Y}_1$ (and hence of $\hat{\xi}_r$) are obtained, those for other quantities are derived by solving Eqs. $(9) \sim (15)$, as follows
\begin{gather}
\hat{\rho_1}=\frac{\rho_0}{\Gamma p_0} \left( \hat{p_1}+\hat{\xi_r} \frac{\partial p_0}{\partial r} \right) - \hat{\xi_r} \frac{\partial \rho_0}{\partial r},
\\
\hat{U}_1 = i \sigma \frac{\hat{Y}_1}{r} ,
\\
\begin{split}
\hat{\Omega}_1 = & \frac{\sigma}{irA} \bigg( - i\frac{m}{r}\hat{P}_1  - \frac{iF B_{\phi 0}}{\Gamma p_0}\hat{p}_1 + \\
& \left[ \frac{F B_{\phi 0}}{\Gamma p_0\sigma} \frac{\partial p_0}{\partial r}- \rho_0\Omega_0 - \rho_0 \frac{\partial}{\partial r}( \Omega_0 r) - \frac{F}{\sigma r} B_{\phi 0}\right] \hat{U}_1 \\
&+  \left[ \left(\frac{\partial B_{\phi0}}{\partial r}+ \frac{B_{\phi0}}{r}\right)+ \frac{F}{\sigma}r\frac{\partial \Omega 0}{\partial r} \right] \hat{B}_{r1}\bigg) ,
\end{split}
\\
\begin{split}
\hat{V}_1 = &  \frac{\sigma}{iA} \bigg( - ik\hat{P}_1 + \frac{iFB_{z0}}{\Gamma p_0}\hat{p}_1 +  \bigg[\frac{FB_{z0}}{\Gamma p_0 \sigma} \frac{\partial p_0}{\partial r} - \\
& \frac{F}{\sigma} \frac{\partial B_{z0}}{\partial r}\bigg]\hat{U}_1  + \bigg[\frac{\partial B_{z0}}{\partial r}+ \frac{F}{\sigma}\frac{\partial V_0}{\partial r}\hat{B}_{r1}\bigg]\bigg) ,
\end{split}
\\
\hat{B}_{r1} =  \frac{F}{\sigma}\hat{U}_1 ,
\\
\begin{split}
\hat{B}_{\phi 1} = \frac{1}{i \sigma} \bigg( & - \hat{U}_1\frac{\partial B_{\phi_0}}{\partial r} +  iFr\hat{\Omega}_1 \\
& - \frac{B_{z0}}{\gamma p_0}\bigg(p_0 - i \sigma \hat{p}_1 + \frac{\partial p_0}{\partial r}\bigg)+\hat{B}_{r1}r\frac{\partial \Omega_0}{\partial r} \bigg) ,
\end{split}
\\
\begin{split}
\hat{B}_{z1} = \frac{1}{i \sigma} \bigg( & - \hat{U}_1\frac{\partial B_{z0}}{\partial r} +iF\hat{V}_1 \\
& - \frac{B_{z0}}{\gamma p_0}\bigg(p_0 - i \sigma \hat{p}_1 + \frac{\partial p_0}{\partial r}\bigg)+\hat{B}_{r1}\frac{\partial V_0}{\partial r} \bigg) ,
\end{split}
\\
\begin{split}
\hat{p}_{1} = & \frac{1}{iS}\bigg( iA\bigg[1+\frac{F^2}{A}\bigg]\hat{P}_1 - \bigg[\sigma \rho_0\bigg(B_{\phi0}r\frac{\partial \Omega_0}{\partial r} \\
& +B_{z0}\frac{\partial V_0}{\partial r}\bigg) +F\bigg(\frac{\partial B_0^2}{2\partial^2 r }+\frac{B_{\phi 0}^2}{r}\bigg)\bigg]\hat{B}_{r1}  \\
& - \bigg[\frac{\sigma \rho_0}{r}B^2_{\phi 0}+ \frac{\sigma \rho_0}{\Gamma p_0}B_0^2\frac{\partial p_0}{\partial r}  - F \rho_0\bigg(2B_{\phi 0}\Omega_0 \\
& + B_{\phi 0} r \frac{\partial \Omega_0}{\partial r} + B_{z 0} \frac{\partial V_0}{\partial r}\bigg) - \frac{\sigma \rho_0}{2} \frac{\partial B_0^2}{\partial^2 r}\bigg]\hat{U}_1 \bigg) .
\end{split}
\end{gather}

Although not employed further in this paper, these functions are needed when we consider which mode is more preferentially induced from a particular initial perturbation (\citealt{Takahashi_2016}).

\subsection{Second-order perturbation of Hamiltonian}
The energy principle is often used to judge the stability of an equilibrium configuration. A self-adjoint Hamiltonian is defined one way or another for the linear perturbation equations and the positive definiteness of the potential involved is adopted as the condition for stability. It has been extensively exploited particularly for configurations in mechanical equilibrium without flows (\citealt{Awad_2022}). The formulation has been known for quite some time even for the stability of steady configurations with flows (\citealt{Frieman_1960}). The Hamiltonian for the perturbation equations is identified with the second-order perturbation of the Hamiltonian for MHD (\citealt{Qin_2013}). 

Unfortunately, the condition that the second-order perturbation of the Hamiltonian is positive-definite is sufficient but not necessary for the stability of a steady flow (or, equivalently, the opposite condition is necessary but not sufficient for instability). The second-order perterbation of the Hamiltonian is still useful in judging the driving mechanism of individual unstable modes. We may determine it by identifying which term in the (second-order variation of) Hamiltonian is most negative. This turns out to be very helpful indeed, particularly when more than one instabilities exist.

In terms of the Lagrangian displacement vector $\boldsymbol{\xi}$ with the $z$- and $\phi$-components incorporated, the second-order perturbation of the Hamiltonian is given as follows (\citealt{Qin_2013}):
\begin{gather}
\begin{split}
\delta ^2 H & =  \int{\rho_0 \boldsymbol{\dot\xi}^2}
+ \rho_0[(\boldsymbol{\xi} \cdot \nabla  \boldsymbol{\xi})\cdot(\boldsymbol{v_0} \cdot \nabla \boldsymbol{v_0}) \\ 
& -(\boldsymbol{v_0} \cdot \nabla \boldsymbol{\xi})^2] d^3x + \delta^2W ,
\end{split}
\\
\begin{split}
\delta ^2 W = & \int{ \bigg( (\nabla \cdot \boldsymbol{\xi})(\boldsymbol{\xi}\cdot\nabla p_0) }
+ \rho_0(\partial_\rho p)_s(\nabla\cdot\boldsymbol{\xi})^2 \\
+ & \boldsymbol{B_1}\cdot(\nabla\times\boldsymbol{B_0}) \times\boldsymbol{\xi} +|\boldsymbol{B_1}|^2 \bigg)d^3x,
\end{split}
\end{gather}
where the first three terms on the right-hand side of Eq. (63) originate from the perturbation to the kinetic energy; the last term, which is independent of the velocity $v_0$ and called the potential, is given by Eq. (64).

The third term on the right hand side of Eq. (63), which is negative-definite, has a much greater modulus than the potential terms for supersonic flows. It is mostly canceled by the first term, though, as may be understood from Eq. (69). This indicates that we need to take all three terms into account in evaluating the velocity-dependent terms. They are responsible for shear-driven instabilities such as KHI or MRI. For our background models, the latter two terms are written as
\begin{equation}
\begin{split}
\rho_0 [(\boldsymbol{\xi} \cdot & \nabla \boldsymbol{\xi})\cdot(\boldsymbol{v_0} \cdot \nabla \boldsymbol{v_0})-\rho_0(\boldsymbol{v_0} \cdot \nabla \boldsymbol{\xi})^2] = \\
& \rho_0 V_0^2 \bigg\{ \bigg(\frac{\partial \xi_r}{\partial z}\bigg)^2 + \bigg(\frac{\partial \xi_\phi}{\partial z}\bigg)^2 +\bigg(\frac{\partial \xi_z}{\partial z}\bigg)^2 \bigg\} \\
& + \rho_0 \frac{\Omega_0^2}{r^2} \bigg\{ \bigg(\frac{\partial \xi_r}{\partial \phi}\bigg)^2 + \bigg(\frac{\partial \xi_\phi}{\partial \phi}\bigg)^2  +\bigg(\frac{\partial \xi_z}{\partial \phi}\bigg)^2\\
& + \xi_r^2 + \frac{\partial\xi_r}{\partial \phi}\xi_\phi -2\frac{\partial\xi_\phi}{\partial \phi}\xi_r + r\frac{\partial\xi_r}{\partial r}\xi_r - r\frac{\partial\xi_r}{\partial z}\xi_z \bigg\} \\
& + 2\rho\frac{\Omega_0V_0}{r}\bigg\{ \frac{\partial\xi_r}{\partial \phi}\frac{\partial\xi_r}{\partial z} + \frac{\partial\xi_\phi}{\partial \phi}\frac{\partial\xi_\phi}{\partial z} + \frac{\partial\xi_z}{\partial \phi}\frac{\partial\xi_z}{\partial z} \\
& - \frac{\partial\xi_r}{\partial z}\xi_\phi  + \frac{\partial\xi_\phi}{\partial z}\xi_r \bigg\}.
\end{split}
\end{equation}
It consists of three parts, each proportional to $V_0^2$, $\Omega_0^2$ and $\Omega_0 V_0$, respectively. They will be evaluated separately. On the other hand, $\dot{\boldsymbol{\xi}}$ {in Eq. (63)} is evaluated via Eq. (69) given below.

The instability associated with the first term of Eq. (64) is called pressure-driven. The second term, which is positive-definite, is a stabilizing effect of compressibility. For our background models, these terms are given concretely as
\begin{gather}
\begin{split}
    (\nabla\cdot\boldsymbol{\xi}) & (\boldsymbol{\xi}\cdot\nabla p_0) = \\
    & \frac{\partial p_0}{\partial r} \bigg\{ \frac{\xi_r}{r}\bigg(\frac{\partial( r \xi_r)}{\partial r}\bigg) + \frac{\xi_r}{r}\frac{\partial \xi_\phi}{\partial \phi} + \xi_r\frac{\partial \xi_z}{\partial z}\bigg\},
\end{split}
\\
\begin{split}
    \rho_0 (\partial_\rho & p)_s(\nabla\cdot\boldsymbol{\xi})^2= \\
    & \rho_0(\partial_\rho p)_s \bigg\{ \frac{\xi_r}{r}\bigg(\frac{\partial( r \xi_r)}{\partial r}\bigg) + \frac{\xi_r}{r}\frac{\partial \xi_\phi}{\partial \phi} + \xi_r\frac{\partial \xi_z}{\partial z}\bigg\}^2. 
\end{split}
\end{gather}

The instability that is induced by the third term in Eq. (64) is referred to as current-driven. The fourth term, which is positive-definite and hence stabilizing, is added to the third in comparing the relative importance of different terms. The third term is written for our background models as
\begin{equation}
\begin{split}
    \boldsymbol{B_1}\cdot(\nabla &  \times\boldsymbol{B_0})\times\boldsymbol{\xi} = \\
    & - \bigg( \frac{\partial B_{z0}}{\partial r} + \bigg( \frac{\partial B_{\phi 0}}{\partial r} + \frac{ B_{\phi 0}}{r}\bigg)\xi_\phi\bigg)B_{r1} \\
    & + \bigg( \frac{\partial B_{\phi0}}{\partial r} + \frac{ B_{\phi 0}}{r} \bigg)\xi_r B_{\phi1} + \frac{\partial B_{z0}}{\partial r}\xi_r B_{z1}.
\end{split}
\end{equation}

We use the second-order perturbation of the Hamiltonian to identify the mechanism of the individual unstable modes that we find in the Laplace-Fourier linear analysis. The Lagrangian displacement vector is derived from the eigenfunctions of the velocity perturbation according to the following relation:
\begin{gather}
\delta \boldsymbol{v} = \dot {\boldsymbol{\xi}} -\boldsymbol{\xi} \cdot \nabla \boldsymbol{v}+\boldsymbol{v} \cdot \nabla \boldsymbol{\xi}.
\end{gather}
In the cylindrical coordinates, Eq. (69) is written component-wise as
\begin{gather}
 U_1 = i\sigma \xi_r,
\\
 \Omega_1r= i\sigma \xi_{\phi}- \xi_{r} r\frac{d\Omega_0}{dr},
\\
 V_1=  i\sigma \xi_z - \xi_{r} \frac{dV_0}{dr}.
\end{gather}

Before proceeding, we mention another approach  to the identification of the mechanism of various modes. It also utlizes the Lagrangian displacement $\boldsymbol{\xi}$ and was originally presented by \citet{Frieman_1960}. \citet{Bodo_2016} applied the method to a concrete model. The equation of motion for $\boldsymbol{\xi}$ is written as
\begin{equation}
\rho_0\omega^2{\boldsymbol{\xi}} - 2\rho_0i\omega(\boldsymbol{v_0\cdot \nabla}){\boldsymbol{\xi}} + \boldsymbol{G[\xi]} =0,
\end{equation}
where $\omega$ ($= - i s$) is the (complex) mode frequency and $\boldsymbol{G[\xi]}$ contains terms corresponding to different forces acting on the perturbation (see Eq. (26) in \citealt{Frieman_1960} for the concrete expression). Multiplying the complex-conjugated $\boldsymbol{\xi}^*$ and integrating the product over space, we recast Eq. (73) into the following quadratic equation in $\omega$:
\begin{equation}
\mathcal{A}\omega^2 - 2\mathcal{B}\omega + \mathcal{F} =0,
\end{equation}
with the following notations:
\begin{gather}
\mathcal{A}=\int^\infty_0d\boldsymbol{r}\ (\rho_0\boldsymbol{\xi^*\cdot\xi}), \\
\mathcal{B}=\int^\infty_0d\boldsymbol{r}\ (2\rho_0iF\boldsymbol{\xi^*\cdot\xi}), \\
\mathcal{F} = \int^\infty_0d\boldsymbol{r}\ \boldsymbol{\xi^*\cdot G[\xi]}.
\end{gather}
Then we obtain $\omega$ as
\begin{equation}
\omega = \frac{\mathcal{B}\pm\sqrt{\mathcal{B}^2-\mathcal{AF}}}{\mathcal{A}}.
\end{equation}
Since $\mathcal{A}$ is positive-definite, this expression indicates that the forces in $\boldsymbol{G[\xi]}$ that give positive contributions to $\mathcal{F}$ drive an instability (\citealt{Bodo_2016}).

\subsection{Models}

\begin{deluxetable*}{lccccccccc}
\tablenum{1}
\tablecaption{Models \label{tab1}}
\label{tablemodel}
\tablehead{\colhead{Model Name} & \colhead{$C_s$} & \colhead{$M_J$} & \colhead{$\zeta$} & \colhead{$B_{\phi J}$} & \colhead{$B_{z J}$} & \colhead{$\kappa$}  & \colhead{$\Omega_J$} & \colhead{Rotation law} & \colhead{$\xi$}}

\startdata
$\rm{DC}$ &  $\sqrt{10}$  & $ 1\sqrt{10} $ &  1 &  0.1  &  0  &  1  &  0 & no rotation & $\infty$ (discontinuous boundary) \\
$\rm{DCV4}$ &  $\sqrt{10}$  &  $4/\sqrt{10}$  &  1 &  0.1  &  0  &  1  &  0 & no rotation & $\infty$ \\
$\rm{DCd}$ &  $\sqrt{10}$  &  $1\sqrt{10}$  &  10 &  0.1  &  0  &  1  &  0 & no rotation & $\infty$ \\
$\rm{DCt}$ &  $\sqrt{10}$  &  $1\sqrt{10}$  &  0.1 &  0.1  &  0  &  1  &  0 & no rotation & $\infty$ \\
$\rm{CC}$ &  $\sqrt{10}$  & $4\sqrt{10}$  &  1 &  0.1  &  0  &  1  &  0 & no rotation & 10 \\
$\rm{DK}$ &  $\sqrt{10}$  &  $4\sqrt{10}$  &  1 &  0  &  0  &  1  &  0 & no rotation & $\infty$ \\
$\rm{CK}$ &  $\sqrt{10}$  & $4\sqrt{10}$  &  1 &  0  &  0  &  1  &  0 & no rotation & 10 \\
$\rm{DC\Omega}$ &  $\sqrt{10}$  & $4\sqrt{10}$  &  1 &  0.1  &  0  &  1  & 1 & Keplerian & $\infty$ \\
$\rm{DCB_z}$ &  $\sqrt{10}$  & $4\sqrt{10}$  &  1 &  0.1  &  0.01  &  1  & 0 & no rotation & $\infty$ \\
\hline
$\rm{DM_z K}$ &  $\sqrt{10}$  &  $1/\sqrt{10}$  &  1 &  0  &  0.01  &  1  &  1 & Keplerian ($\propto r^{-3/2}$) & $\infty$ \\
$\rm{DM_z KV01}$ & $ \sqrt{10}$  &  $0.1/\sqrt{10}$  &  1 &  0  &  0.01  &  1  &  1 & Keplerian & $\infty$ \\
$\rm{DM_z KV4}$ &  $\sqrt{10}$  &  $4/\sqrt{10}$  &  1 &  0  &  0.01  &  1  &  1 & Keplerian & $\infty$ \\
$\rm{DM_z Kk2}$ &  $\sqrt{10}$  &  $1/\sqrt{10}$  &  1 &  0  &  0.01  &  2  &  1 & Keplerian & $\infty$ \\
$\rm{DM_z K\Omega2}$ &  $\sqrt{10}$  &  $1/\sqrt{10}$  &  1 &  0  &  0.01  &  1  &  2 & Keplerian & $\infty$ \\
$\rm{DM_z KB_z02}$ &  $\sqrt{10}$  &  $1/\sqrt{10}$  &  1 &  0  &  0.002  &  1  &  1 & Keplerian & $\infty$ \\
$\rm{DM_z Kk2}$ &  $\sqrt{10}$  &  $1/\sqrt{10}$  &  1 &  0  &  0.01  &  2  &  1 & Keplerian & $\infty$ \\
$\rm{DM_z R}$ &  $\sqrt{10}$  &  $1/\sqrt{10}$  &  1 &  0  &  0.01  &  1  &  1 & Rayleigh-marginal ($\propto r^{-2}$) & $\infty$ \\
$\rm{DM_\phi K}$ &  $\sqrt{10}$  &  $1/\sqrt{10}$  &  1 &  0.02  &  0  &  1  &  1 & Keplerian & $\infty$ \\
$\rm{DM_\phi KV01}$ &  $\sqrt{10}$  &  $0.1/\sqrt{10}$  &  1 &  0.02  &  0  &  1  &  1 & Keplerian & $\infty$ \\
$\rm{DM_\phi KV4}$ &  $\sqrt{10}$  &  $4/\sqrt{10}$  &  1 &  0.02  &  0  &  1  &  1 & Keplerian & $\infty$ \\
$\rm{DM_\phi K\Omega2}$ &  $\sqrt{10}$  &  $1/\sqrt{10}$  &  1 &  0.02  &  0  &  1  &  2 & Keplerian & $\infty$ \\
$\rm{DM_\phi KB_\phi 4}$ &  $\sqrt{10}$  &  $1/\sqrt{10}$  &  1 &  0.04  &  0  &  1  &  1 & Keplerian & $\infty$ \\
$\rm{DM_\phi R}$ &  $\sqrt{10}$  &  $1/\sqrt{10}$  &  1 &  0.02  &  0  &  1  &  1 & Rayleigh-marginal & $\infty$ \\
\hline
$\rm{HH211}$ &  $0.424$  & $4.72$  &  10 &  0.4  &  0  &  2  & 0 & no rotation & $\infty$ \\
L1-0 &  $0.5$  & $1$  &  1.4 &  5  &  2.132  &  3  & 4 & Keplerian & $\infty$ \\
L1-90 &  $1$  & $-1$  &  1.3  &  1.8  &  1.1/0.85  &  4  & 4 & Keplerian & $\infty$ \\
\enddata
\tablecomments{$C_s$: the sound velocity at the jet center, $M_J$: the Mach number also at the jet center, $\zeta$: the density ratio between the jet and the ambient matter, $\xi$: the width of the jet boundary. Note that the jet velocity, or the Mach number, is actually a variable and those shown here are the base values for the models. See the text for others. The last three models are meant for a comparison either with observations of a protostellar jet or with realistic MHD simulations of CCSN jets. In models L1-0 and L1-90, the $B_{\phi}$ profile is modified so that it has a peak not at the jet boundary ($r=1$) but at $r=0.3$. In model L1-90, $B_z$ is discontinuous at the jet boundary: $B_z = 1.1$ inside the jet and $B_z = 0.85$ in the ambient matter.}
\end{deluxetable*}

In this subsection, we give the concrete background settings we adopt in this paper. We consider steady, infinitely long, cylindrical jets surrounded by ambient matter. Axisymmetry is assumed for the entire system and all quantities are functions of $r$, the distance from the z-axis, alone. The density profile is either piecewise constant or continuous with a narrow transition region. In the former case, the jet and ambient matter are individually uniform, having different constant densities. In the latter case, the density profile is given by 
\begin{equation}
 \rho = \rho_{J}\tanh (-\xi(r-1)+\rho_{c}),
\end{equation}
where $\xi$ is the parameter to control the width of the transition layer; $\rho_J$ and $\rho_c$ are constants to specify the asymptotic values. We call the radius $r=1$ the jet radius in both cases. The discontinuity is located there in the former whereas the transition layer is centered at this radius in the latter.

The z-component of velocity, $V$, is also either piecewise constant or continuous. In the former case, the ambient matter is at rest. In the latter case, the radial profile is given by 
\begin{equation}
V = V_{J}\tanh (-\xi(r-1)+V_{c}) ,
\end{equation}
where $\xi$ is common to the density profile and the constants $V_J$ and $V_c$ are chosen so that the velocity approaches zero as $r$ goes to infinity.

We consider again two radial profiles of $B_z$, the z-component of magnetic field. In the first case, $B_z$ is constant over the whole region. In the second case, we assume a decaying $B_z$ as follows:
\begin{equation}
B_z = \frac{B_{zJ}}{\kappa r^a+1},
\end{equation}
where $a=1,2$ and $\kappa$ is a constant. 

As for {$B_{\phi}$}, the azimuthal component of magnetic field, we assume $B_{\phi}\propto r$ at $r \to 0$ and $B_{\phi}\propto r^{-n}$ at $r\to \infty$ with $n=1,2$, and take the following radial profile:
\begin{equation}
B_{\phi}=\left \{
\begin{aligned}
& B_{\phi J}\sin (\frac{\pi}{2}r)  \ \  (r<1) , \\
& \frac{B_{\phi J}}{\kappa(r-1)^n+1} \ \   (r>1).
\end{aligned}
\right.
\end{equation}

The angular velocity $\Omega$ is constant at $r\to 0$, and it should approach $0$ as $r$ goes to infinity. We hence adopt the following angular velocity profile:
\begin{equation}
\Omega = \frac{\Omega_{J}}{\kappa r^b+1},
\end{equation}
with $\Omega_{J}$ and $\kappa$ being constants; we choose $b=3/2$ (Keplerian-type rotation) or $b=2$ (Rayleigh-marginal rotation).

Finally, the matter pressure is set so that the mechanical equilibrium should hold:

\begin{equation}
\frac{dp}{dr}=-\frac{B_{\phi}^2}{r}-\frac{d\frac{1}{2}B^2}{dr} + \rho\Omega^2 r .
\end{equation}
The adiabatic constant is fixed to $5/3$ in this study.

In Table \ref{tablemodel}, we summarize for all models the constants that appear in the radial profiles given above: $C_s = \sqrt{{\Gamma p_J}/{\rho_J}}$ is the sound velocity; $M_J = V_J/C_s$ is the Mach number; $\zeta = \rho_J/\rho_A$ is the ratio of the density of jet to that of ambient matter. The initial letter $D$ in the model name indicates that the jet boundary is sharp with $V$ and $\rho$ discontinuous, whereas letter $C$ is used for the models that have a continuous jet profile with $\xi=10$. The second letter denotes the instability expected naively for the model: $C$ stands for the current-driven instability; $K$ means the Kelvin-Helmholtz instability; $M_\phi$, $M_z$ represent the magneto-rotational instability induced by $B_\phi$ and $B_z$, respectively. The subsequent letters indicate the values of parameters employed. Note that the real natures of unstable modes may not coincide with those suggested by the model names. They are determined by the Hamiltonian analysis.

\section{Results} \label{sec:floats}
\subsection{Poles on the complex plane}
\begin{figure}
    \centering
    \includegraphics[scale=0.16]{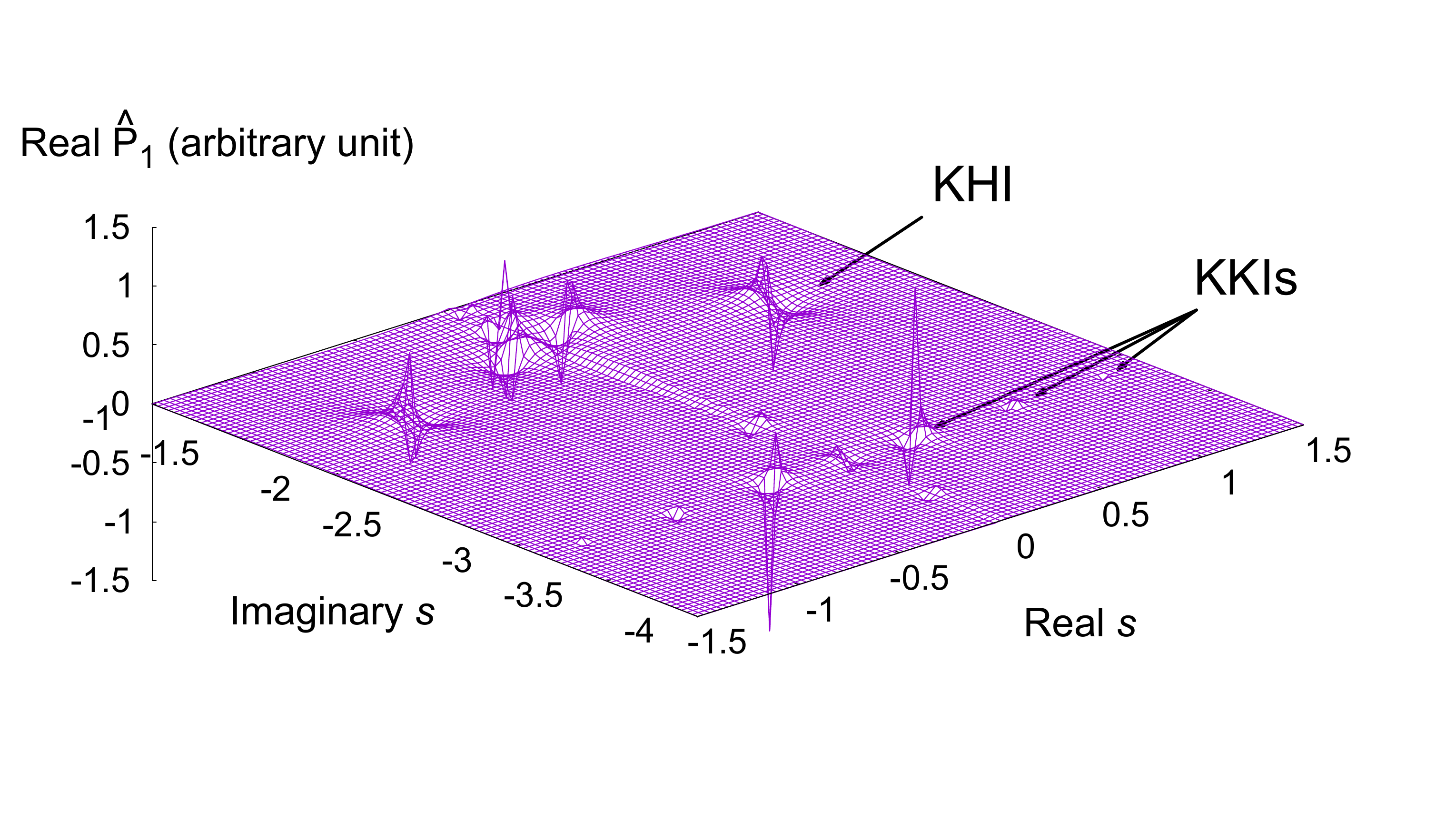}
    \caption{The real part of $\hat{P}_1$ in the complex $s$ plane for model $\rm{DC}$ with $m=1$ and $k=2\pi$. The unit of the vertical axis is arbitrary. There are a couple of unstable and stable modes that exist. See the text for details.}
    \label{KHplot}
\end{figure}

We first look at how unstable modes are obtained in the Laplace-Fourier linear analysis, using model DC as a fiducial case. In this model, the jet is subsonic and non-rotating with the magnetic field having a toroidal component alone. We consider the perturbation with $m=1$ and $k=2\pi$. As for the initial condition, $I_{*}^{Y}$ and $I_{*}^{P}$, we set $B_{\phi *}=0.01$, which is sufficient in this case. The choice is rather arbitrary but does not affect the result as mentioned earlier. For each $s$ given, Eqs. (18) and (19) are solved to obtain $\hat{Y}_1$ and $\hat{P}_1$ as a function of $r$ (see Section 2.2). In Figure \ref{KHplot}, we present the real part of $\hat{P}_1$ at the jet boundary $r=1$ on the complex $s$ plane as a 3D plot. 

One recognizes that there are several points, where a pair of peaks and dips occur side by side. They are actually singular points (first-order poles), at which condition (54) is violated and there is a nontrivial solution for the homogeneous equations without $I_{*}^{Y}$ and $I_{*}^{P}$ for the given boundary conditions. This homogeneous solution is nothing but the eigenfunction of an unstable mode. In this particular model, several unstable modes exist simultaneously. As indicated in the figure, they are driven by different mechanisms. Some of them are higher overtones with different numbers of radial nodes. In the Laplace analysis, the growth rate and oscillation frequency of an unstable mode are given as the real and imaginary part of the pole location. Those singular points with positive values of the real part of $s$ give instability. Unlike the normal modes of stationary configurations, they are in general overstabilizing modes, i.e., an oscillation accompanies the exponential growth of their amplitude. Note that to each pole there is another with the opposite sign of Re $s$, which is clearly seen in Figure~\ref{KHplot}.

In the same figure, the pole labeled with KHI is located at $s\approx 1.1 - 1.9i$, hence having the growth rate of $\approx1.1$ and the oscillation frequency of $\approx 1.9$. This mode is driven by the shear motion, which is why it is labeled with KHI. Besides this KH mode, there are three other poles, that are current-driven instead (hence labeled with KKIs). They have similar values $\approx 3.3$ of the imaginary part (and hence the oscillation frequencies). They are the fundamental mode and two overtones, among which the former has the largest growth rate of $\approx 1.4$. There is yet another pole at $s\approx 0.1-1.5i$, which is probably the first overtone of the KH mode mentioned above. There are many other poles on the imaginary axis with no real part. They are purely oscillating modes. One also finds some discontinuities possibly associated with branching singularities.

The finite heights (depths) of the peaks (dips) at the poles are an artifact of the finite resolution in $s$ in the figure and should be infinite actually. Those finite heights or depths still reflect the residues at the poles, which are in turn dependent on the initial condition, $I_{*}^{Y}$ and $I_{*}^{P}$. In fact, the residue of each pole is given by the overlap integral of the initial perturbation with the eigenfunction for pole (\citealt{Takahashi_2016}); the higher (deeper) {the} peak (dip) is, the more strongly the mode is induced more strongly by the given initial perturbation. Although this is important information on its own, we will not consider it in this paper. Note that the vertical axis of Figure \ref{KHplot} is adjusted so that the peaks and dips could be observed well; other features are suppressed and indiscernible although they are existent actually. Incidentally, KHIs for the same configuration were studied by \citet{Appl_1992}; we confirm that their dispersion relation (see Eq. (79) in \citealt{Appl_1992}) obtained from the Fourier analysis is consistent in high accuracy with the pole positions we find here.

\subsection{Hamiltonian analysis}

In Section 3.1, we employ the second-order perturbation of the MHD Hamiltonian, $\delta^2 H$ (Eq. (63) in Section 2.4), to identify the driving mechanism of the individual unstable modes in Figure \ref{KHplot}. This method is very useful and sometimes indispensable indeed for mode identification as we demonstrate in the following sections.

We divide $\delta^2 H$ into three parts: the shear part, $\delta^2H_{\rm sh}$, corresponds to the sum of the first three terms on the right-hand side of Eq. (63):
\begin{equation}
\delta^2H_{\rm sh} =\int \big( \rho_0 \boldsymbol{\dot\xi}^2 + \rho_0[(\boldsymbol{\xi} \cdot \nabla  \boldsymbol{\xi})\cdot(\boldsymbol{v_0} \cdot \nabla \boldsymbol{v_0})
-(\boldsymbol{v_0} \cdot \nabla \boldsymbol{\xi})^2]\big)d^3x;
\end{equation}
the second part, $\delta^2 H_{\rm pr}$, is the pressure-related terms given by the first two terms in Eq. (64):
\begin{equation}
\delta^2H_{\rm pr} =\int\big( (\nabla \cdot \boldsymbol{\xi})(\boldsymbol{\xi}\cdot\nabla p_0) + \rho_0(\partial_\rho p)_s(\nabla\cdot\boldsymbol{\xi})^2\big)d^3x;
\end{equation}
the third part is the terms that involve the magnetic field, that is, the sum of the last two terms in Eq. (64):
\begin{equation}
\delta^2H_{\rm B} = \int \big(\boldsymbol{B_1}\cdot(\nabla\times\boldsymbol{B_0}) \times\boldsymbol{\xi} +|\boldsymbol{B_1}|^2 \big)d^3x.
\end{equation}
As explained in Section 2.4, $\delta^{2} H$ should fail to be positive-definite if the system is unstable. The important thing here is that the eigenfunction for an unstable mode does not give a negative value to $\delta^2 H$. It is vanishing instead. This is understandable if one recalls that $\delta^2 H$ is a conserved quantity and hence constant in time whereas all its terms grow exponentially in time. This is possible only if the sum of these terms vanishes. Since the first term of Eq. (63), which may be interpreted as the kinetic energy of the perturbation, is positive definite, there should be negative term(s) that would compensate it. If $\delta^2 H_{\rm {sh}}$ is the most negative term in $\delta^{2} H$, then we judge that the instability is shear-driven, i.e., either KHI or MRI. They are distinguished further according to which term in Eq. (65) is the most negative: if it is the term proportional to $V_0^2$, the instability is KHI, whereas it is MRI if the most negative term is the one proportional to $\Omega_{0}^{2}$ (see Eq. (65)). If $\delta^2 H_{\rm pr}$ is the dominant negative contributor to $\delta^2 H$ instead, we consider that the instability is pressure-driven. The instability is called current-driven when $\delta^2 H_{\rm B}$ overwhelms the other two terms as the negative contributor to $\delta^2 H$. Among them are the sausage and kink modes.

In the following, we will pay particular attention to how the nature of unstable modes changes along a one-parameter sequence. We demonstrate that the mode nature can change continuously from one to another; there also occurs an exchange of their characteristics between two modes after they come close to each other in the complex plane.

\subsubsection{Mode exchange of KHI and KKI}

\begin{figure}
\centering
\subfigure[]{\includegraphics[scale=0.16]{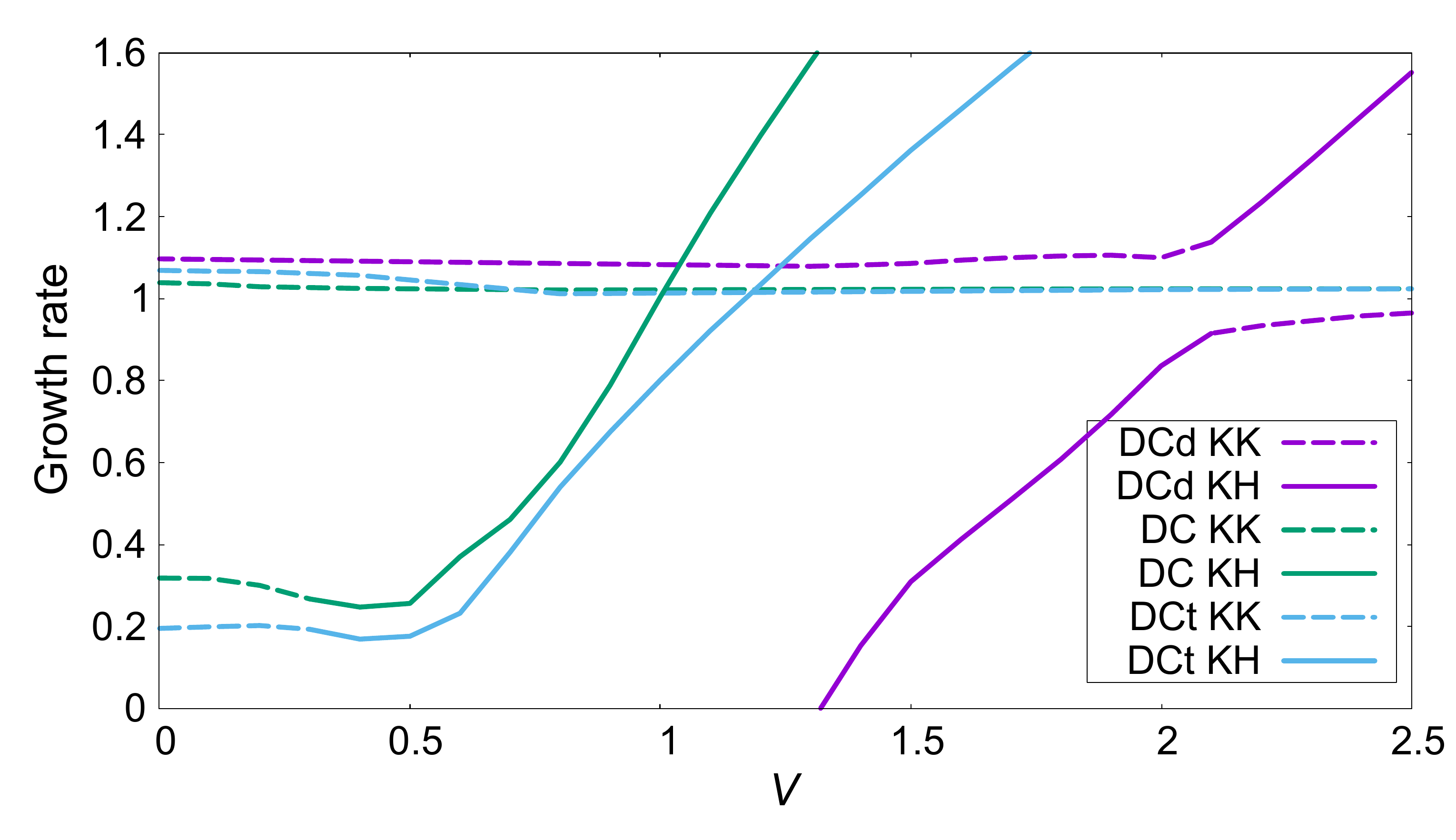}\label{KHv}}
\subfigure[]{\includegraphics[scale=0.165]{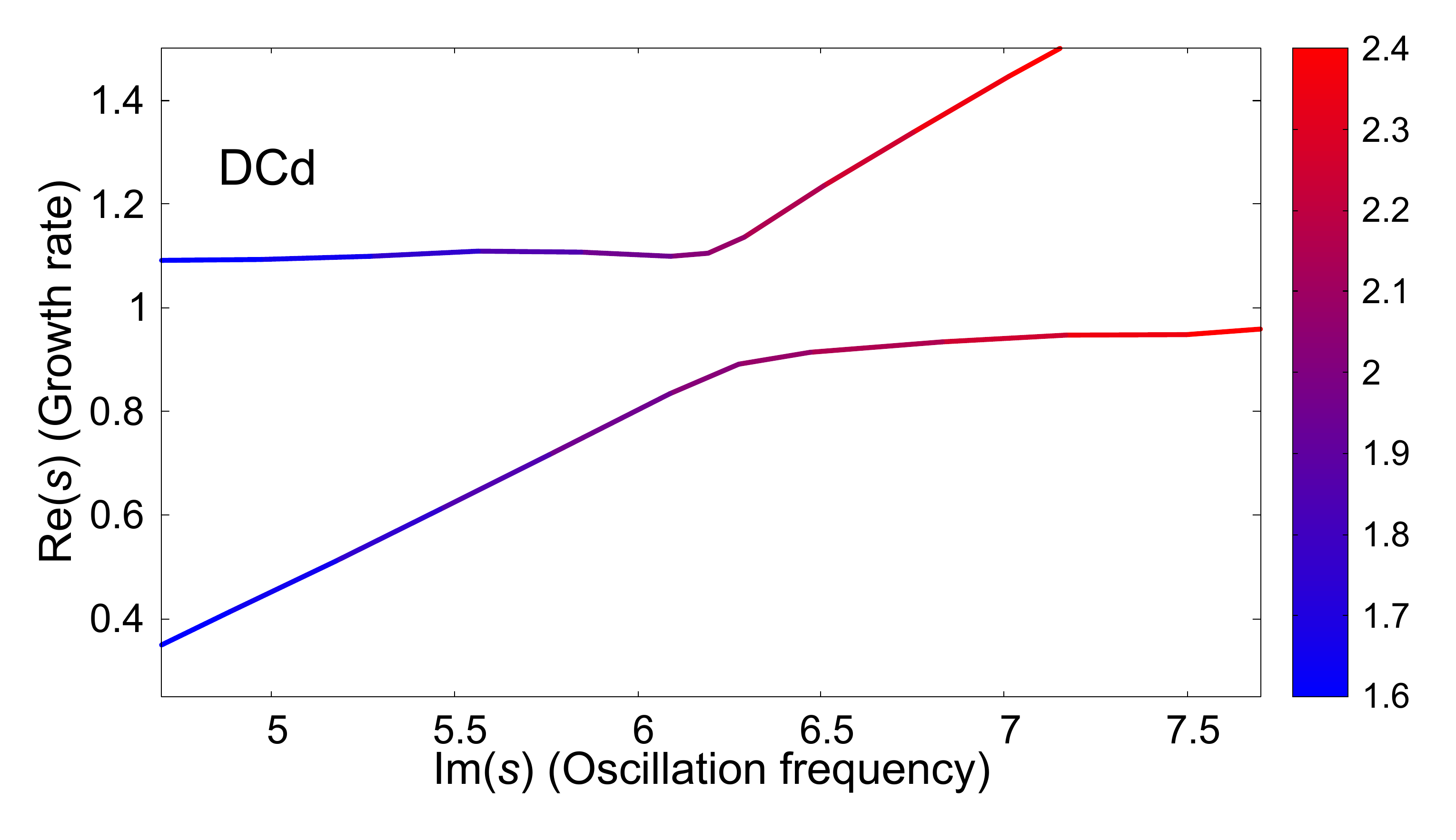}\label{KKKHGroOsc}}
    \caption{(a) Growth rates of the unstable modes with $k = \pi$ and $m = 1$ in models $\rm{DC}$ (green), $\rm{DCd}$ (purple), and $\rm{DCt}$ (blue) as a function of the jet velocity. The Kelvin-Helmholtz modes are shown with solid lines whereas the kink modes are presented with dashed lines. (b) Trajectories of the two unstable modes in model $\rm{DCd}$ that exchange their mode natures ($\rm{DCd}$ KK and $\rm{DCd}$ KH in panel (a)). The line colors indicate the jet velocity $V$: bluish (reddish) colors denote lower (higher) velocities.}
\end{figure}

\begin{figure}
\centering
\subfigure[]{\includegraphics[scale=0.16]{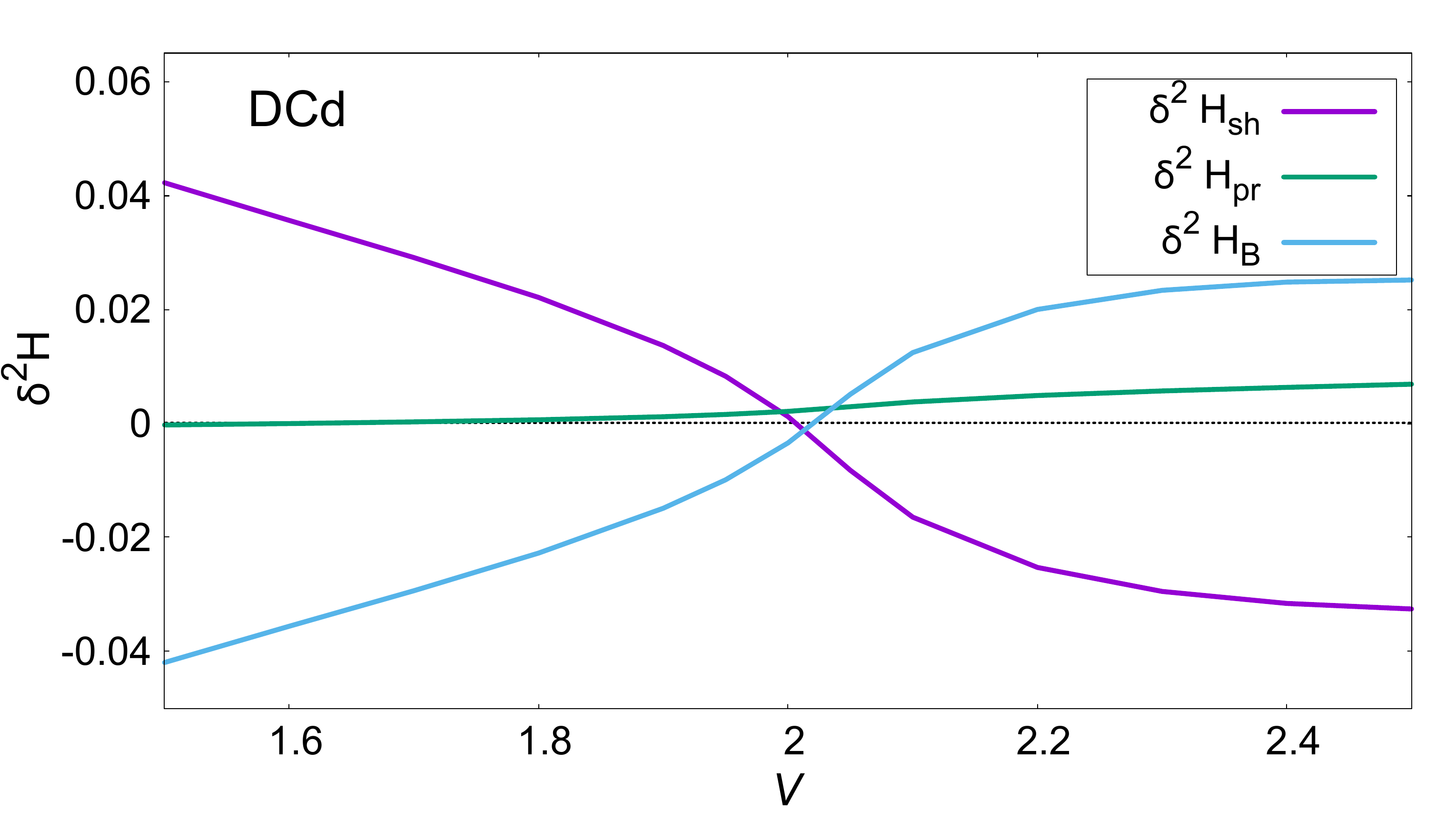}\label{KHHamil1}}
\subfigure[]{\includegraphics[scale=0.16]{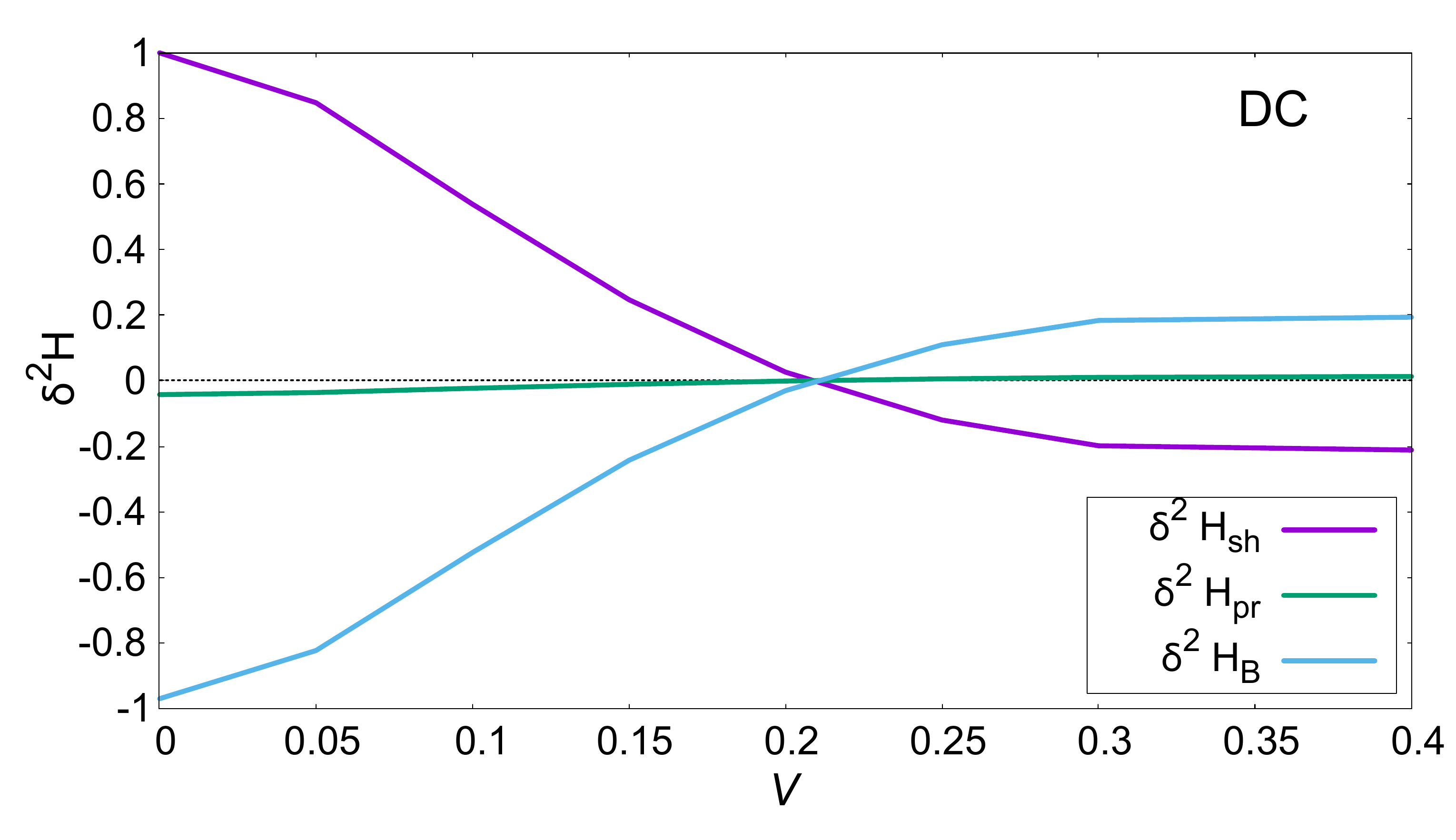}\label{KHHamil2}}
    \caption{The three components, $\delta^2 H_{\rm sh}$ (purple), $\delta^2 H_{\rm pr}$ (green), and $\delta^2 H_{\rm B}$ of the second order perturbation of the Hamiltonian $\delta^2 H$ for (a) the first mode in model $\rm{DCd}$ and (b) model $\rm{DC}$ as a function of $V$. They are normalized by the kinetic energy $\int{\rho_0 \boldsymbol{\dot\xi}^2} d^3x $.}
\end{figure}

\begin{figure}
\centering
\subfigure[]{\includegraphics[scale=0.16]{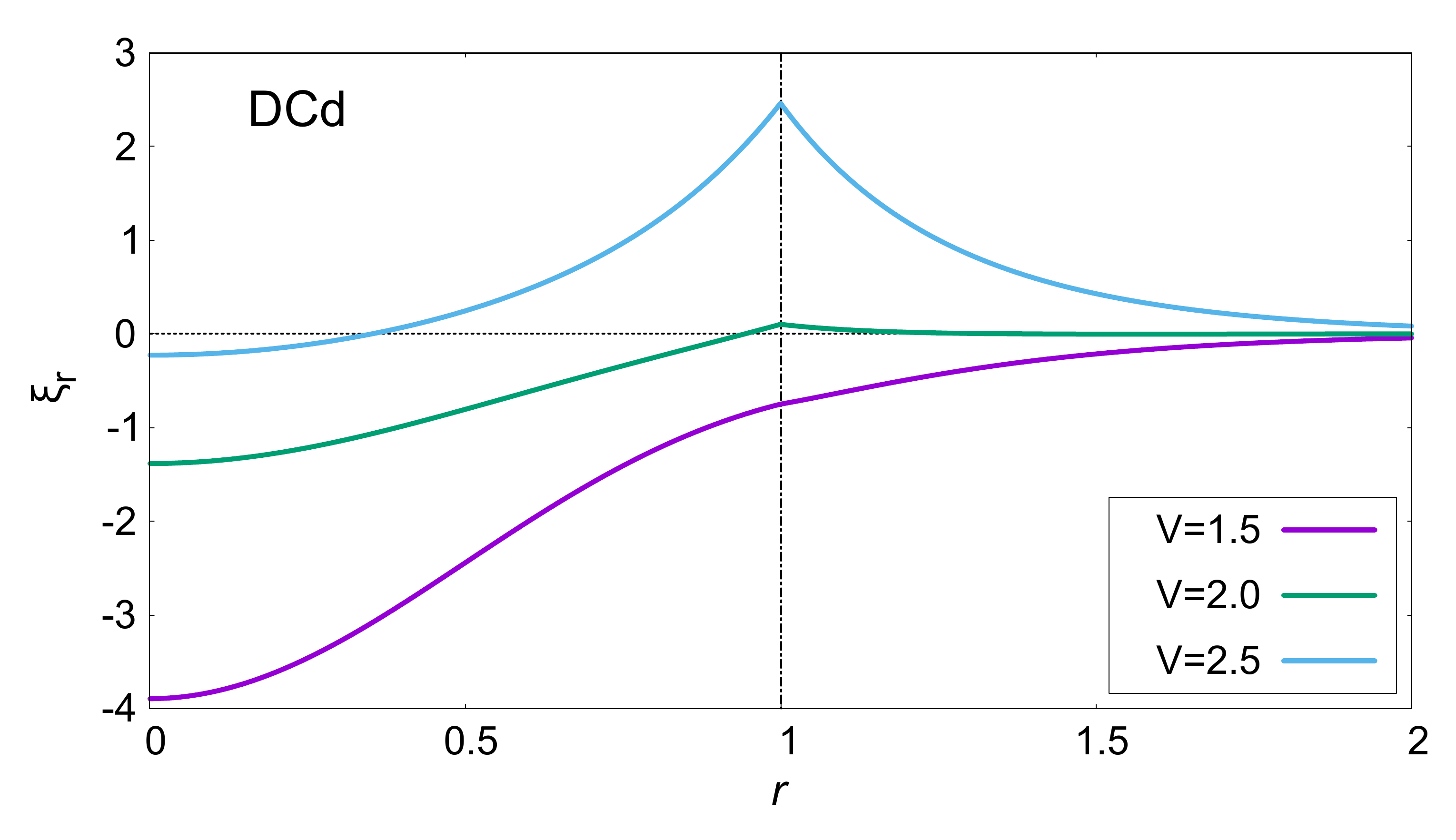}\label{KHeigen1}}
\subfigure[]{\includegraphics[scale=0.16]{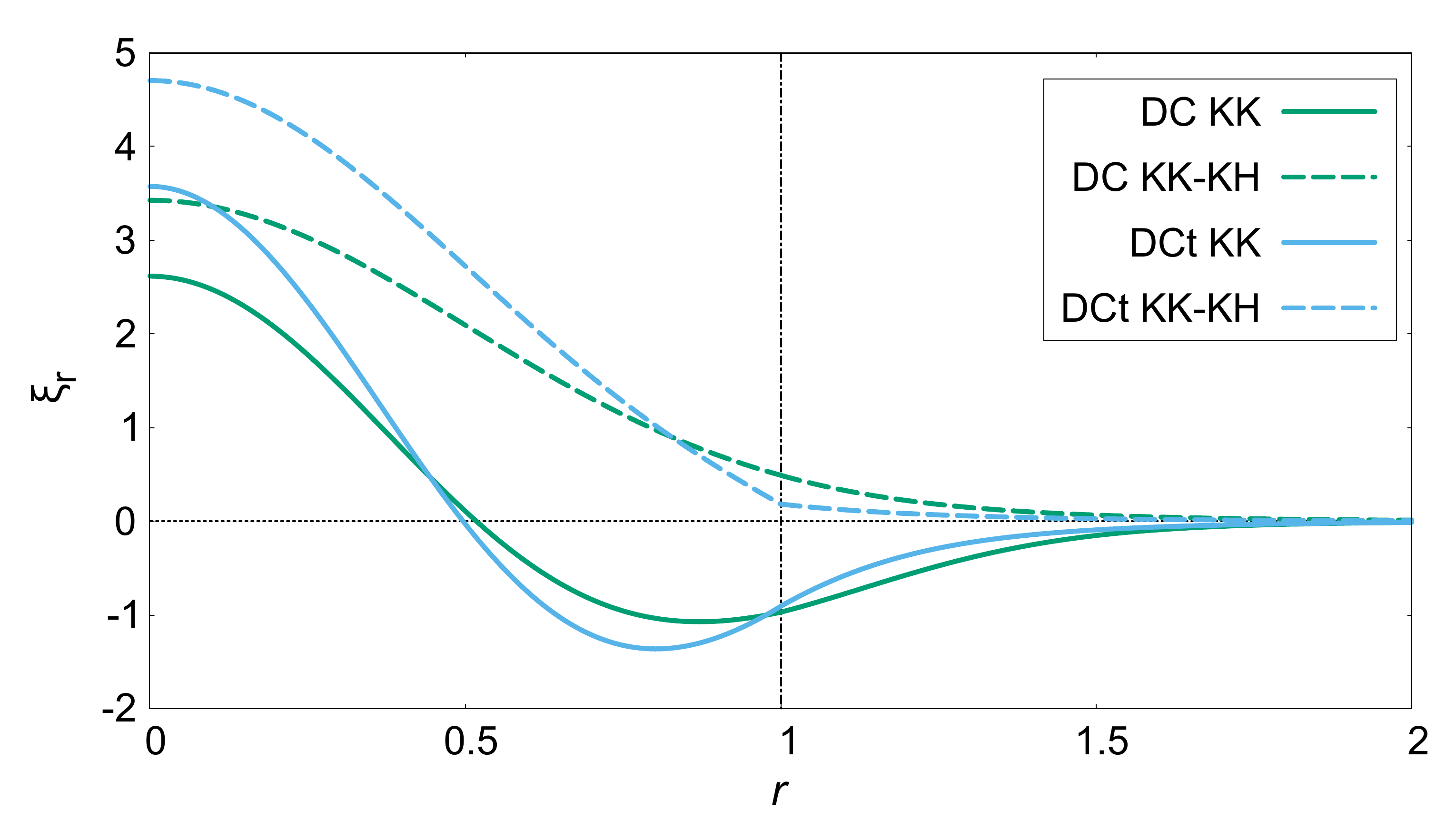}\label{KHeigen2}}
\subfigure[]{\includegraphics[scale=0.16]{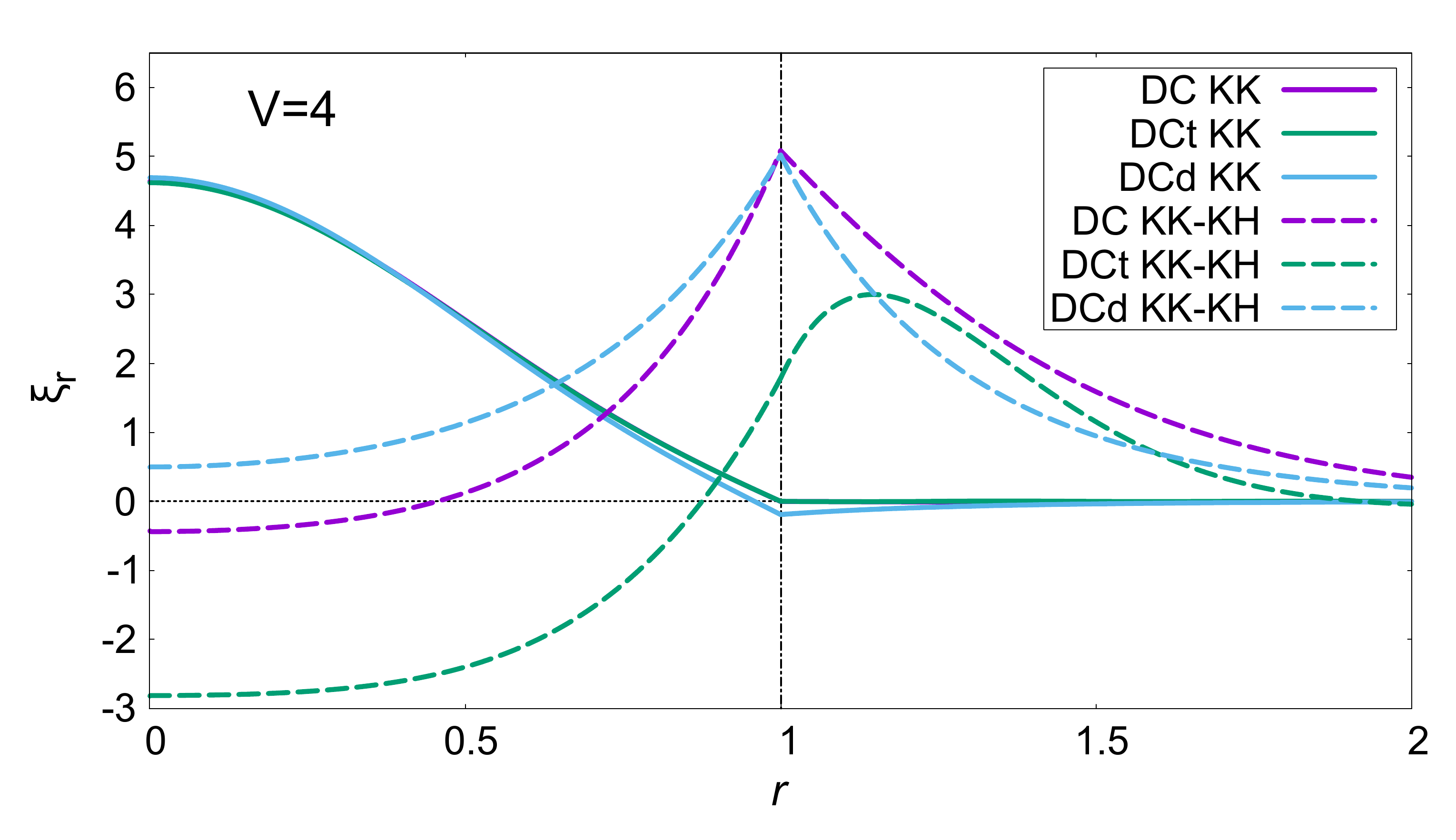}\label{KHeigen3}}
    \caption{The eigenfunctions of the radial displacement $\xi_r$ as functions of the cylindrical radius $r$ for different jet velocities: (a) the first unstable mode in model $\rm{DCd}$ (see Fig. 2a); (b) the two unstable modes, i.e., the fundamental mode (dashed) and the first overtone mode (solid) for $\xi_r$ model $\rm{DC}$ (green) and $\rm{DCt}$ (blue) at $V = 0.1$ (c) the first (solid) and second (dashed) unstable modes of models $\rm{DC}$ (purple), $\rm{DCt}$ (green) and $\rm{DCd}$ (blue) at $V = 4$. Note that the purple solid line is overlapped with the green solid line.}
    \label{F4}
\end{figure}

We will first study the pair of KHI and KKI. We take model $\rm{DC}$ again but pay attention to the modes with $k=\pi$ and $m=1$. We also consider models DCd and DCt, which differ from model DC only in the density ratio between the jet and the ambient matter: it is 10 times larger (smaller) for the former (latter). The jet is non-rotating and has only the toroidal magnetic field in all these models. We focus on a pair of two unstable modes in each model. We will study their natures based on the second-order perturbation of Hamiltonian as we explained. We are particularly interested in how they change along a continuous sequence.

In Figure \ref{KHv}, we show the growth rates of those pairs for the three models as a function of the jet velocity. Those that are judged to be KK modes in the Hamiltonian analysis are exhibited with dashed lines whereas those regarded as KH modes are displayed with solid lines. Most of the lines consist of horizontal and slanted portions, which are smoothly connected with each other. Intriguingly, the two portions correspond to different mechanisms: on the horizontal part, the instability is mostly current-driven whereas the shear feeds the mode on the slanted part.

This is particularly apparent in model DCd, which is presented in purple in Figure \ref{KHv}. One of the two modes is unstable for all jet velocities whereas the other mode becomes unstable at high jet velocities, $V \gtrsim 1.3$. The growth rate of the first mode is almost independent of the jet velocity at small values of $V$. This is expected for the KK mode. The Hamiltonian analysis indicates unambiguously that this is a KK mode indeed, with $\delta^2 H_{\rm B}$ being the most negative. Recall that the KK modes are shown with dashed lines in the figure.

On the other hand, once it becomes unstable, {the second mode} has a growth rate that increases monotonically with the jet velocity. Such behavior is expected for the KH mode. It is again confirmed by the Hamiltonian analysis: this time $\delta^2 H_{\rm sh}$ is the most negative. The KH modes are shown with solid lines in the figure. 

The situation changes drastically at $V \approx 2$, where the growth rates of the two modes come close to each other. The growth rate of the first mode starts to go up with the jet velocity whereas the growth rate of the second mode levels off. The natures of the two modes look exchanged there: the first mode is a KH mode now while the second mode becomes a KK mode. It is indeed the case as the Hamiltonian analysis clearly demonstrates. In Figure \ref{KHHamil1} we show $\delta^2 H_{\rm sh}$, $\delta^2 H_{\rm B}$ and and $\delta^2 H_{\rm pr}$ for the first mode as a function of the jet velocity $V$. We normalize them by the kinetic term: $\int{\rho_0 \boldsymbol{\dot\xi}^2} d^3 x$. The position of the most negative term occupied by $\delta^2 H_{\rm B}$ at small values of $V$ is taken over by $\delta^2 H_{\rm sh}$ at $V \gtrsim 2.0$. The opposite is true for the second mode. Note that $\delta^2 H_{\rm sh}$, $\delta^2 H_{\rm pr}$ and $\delta^2 H_{\rm B}$ are zero-sum in $\delta^2 H$.

The exchange of the mode natures finds further support from {their} eigenfunctions. We plot in Figure \ref{KHeigen1} the radial displacement $\xi_r$ in the first mode as a function of the cylindrical radius for {three} jet velocities of $V=1.5$ (purple), $2.0$ (green) and $2.5$ (blue). They correspond to the horizontal, transitional and slanted parts of the growth-rate curve for the same mode in Figure \ref{KHv}. At $V = 1.5$, the displacement is the largest at the jet center, indicating that this is a KK mode; at $V = 2.5$, on the other hand, the peak of $\xi_r$ shifts to the jet boundary and the displacement of the jet center is much diminished, just as expected for the KH mode. The eigenfunction at $V=2.0$ is of intermediate nature with the displacement declining at the center and developing at the jet boundary.

We plot in Figure \ref{KKKHGroOsc} the trajectories of the two modes in the complex $s$ plane. The line colors indicate the jet velocity: bluish colors denote lower velocities while reddish ones represent higher velocities. The real and imaginary axes correspond, respectively, to the growth rate and the oscillation frequency of the mode. The close encounter of the the two modes coincides with the change (actually the exchange) of the mode natures.

The picture is different for models $DC$ and $DCt$. There exist two unstable modes all the way down to $V=0$. They are both horizontal at low jet velocities $V \lesssim 0.5$ and the Hamiltonian analysis indicates that they are both KK modes. One of the two modes, which we refer here to as the first mode,  remains horizontal up to high jet velocities $V\approx 2.5$. The other mode, or the second mode, changes its nature at $V \approx 0.5$, above which the Hamiltonian analysis indicates that it is a KH mode. In fact, the growth-rate curve is bent upward and increases with the jet velocity as expected. We plot the three components of the second variation of the MHD Hamiltonian for the second mode in Figure \ref{KHHamil2}. Interestingly, the transition happens around $V\approx 0.2$, much earlier than the growth-rate curves suggest. The growth rate of the second mode increases monotonically at $V \gtrsim 0.5$ and takes over the growth rate of the first mode, which is insensitive to the jet velocity. 

This time there is no exchange of the mode natures. This may seem to contradict the previous result for model $\rm{DCd}$ but actually it does not. The positions of these two modes in the complex $s$ plane do not come close to each other at all. The oscillation frequencies (or the imaginary parts of $s$ at the poles) are quite different between the two modes unlike in model DCd, in which both the growth rates and oscillation frequencies are simultaneously similar for the two modes. It seems that a close encounter is not necessarily a driver of the mode change. In fact, we will see another such example later for an MRI, which turns into {a} KHI continuously without close encounters.

We plot the eigenfunctions $\xi_r$ for models $\rm{DC}$ and $\rm{DCt}$ at $V=0.1$ in Figure \ref{KHeigen2}. All four unstable modes have peaks at the jet center, indicating that they are all KK modes in agreement with the Hamiltonian analysis given above. It is apparent from the absence or presence of a radial node that the first mode is the fundamental mode and the second mode is the first overtone. 

In Figure \ref{KHeigen3}, on the other hand, we plot $\xi_r$ at $V=4$. The first (solid) and second (dashed) unstable modes of models $\rm{DC}$ (purple), $\rm{DCt}$ (green) and $\rm{DCd}$ (blue) {are presented}. Having the greatest radial displacement at the jet center, the first modes are of KK-mode nature, as the Hamiltonian analysis indicates. They are similar to one another although one of them (model $\rm{DCd}$) is the first overtone having one radial node. On the other hand, the second modes are of KHI nature, having the largest amplitude of $\xi_r$ at the jet boundary. This conforms to the conclusion of the Hamiltonian analysis again.

Incidentally, we do not find any unstable sausage mode{, i.e.,} the current-driven instability with $m=0${,} for the above three models. We summarize the results of some Hamiltonian analyses for {the} modes with $m=0$ in Table \ref{tableKHKK}. For comparison, we include the results for the KK modes {or} the current-driven instability with $m=1$. The perturbed current gives positive values to $\delta^2 H_{\rm B}$ for the axisymmetric modes in these models whereas $\delta^2H_{\rm sh}$ is always a negative contributor. The unstable axisymmetric modes are hence all KH modes.

\begin{deluxetable}{lccclll}
\tablenum{2}
\tablecaption{The Hamiltonian analysis of the modes with $m=0, 1$ in models $\rm{DC}$ and $\rm{DCV4}$.}
\label{tableKHKK}
\tablehead{
\colhead{Model} & \colhead{$m$} & \colhead{$k$} & \colhead{Mode} & \colhead{$\delta^2H_{\rm sh}$} & \colhead{$\delta^2H_{\rm pr}$} & \colhead{$\delta^2H_{\rm B}$} 
}
\startdata
$\rm{DC}$ & $0$ & $\pi$& KH & -0.166134 & 0.0379143 & 0.125261  \\
$\rm{DC}$ & $0$ & $4\pi$ & KH & -0.0493603 & 0.0252706 & 0.0114536\\
$\rm{DCV4}$ & $0$ & $\pi$ & KH & -0.333822 & 0.290655 & 0.0394584 \\
$\rm{DCV4}$ & $1$ & $\pi$& KK & 0.00646382 &  -4.35964e-05 & -0.00642571  \\
$\rm{DCV4}$ & $1$ & $\pi$ & KH & -0.31611 & 0.0483725 & 0.263843  \\
\enddata
\tablecomments{The Hamiltonian is normalized by the kinetic energy $\int{\rho_0 \boldsymbol{\dot\xi}^2} d^3x $.}
\end{deluxetable}

\subsubsection{Mode change of MRI in rotational jets}
Next, we consider unstable modes in rotational jets. In addition to KHI and KKI, MRI can exist. Since it feeds on the shear in rotation, $\delta^2 H_{\rm sh}$ is expected to be the most negative contributor to the Hamiltonian for MRI. In order to distinguish it from KHI, which is also driven by shear motions, we look {into} $\delta^2 H_{\rm sh}$ more in detail. It is in fact composed of three terms, proportional to $V_0^2$, $\Omega_0^2$ and $V_0 \Omega_0$, respectively (see Eq. (65)). If the term proportional to $V_0^2$ is the most negative, then the mode is judge to be {a} KHI whereas the term proportional to $\Omega_0^2$ is the most negative contributor, we regard it as an MRI. The cross-term proportional to $V_0 \Omega_0$ becomes negative always in association with one of the above two terms when it is negative. MRI occurs for both the poloidal $(B_z)$ and toroidal ($B_{\phi})$ magnetic fields. They are stretched in the radial direction by matter motions in the instability (\citealt{Balbus_1998}).

\begin{figure}
\centering
\subfigure[]{\includegraphics[scale=0.16]{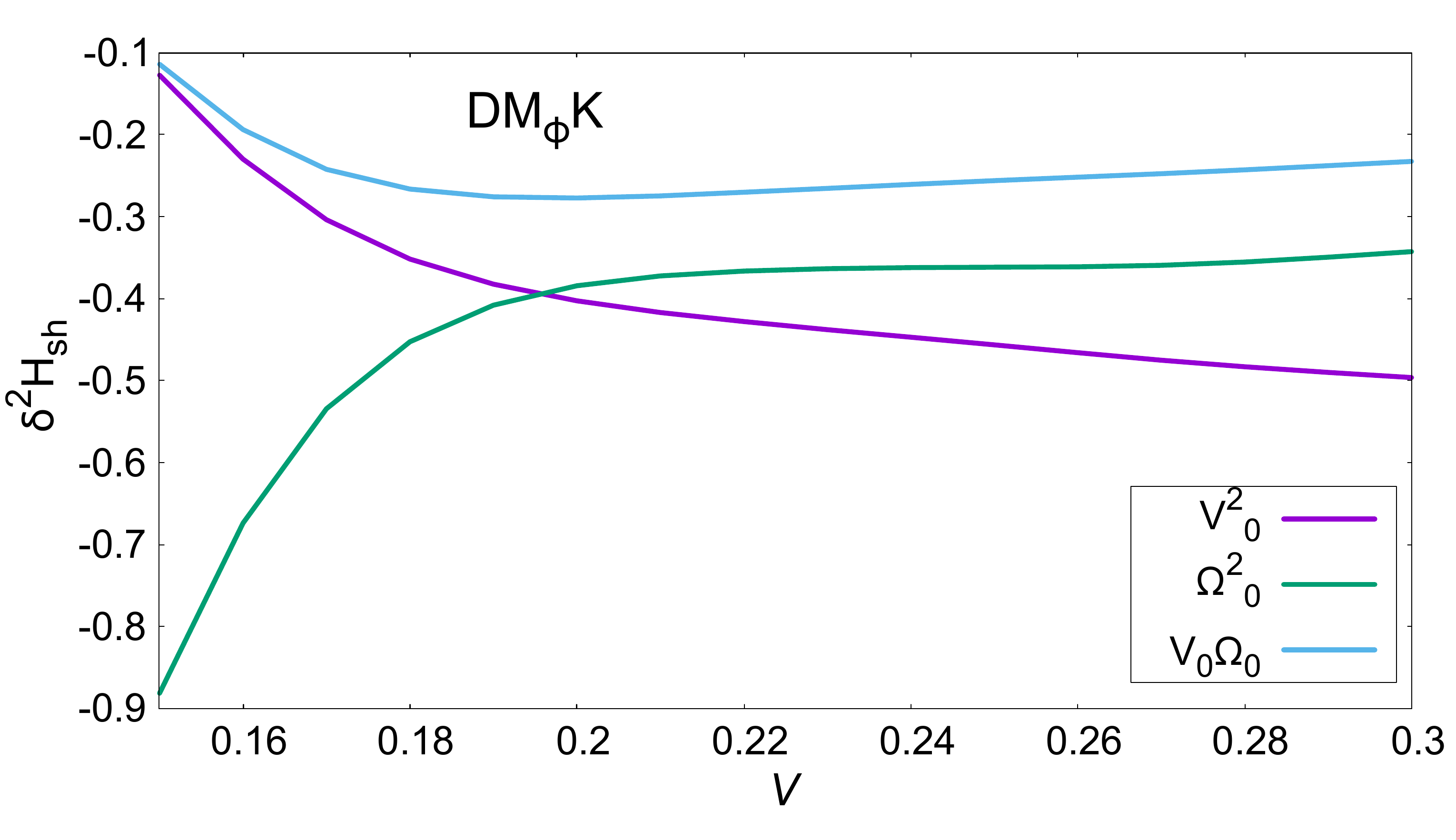}\label{MRKHHamil}}
\subfigure[]{\includegraphics[scale=0.16]{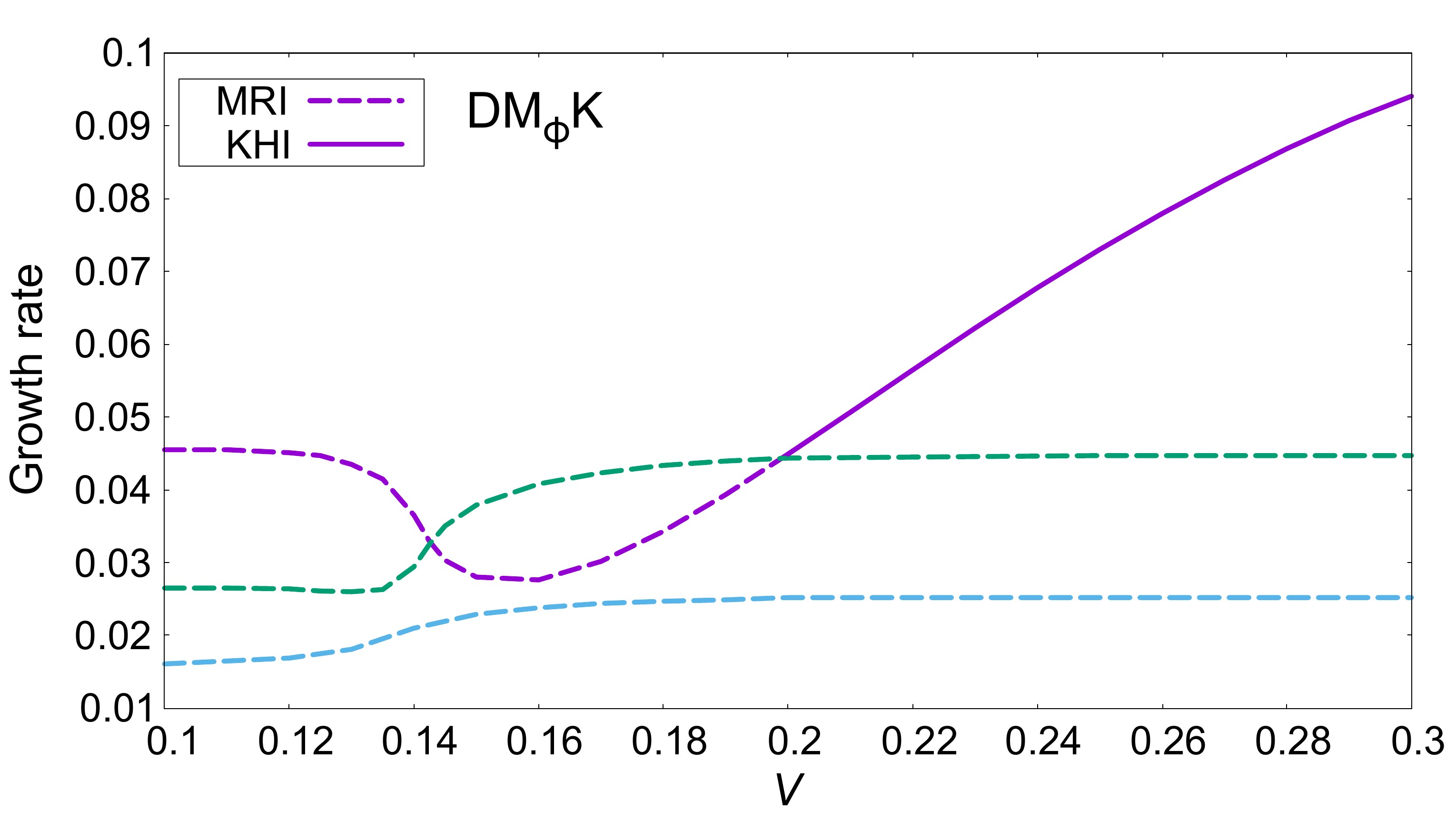}\label{MRKHg}}
\subfigure[]{\includegraphics[scale=0.16]{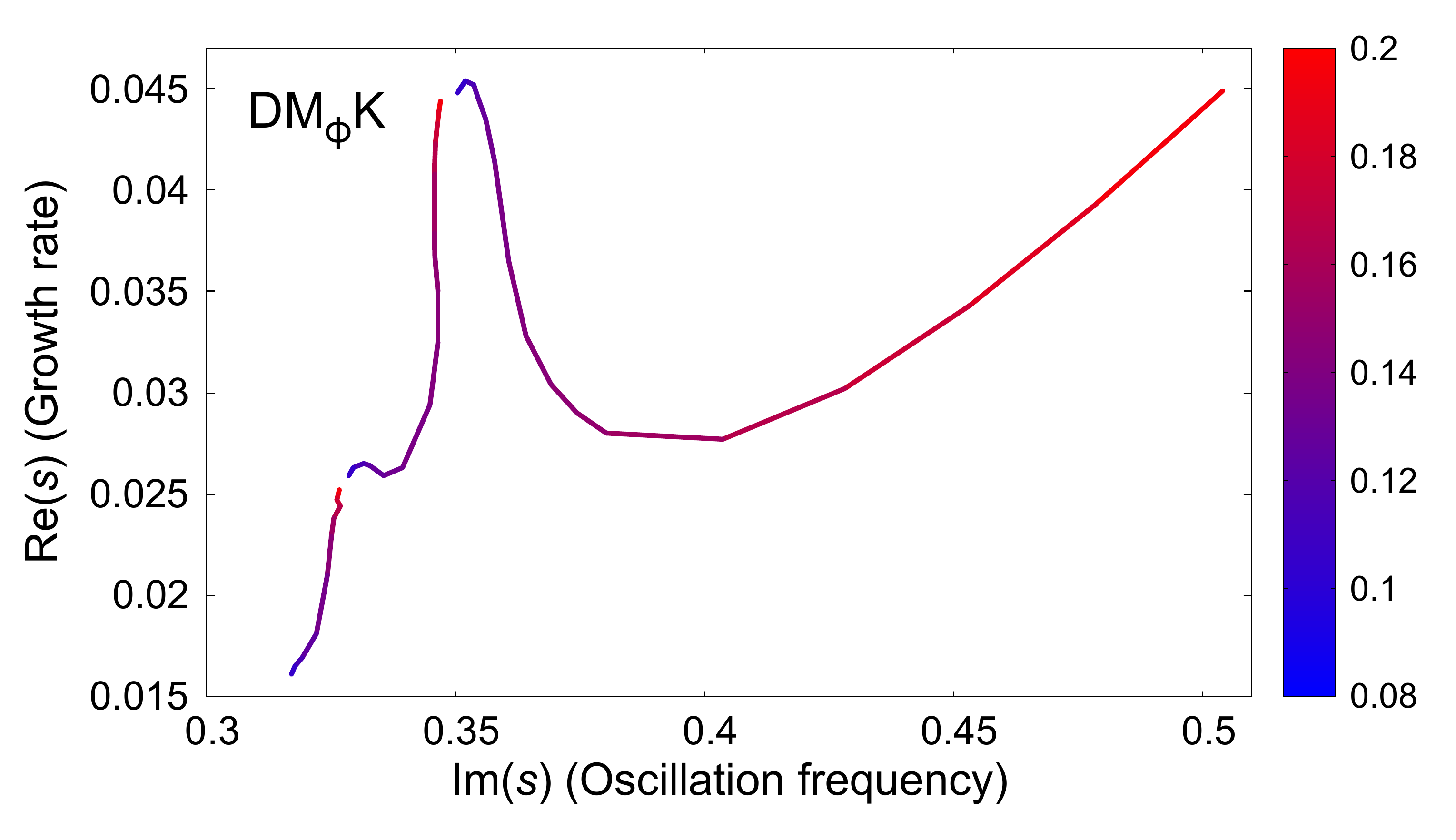}\label{MRKHGrOs}}
    \caption{(a) The three components of $\delta^2 H_{\rm{sh}}$ as functions of the jet velocity $V$ for model $\rm{DM_\phi K}$: the terms proportional $V_0^2$ (purple), $\Omega_0^2$ (green), and $V_0\Omega_0$ (blue), respectively (see Eq. (65)); they are normalized by the kinetic energy $\int{\rho_0 \boldsymbol{\dot\xi}^2} d^3x$. (b) The growth rates of the three most unstable modes with $m = 1$ and $k = \pi$ for model $\rm{DM_\phi K}$ as functions of the jet velocity $V$; the purple line corresponds to the mode in panel (a). (c) The trajectories of the same three unstable modes on the complex $s$ plane as the jet velocity $V$ changes. The line color indicates the jet velocity: the bluish (reddish) colors correspond to lower (higher) velocities.}
    \label{F5}
\end{figure}

\begin{figure}
\centering
\subfigure[]{\includegraphics[scale=0.16]{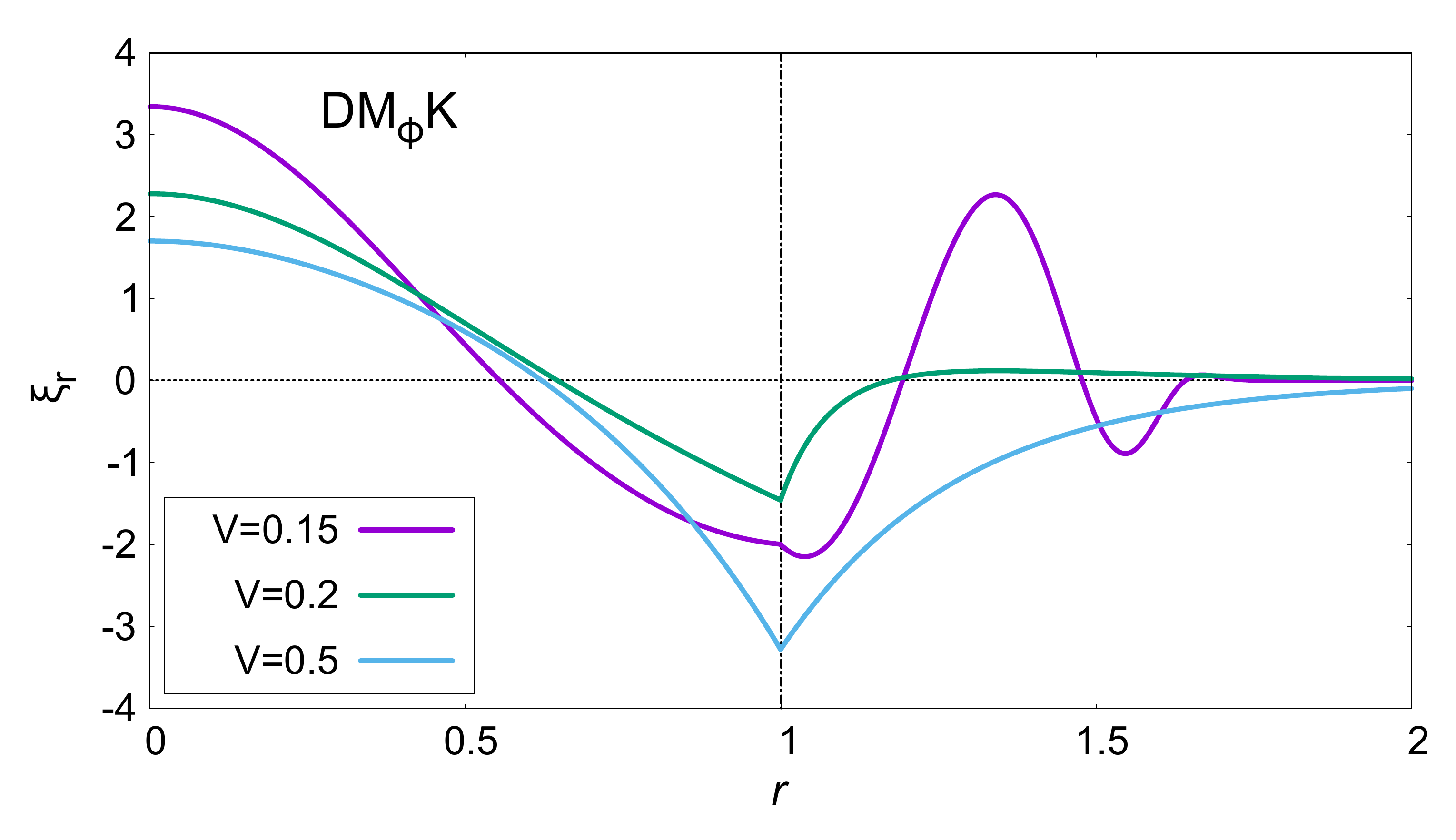}\label{MRKHeigen2}}
\subfigure[]{\includegraphics[scale=0.16]{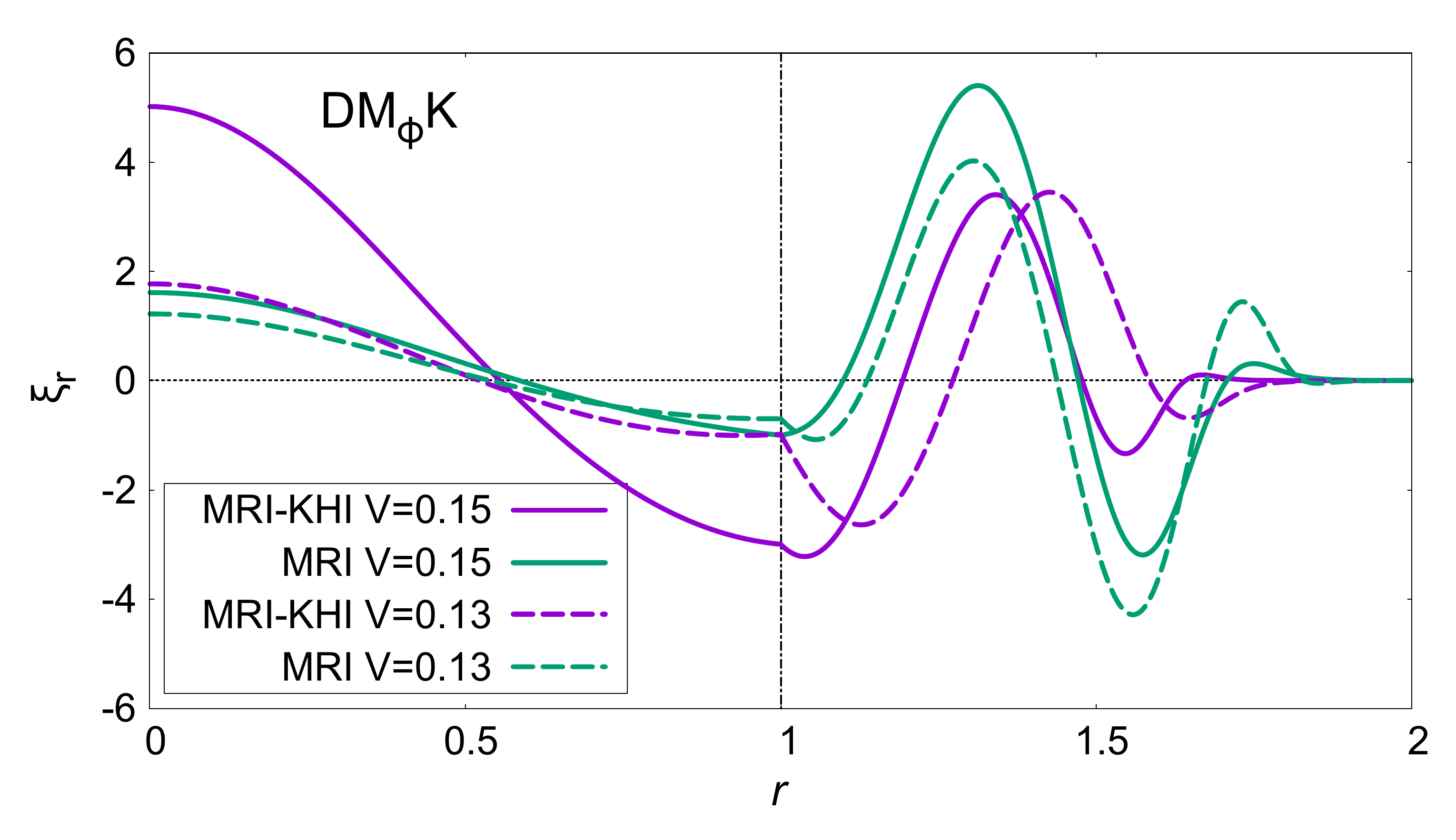}\label{MRKHeigen3}}
    \caption{(a) The eigenfunctions of the radial displacement $\xi_r$ for the MRI-KH transition mode at three different values of $V$. (b) Comparison of $\xi_r$ between the MRI-KH transition mode and the fundamental MRI mode at the jet velocities designated in the legend.}
    \label{F6}
\end{figure}

As their model names indicate, models $\rm{DM_\phi K}$ and $\rm{DM_z K}$ are meant for the study of MRI. All these models have a sharp jet boundary with the density also being discontinuous there. The angular velocity is nonuniform and given in Eq. (83). Other parameters are summarized in Table \ref{tableKHKK}.

The results of the Hamiltonian analysis are presented summarily in Table \ref{tableMRI}. We find that there is a KHI in addition to an MRI in these models. For both modes, $\delta^2H_{\rm sh}$ is the most negative. They are shear-driven instabilities indeed. They have different contents as exhibited in Table \ref{tableMRI}: the term proportional to $V_0^2$ is the most negative in all KH modes and the other terms proportional to $\Omega_0^2$ and $V_0\Omega_0$, respectively, are negligible although negative; for MRI, on the other hand, the term proportional to $\Omega_0^2$ is the dominant negative contributor to compensate the kinetic terms in $\delta^2 H$. As a matter of fact, we define KHI and MRI just in this way.

A mode change occurs also in the rotational jet but between KHI and MRI in this case (of course, the mode exchange between KHI and KKI also exists just in the same way as in the non-rotational case, as long as the rotation is not very fast). We take model $\rm{DM_\phi K}$ with $m=1$ and $k=\pi$ as the canonical case. We show in the top panel of Figure \ref{F5} the three components of $\delta^2 H_{\rm sh}$ as a function of the jet velocity for one of the two unstable modes in this model sequence. The term proportional to $\Omega_0^2$ is the most negative at low jet velocities, $V\lesssim 0.2$. This is hence an MRI. As the jet velocity becomes higher, however, the modulus of this term decreases monotonically whereas that of the term proportional to $V_0^2$ increases steadily. At $V \gtrsim 0.2,$ the latter takes over the place of the former as the most negative term and the mode changes to KHI continuously. 

The mode transition is evident in the middle panel of the same figure, where the growth rate is exhibited as a function of the jet velocity not only for this mode but also for two other unstable modes that are simultaneously existent. They are actually the three most unstable modes with $m = 1$ and $k = \pi$. The purple line corresponds to the mode in the top panel. This mode has nearly no velocity dependence at $V\lesssim0.12$, the fact conforming to its MRI nature revealed in the Hamiltonian analysis above. As the jet velocity becomes higher, the growth rate decreases a bit at $V \approx 0.14 \sim 0.16$. Then it rises monotonically{. The mode nature} changes at $V \approx 0.2$ from MRI to KHI. The decrease of the growth rate and the change of mode nature coincide with the crossing in the growth rate with another mode denoted by the green line. However, this is not a close encounter of these two modes in the complex $s$ plane as demonstrated in the bottom panel of Figure \ref{F5}, where the trajectories of the three unstable modes are shown in the complex $s$ plane. This is hence another example of the mode change that is not associated with any close encounter. The other two unstable modes remain MRI's for all jet velocities although they show an increase in their growth rates at $V \approx 0.14$.

The change of the mode nature is also confirmed by the eigenfunctions in the top panel of Figure \ref{F6}. They are the eigenfunctions of the mode in Fig. \ref{KHeigen1}, presented for three different jet velocities as a function of radius. We expect naively from the difference in the radial profiles of the jet velocity and the angular velocity that the eigenfunction of KHI is confined to the vicinity of the jet boundary whereas that of MRI is more extended both inward and outward. This picture is qualitatively correct but not so clear-cut. With the eigenfunctions alone, we could not identify the mechanism of the instability unambiguously. The Hamiltonian analysis is hence indispensable. For comparison, we plot in {Fig. \ref{MRKHeigen3}} for one of the other two unstable modes, i.e., the one corresponding to the green line in Figure \ref{MRKHg}. Although the radial profile of this mode changes a bit across $V\approx 0.14$, it remains of MRI nature and the difference from the previous mode that shows the mode change is evident.

Unfortunately, we do not find a mode change between MRI and KKI for the models that we investigate in this paper. The rotation tends to stabilize the KK mode whereas MRI feeds on (differential) rotation. In our models with rotation, KK mode {is} stabilized before an MRI is induced as we increase the angular velocity (see Figures \ref{KKIomega}, \ref{fig:MRIBzomega} and \ref{fig:MRIBphiomega} below). For reference, we give {in Table \ref{tableKKrotation}} some results of the Hamiltonian analysis of the KK modes in model $\rm{DCV4}${, which has} a Keplerian rotation. Nonetheless, we still suppose that the mode exchange between MRI and KKI should happen if we explore an appropriate parameter region. The confirmation of this expectation will be a future study. 

\begin{deluxetable*}{ccccllllll}
\tablenum{3}
\tablecaption{Hamiltonian analysis of various modes in rotational models}
\label{tableMRI}
\tablehead{
\colhead{Model} & \colhead{$m$} & \colhead{$k$} & \colhead{Mode}  &  \colhead{$V_0^2$} &  \colhead{$\Omega_0^2$} &  \colhead{$V_0\Omega_0$}  & \colhead{$\delta^2H_{\rm pr}$} & \colhead{$\delta^2H_{\rm B}$}  
}
\startdata
$\rm{DM_\phi K}$ & $1$ & $\pi$ & KH & -1.148 & -0.00439769 & -0.0580996 & 0.210938 & 0.00144683 \\
-& $1$ & $\pi$ & MRI & -1.01794e-04 & -1.12190 & -4.5228e-05 & -1.14326e-04 & 0.122001 \\
\hline
$\rm{DM_z K}$& $1$ & $\pi$ & KH & -1.14735 & -0.00438821 & -0.0579413 & 0.210163 & 0.0013061 \\
- & $1$ & $\pi$ & MRI & -0.00158019 & -1.78638 & -2.3163e-04 & 7.18829e-05 & 0.788083 \\
-& $0$ & $\pi$ & KH & -1.26916 & -0.00201216 & 0.0104968 & 0.261033 & 0.00141637 \\
-& $0$ & $\pi$ & MRI & -0.0164456 & -8.16879 & 0.00144415 & -5.26569e-04 & 7.18462  \\
\hline
\hline
$\rm{DM_\phi K}$& $1$ & $\pi$ & MRI & -4.71292e-05 & -1.14829 & -7.03113e-06 & -0.00117370 & 0.1484 \\
-& $1$ & $\pi$ & MRI-$1st$ & -1.29382e-05 & -1.10283 & -5.75644e-6 & 2.12562e-05 & 0.102692 \\
-& $1$ & $\pi$ & MRI-$2nd$ & -2.86979e-06 & -1.10009 & -1.27738e-06 & 6.73059e-05 & 0.099736\\
-& $3$ & $\pi$ & MRI & -9.64262e-05 & -1.107144 & -7.26191e-05 & -3.84688e-05 & 0.106916 \\
-& $2$ & $2\pi$ & MRI & -2.75533e-05 & -1.14966 & -7.2599e-06 & -1.59658e-04 & 0.14976 \\
$\rm{DM_\phi KV4}$ & $2$ & $\pi$ & MRI & -8.86795e-06 & -1.12224 & -1.28015e-06 & -1.11798e-04 & 0.121779 \\
$\rm{DM_\phi K\Omega2}$ & $2$ & $\pi$ & MRI & -1.54047e-04 & -1.03537 & -1.24679e-04 & -2.16802e-05 & 0.0354226 \\
$\rm{DM_\phi K B_\phi4}$& $2$ & $\pi$ & MRI & -2.06236e-06 & -1.13635 & -9.35704e-07 & -2.72639e-04 & 0.13662 \\
\hline
$\rm{DM_z K}$& $2$ & $\pi$ & MRI & -3.61558e-04 & -1.16921 & -1.59177e-04 & 3.22081e-05 & 0.168862 \\
-& $2$ & $\pi$ & MRI-$1st$ & -0.00120331 & -2.15709 & -1.78076e-04 & 1.47021e-04 & 1.15812 \\
-& $2$ & $\pi/2$ & MRI & -0.00118239 & -1.1742 & -5.55592e-05 & 3.21092e-05 & 0.175102 \\
$\rm{DM_z KV4}$ & $1$ & $\pi$ & MRI & -1.77389e-04 & -1.78553 & -1.42908e-05  & 
2.55377e-04 & 0.785468  \\
$\rm{DM_z K\Omega 2}$& $1$ & $\pi$ & MRI & -7.47988e-04 & -1.14438 & -1.00057e-04 & 1.09791e-05 & 0.145138  \\
$\rm{DM_z KB_z02}$& $1$ & $\pi$ & MRI & -8.05019e-06 & -1.0184 & -1.15889e-06 & 3.82835e-07 & 0.0182727 \\
\enddata
\tablecomments{Each component of $\delta^2 H$ is normalized by the kinetic energy $\int{\rho_0 \boldsymbol{\dot\xi}^2} d^3x $.}
\end{deluxetable*}

\begin{deluxetable}{cccccc}
\tablenum{4}
\tablecaption{Hamiltonian of KK modes with rotation}
\label{tableKKrotation}
\tablehead{
\colhead{Model} & \colhead{mode} & \colhead{$\Omega_0$} & \colhead{$\delta^2H_{\rm sh}$} & \colhead{$\delta^2H_{\rm pr}$} & \colhead{$\delta^2H_{\rm B}$} 
}
\startdata
$\rm{DC\Omega}$ & KK & 0 & 0.00646371 & 4.37607e-05 & -0.00642571 \\
$-$ & KK & 1 & 0.00618856 & 2.00207e-05 & -0.0062088 \\
$-$ & KK & 1.9 & 0.00408031 & 4.19377e-05 & -0.00430847 \\
$\rm{DM_\phi K}$ & MRI & 1 & -0.148859 & -0.000117641 & 0.148351 \\
$\rm{DM_\phi K\Omega 2}$ & MRI & 2 & -0.0392224 & -1.0238e-05 & 0.0376074 \\
\enddata
\tablecomments{Second-variation of the Hamiltonian of KK mode with a Keplerian rotation in $DC$ and $DM_\phi K$ with $m=1$, $k=\pi$. }
\end{deluxetable}

\subsection{Parameter dependence}

\subsubsection{Effect of a finite transition region}
\begin{figure}
    \centering
    \includegraphics[scale=0.16]
    {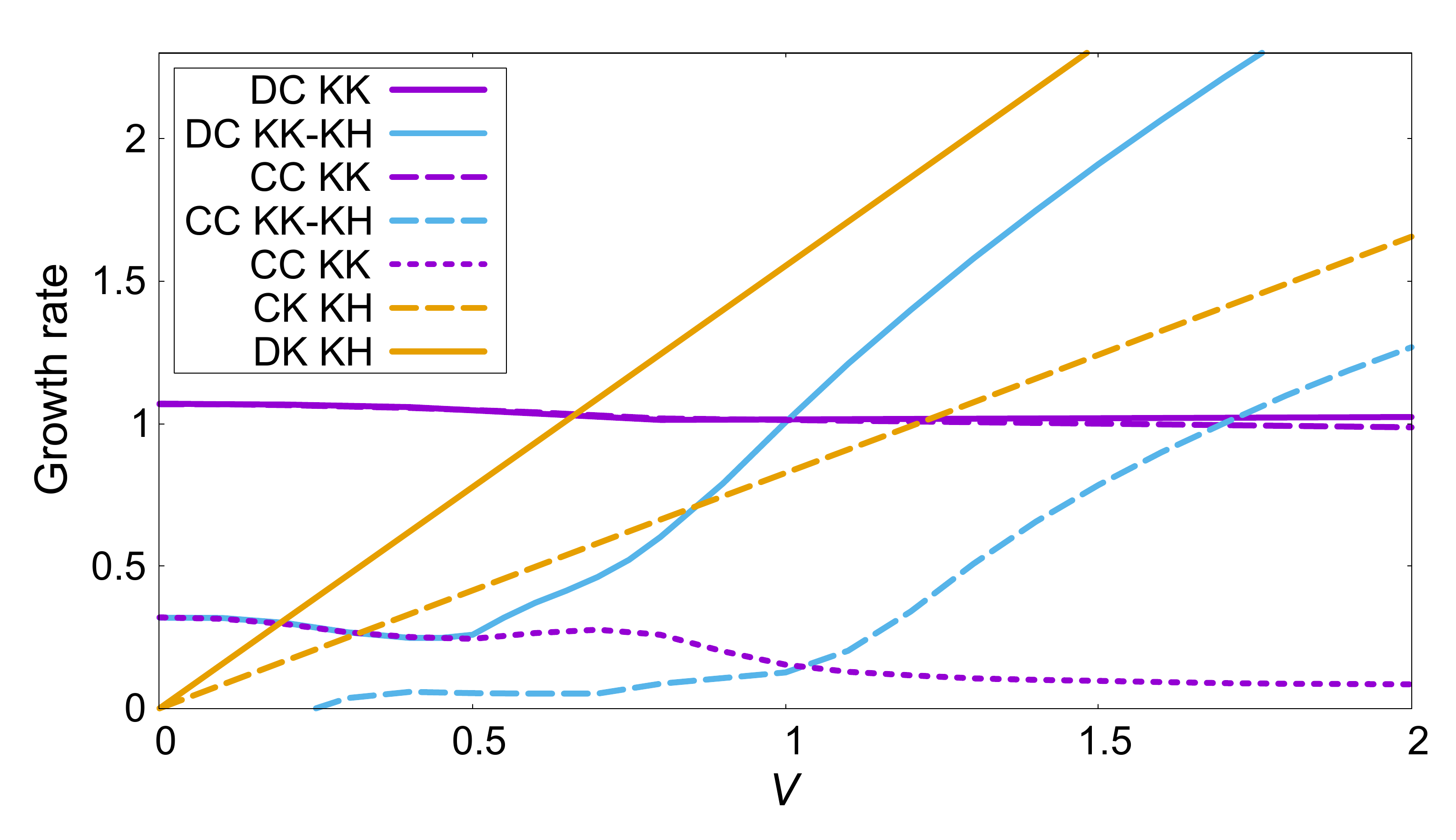}
    \caption{The growth rates of the KK and KH modes with $k = \pi$ and $m = 1$ as functions of the jet velocity for different velocity profiles. Line types and colors distinguish models and modes, respectively, as displayed in the legend.}
    \label{KHIcon}
\end{figure}

So far we have considered only the models with a sharp boundary between the jet and the ambient matter. In this section, we present the results of models with smooth jet profiles. Note that in the models meant for MRI, the profiles of angular velocity and magnetic fields are always continuous. We pay attention to the KH and KK modes with $k = \pi$ and $m = 1$. We compare models $\rm{CC}$ and $\rm{CK}$ with their counterparts with a sharp boundary, $\rm{DC}$ and $\rm{DK}$. {These} models are all non-rotating (see Table \ref{tablemodel}). In Figure \ref{KHIcon}, we plot the growth rates of some unstable modes as functions of the jet velocity $V$.

We first study the non-magnetic models: $\rm{CK}$ and $\rm{DK}$, which are exhibited with yellow dashed and solid lines, respectively. The unstable mode found in each model is both KHI: their growth rates rise linearly with $V$. The smooth transition at the jet boundary only reduces the increase rate by a factor of $\sim 2$. The corresponding magnetized models: $\rm{CC}$ and $\rm{DC}$ are presented in the same figure. In these models, the toroidal magnetic field is added. The unstable modes shown with the blue dashed and solid lines are the counterparts to the previous non-magnetized modes denoted with the yellow dashed and solid lines, respectively. They have a similar behavior in their growth rates, i.e., a linear increase with the jet velocity at large values of the jet velocity: $V \gtrsim 1.5$ for the discontinuous model DC and $V \gtrsim 1.0$ for the continuous counterpart model $\rm{CC}$. The growth rate for the continuous model is lower by a factor of $\gtrsim 2$ than that for the discontinuous model. These modes are both KHI.

As the jet velocity is lowered, however, the toroidal magnetic field becomes more important and the two modes change into KKI continuously. The mode changeoccurs also in the continuous model just in the same way as in the discontinuous counterpart. In the same figure, we plot in purple the growth rates of the other three unstable modes we detect in models DC and CC. They are all KK modes and are insensitive to the jet velocity as expected. Interestingly, they do not change their mode nature, i.e., they remain KKI all the way up to large values of $V$. Their growth rates of the most unstable KK modes (long-dashed and solid purple lines) are almost identical to each other whereas we do not find a counterpart in model DC to the second unstable KK mode in model CC (short-dashed purple line). Judging from these results, we may say that the transition width in the radial profiles of density and jet velocity do not affect KK modes very much.

\subsubsection{$k$, $B_z$, $B_{\phi}$ and $\Omega_0$ dependences}

\begin{figure*}
\centering
\subfigure[]{\includegraphics[scale=0.16]{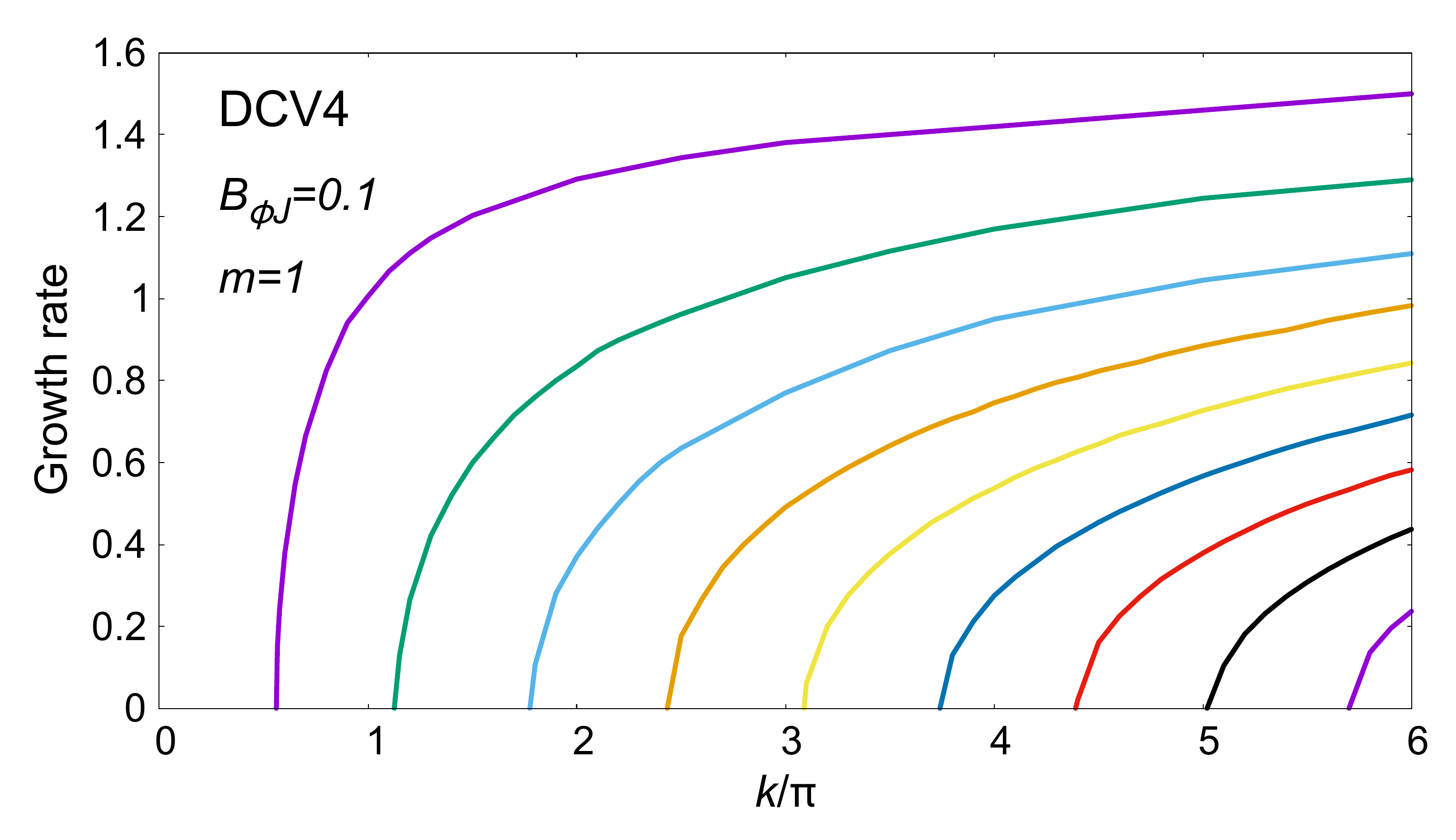}\label{KKIk}}
\subfigure[]{\includegraphics[scale=0.16]{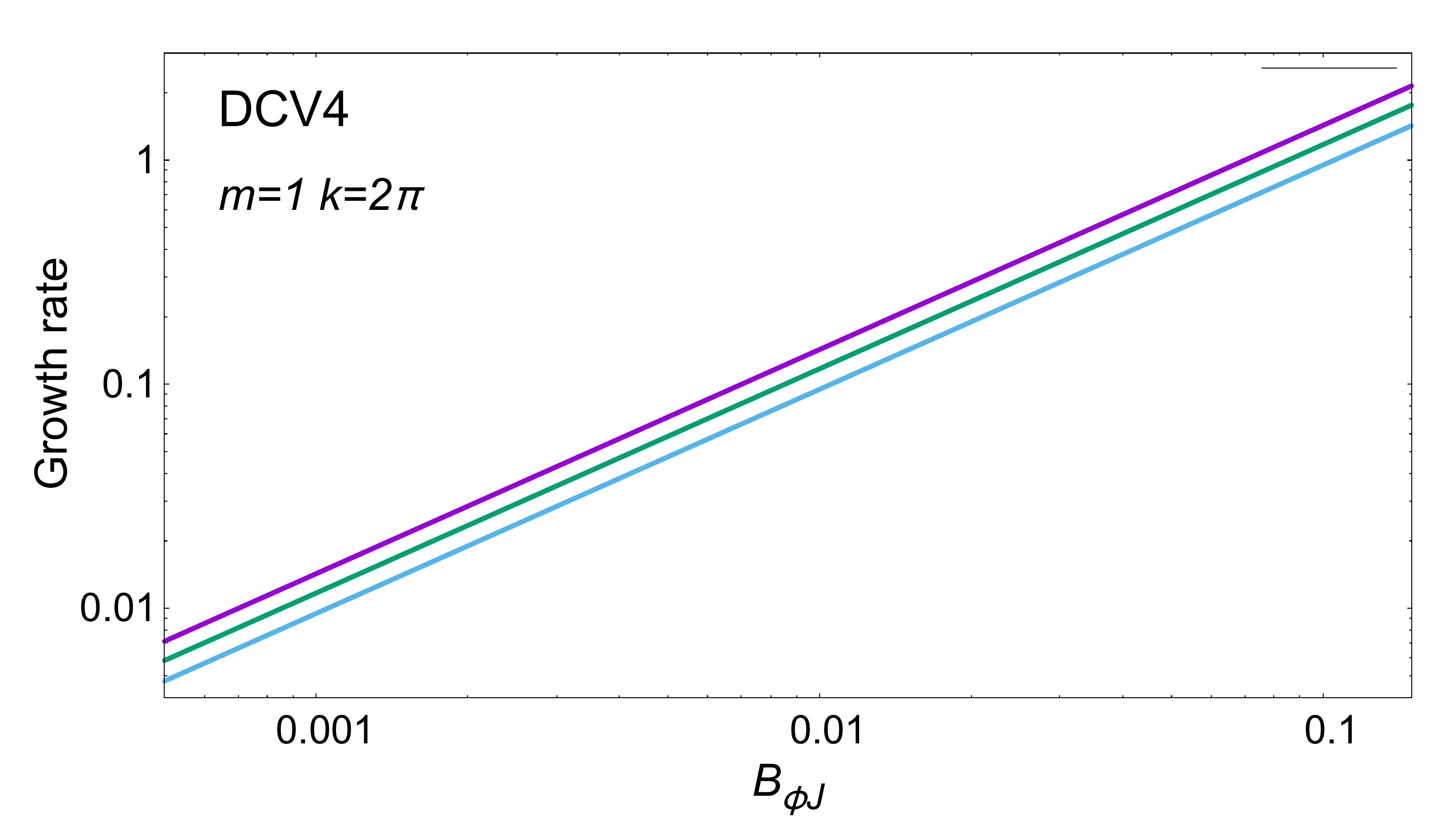}\label{KKIBphi}}
    \hfill
\subfigure[]{\includegraphics[scale=0.16]{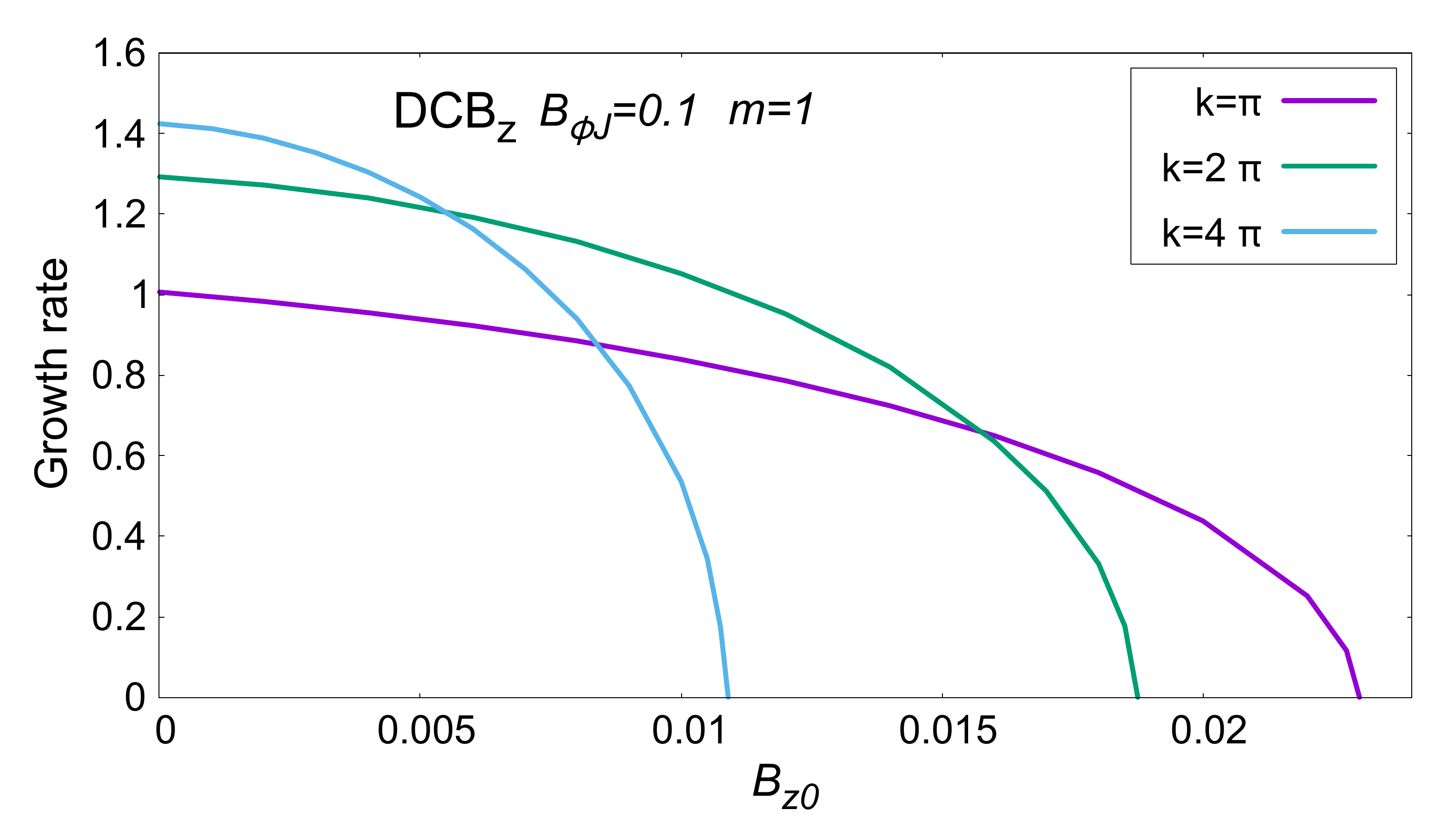}\label{KKIBz}}
\subfigure[]{\includegraphics[scale=0.16]{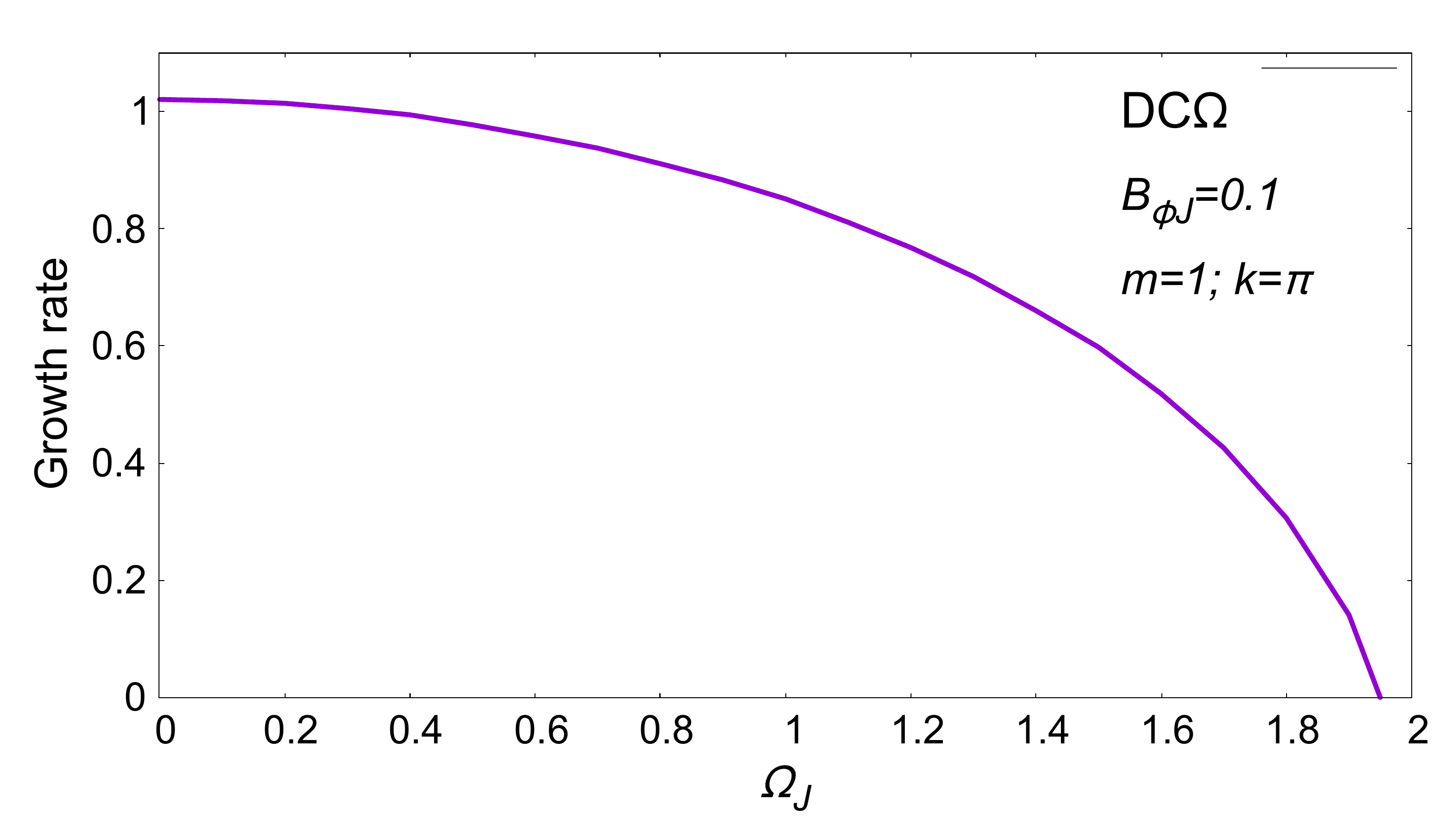}\label{KKIomega}}
    \caption{(a) The growth rates of KK modes as functions of the wave number $k$ for model DCDV4. The fundamental and higher overtones are shown from left to right. (b) The growth rates of the fundamental (purple), first (green), and second (cyan) overtones of KK modes as functions of the toroidal magnetic field $B_{\phi J}$ for model DCV4. (c) The growth rates of the fundamental KK modes with $k=\pi$ (purple), $k=2\pi$ (green), and $k=4\pi$ (cyan) for model DCV4 as functions of $B_{z 0}$. (d) The growth rate of the fundamental KK mode with $k=\pi$ as a function of $\Omega_J$.}
\end{figure*}

Now we shift our attention to other parameter dependences. In Figure \ref{KKIk}, we present the growth rates of KKI's we find in model $\rm{DCV4}$ as functions of the wave number $k$; m is set to 1 and $B_{\phi J}$ is fixed to $0.1$. This model is non-rotating and has only a toroidal magnetic field as in the canonical model $\rm{DC}$ but has a four times larger jet velocity (see Table \ref{tablemodel}). There is no kink mode at $k/\pi \lesssim 0.5$ in this model. The fundamental kink mode becomes unstable first, and overtones follow suit one by one as the wave number increases and there are nine KKI's at $k=6\pi$.

Figure \ref{KKIBphi} shows the growth rates of the fundamental mode (purple) and the first two overtone modes (green and blue) of KKI as functions of $B_{\phi J}$ in model DCV4; $k$ is fixed to $2\pi$ this time. They all show a linear dependence on $B_{\phi J}$ with the same increase rate. The number and identity of unstable overtones have no dependence on the toroidal magnetic field strength.

The stabilization effect of $B_z$ on the KK mode has been studied in many previous papers. We confirm it in Figure \ref{KKIBz} for model $\rm{DCB_z}$. This model is the same as model DCV4 except that the former has non-vanishing poloidal magnetic fields, $B_z$. The purple, green, and blue curves denote the fundamental KK modes with $k=\pi$, $2\pi$, and $4\pi$, respectively; $B_{\phi J}$ is set to $0.1$ here. The growth rates decrease monotonically with $B_z$ for all the modes. The KK modes with larger $k$ are stabilized more strongly by the poloidal magnetic field although they have greater growth rates when $B_z$ is absent. Considering the common existence of poloidal magnetic fields, we expect that the KKI tends to develop with low $k$ values in astrophysical jets.

For KK modes, rotation is another suppressor. We plot in Figure \ref{KKIomega} the growth rate of the fundamental KK mode as a function of the angular velocity $\Omega_J$ for model $\rm{DC\Omega}$. {This} model is the same as the canonical model DC except that a Keplerian-type rotation is added (see Table \ref{tablemodel}). We set $k = \pi$, $m = 1$ and $B_{\phi J} = 0.1$ here. {As} the rotation becomes faster, the growth rate of this mode is reduced monotonically and the mode becomes stable above a threshold value ($\Omega_J \approx 2$). This is in contrast to the behavior of MRI as shown shortly.

\begin{figure}
\centering
\subfigure[]{\includegraphics[scale=0.16]{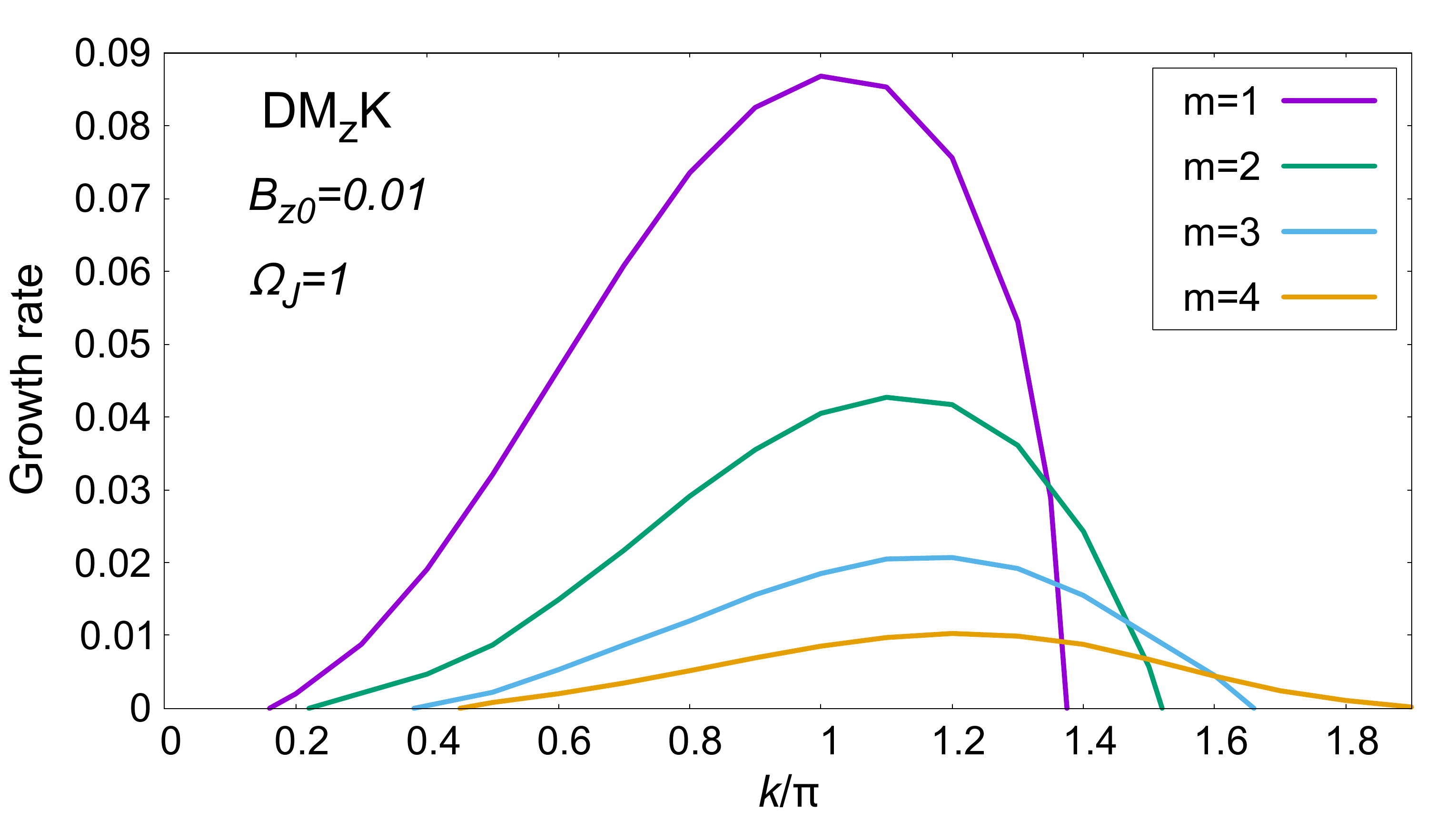}\label{fig:MRIBzmk}}
\subfigure[]{\includegraphics[scale=0.16]{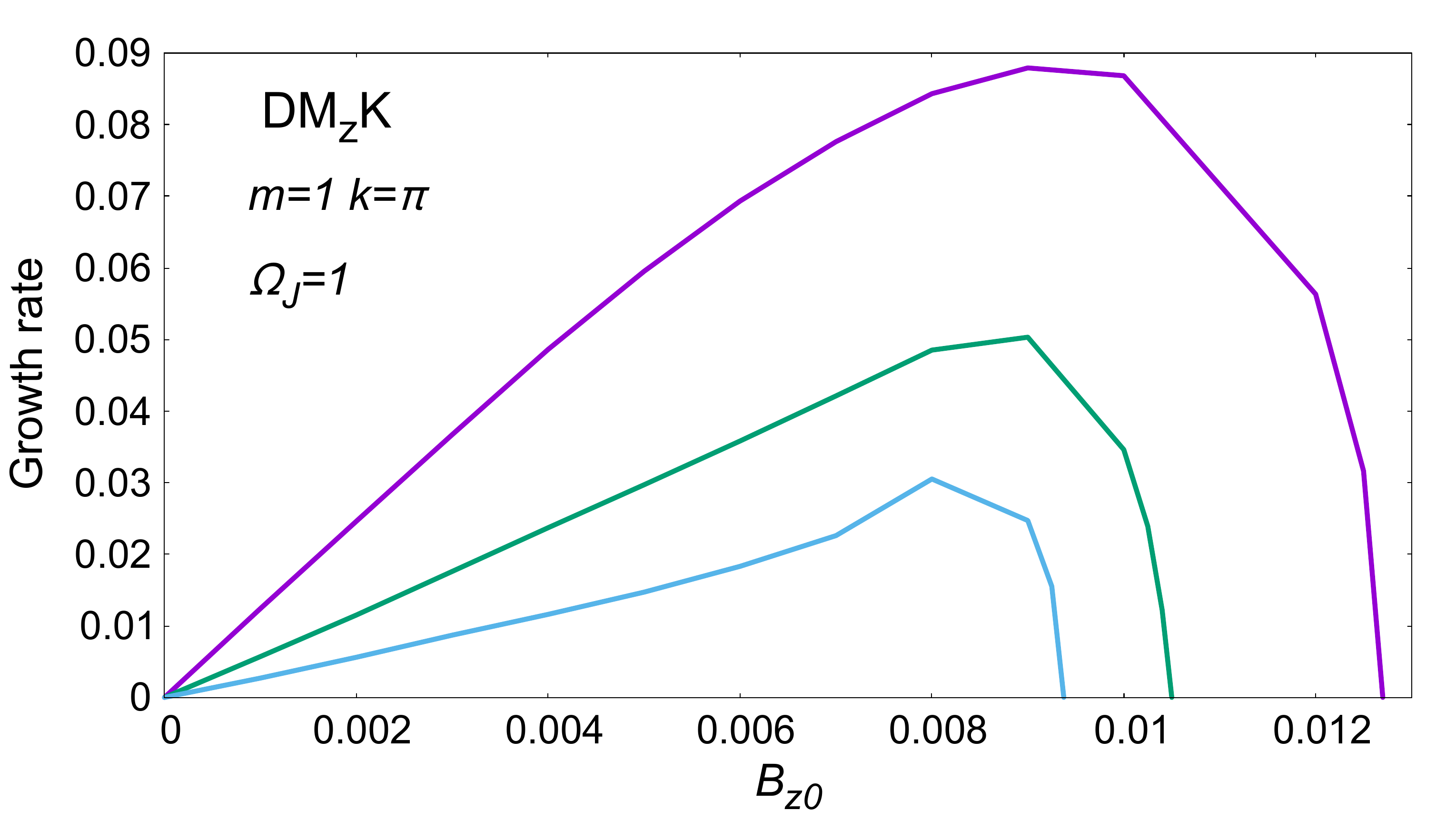}\label{fig:MRIBzBz}}
\subfigure[]{\includegraphics[scale=0.16]{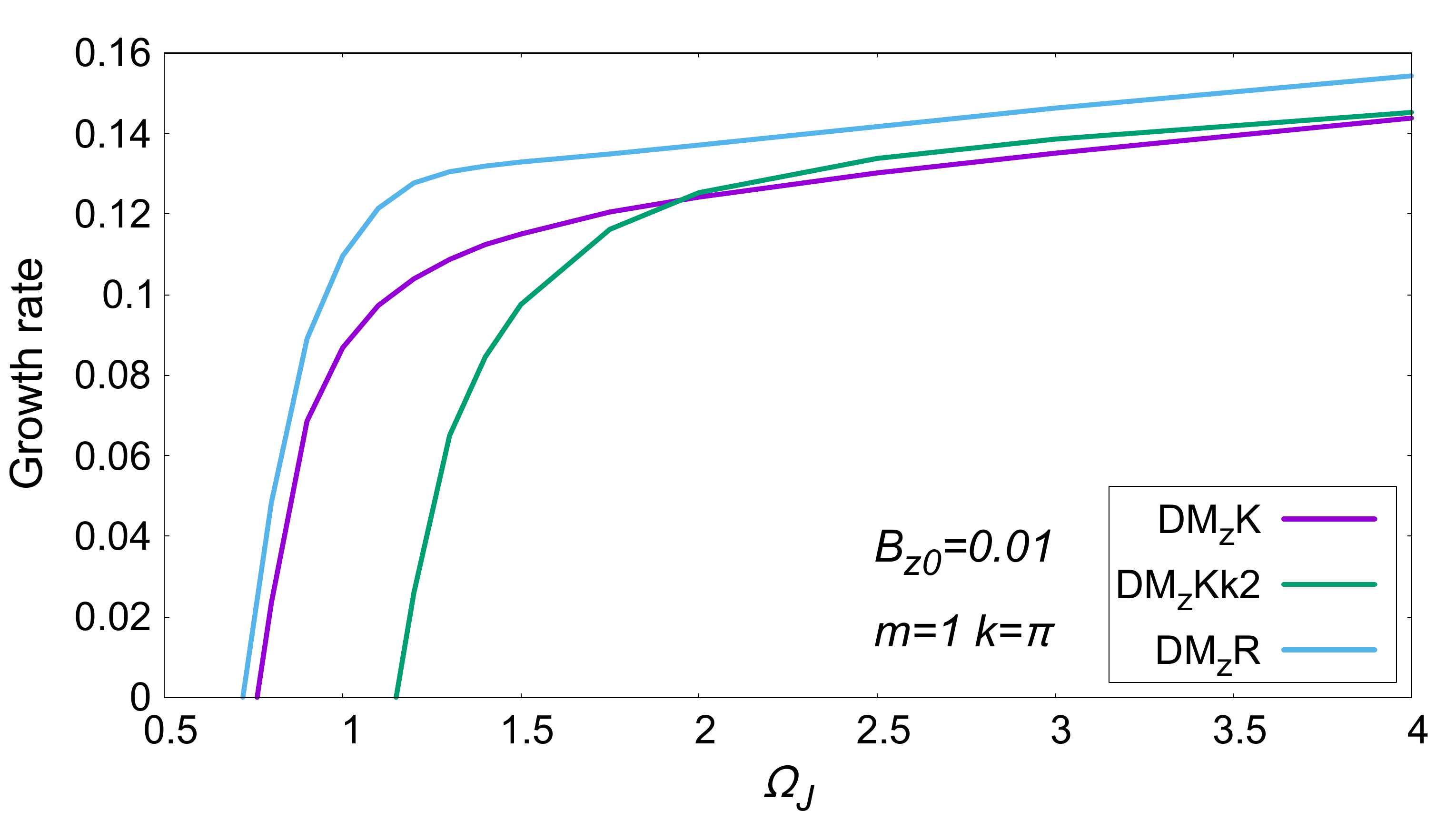}\label{fig:MRIBzomega}}
    \caption{(a) The growth rates of $B_z$-induced MRI's in model $\rm{DM_zK}$ as functions of the wave number $k$. The purple, green, blue, and orange curves correspond to the fundamental MRI modes with $m = 1$, $2$, $3$, and $4$, respectively. (b) The same as (a) but with the horizontal axis $B_{z0}$. The fundamental, the first and second overtones are colored purple, green, and blue, respectively. The wave numbers are $m=1$ and $k=2\pi$. (c) The growth rates as functions of the angular velocity $\Omega_J$ for the Keplerian-type with $\kappa=1$ (model $\rm{DM_zK}$, purple), $\kappa=2$ (model $\rm{DM_zKk2}$, green) and the Rayleigh-merginal rotation (model $\rm{DM_zR}$, blue). The wave numbers are fixed to $m=1$ and $k=\pi$.}
    \label{fig:MRIBz}
\end{figure}

MRI can be induced by both poloidal and toroidal fields. We begin with the former. In Figure \ref{fig:MRIBzmk}, we present the growth rate of $B_z$-induced MRI's for four different azimuthal wave numbers: $m = 1, 2$, $3$, and $4$ as functions of the wave number $k$. We take model $\rm{DM_zK}$, which is the same as the canonical model $\rm{DC}$ but with a poloidal field and a Keplerian-type rotation added (see Table \ref{tablemodel}). At very small values of $k$ they are all stable. As the wave number increases, the mode with $m = 1$ becomes unstable first. Then the modes with $m=2$ and $3$ follow in this order. The Hamiltonian analysis confirms that they are all MRI's. The growth rates of these modes increase with $k$ in the beginning. They are suppressed, however, for larger values of $k$, since the magnetic tension resists the bending of magnetic field lines. {It} is more effective for shorter wavelengths, or larger values of $k$.  As a consequence, the maximum growth rate is obtained at some intermediate value of $k$: $k/\pi \approx 1$ in this model. The MRI with $m = 1$ reaches the maximum earliest, i.e., at the smallest value of $k$. The growth rate at the maximum is largest for the mode with $m = 1$. The same mode is suppressed most quickly at large values of $k$. The $B_z$-induced MRI is stabilized completely at $m>5$ in this model

Figure \ref{fig:MRIBzBz} illuminates the $B_z$ dependence of $B_z$-induced MRI in the same model $\rm{DM_zK}$. We fix the wave numbers to $m=1$ and $k=\pi$. We show the three most unstable modes. The growth rates are vanishing at $B_z = 0$ because they are current-driven. They become nonvanishing once $B_z$ is present. They have a linear dependence on the $B_z$ in the weak field regime. The three modes peak successively without changing the order in the growth rate. At even larger values of $B_z$, MRI is stabilized, since the magnetic tension dominates over the centrifugal force. The growth rate decreases sharply after passing the peak. Each mode becomes stable above a certain critical value of $B_z$.

Rotation is another important factor for MRI. We consider two rotational profiles in this paper: the Keplerian-type rotation $\Omega_0\propto r^{-3/2}$ at large radii and Rayleigh-marginal rotation $\Omega_0\propto r^{-2}$ at large radii. We compare them in Figure \ref{fig:MRIBzomega}, where the growth rates of the most unstable MRI's with $m=1$ and $k = \pi$ are shown for model $\rm{DM_ZK}$ and its two variants: $\rm{DM_ZKk2}$ and $\rm{DM_ZR}$. Model $\rm{DM_ZKk2}$ differs from Model $\rm{DM_ZK}$ in that it has ${\kappa = 2}$ instead of $1$ whereas model $\rm{DM_ZR}$ is different from these two models in that it has the Rayleigh-marginal rotation profile (see Table \ref{tablemodel} for details). The behavior of growth rate is qualitatively the same for all three cases. Each mode becomes unstable once the rotation velocity passes a certain threshold, at which point the centrifugal force overcomes the magnetic tension. This threshold is slightly higher for model $\rm{DM_ZKk2}$, since the angular velocity is smaller at the same value of $\Omega_J$ for this model. Then the growth rate increases sharply with the angular velocity. As the rotation becomes even faster, the growth rates rise much more slowly and show a clear linear dependence on $\Omega_0$ for both the Keplerian-type and Rayleigh-marginal rotation profiles. The growth rate for the Rayleigh-marginal rotation with $\kappa=1$ (blue) is a bit larger than that for the Keplerian-type rotation with the same $\kappa$ (purple). This is simply because the differential rotation, on which MRI, one of the shear-driven instabilities, feeds, is stronger for the former. Our results suggest that the rotation speed itself plays more important roles than the rotation profile in setting the growth rate of MRI although the differential rotation is definitely needed for MRI to occur.

\begin{figure}
\centering
\subfigure[]{\includegraphics[scale=0.16]{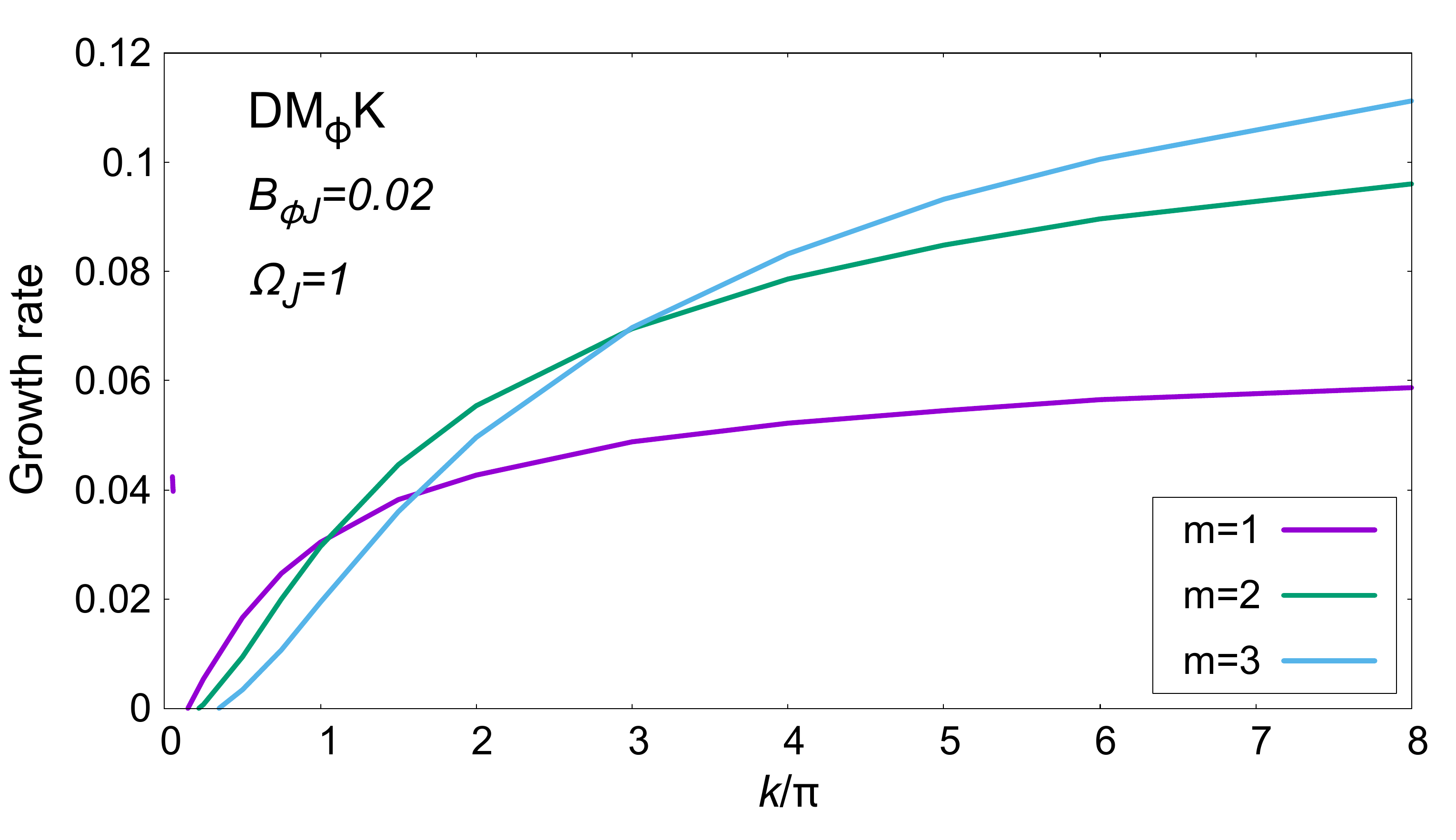}\label{fig:MRIBphimk}}
\subfigure[]{\includegraphics[scale=0.16]{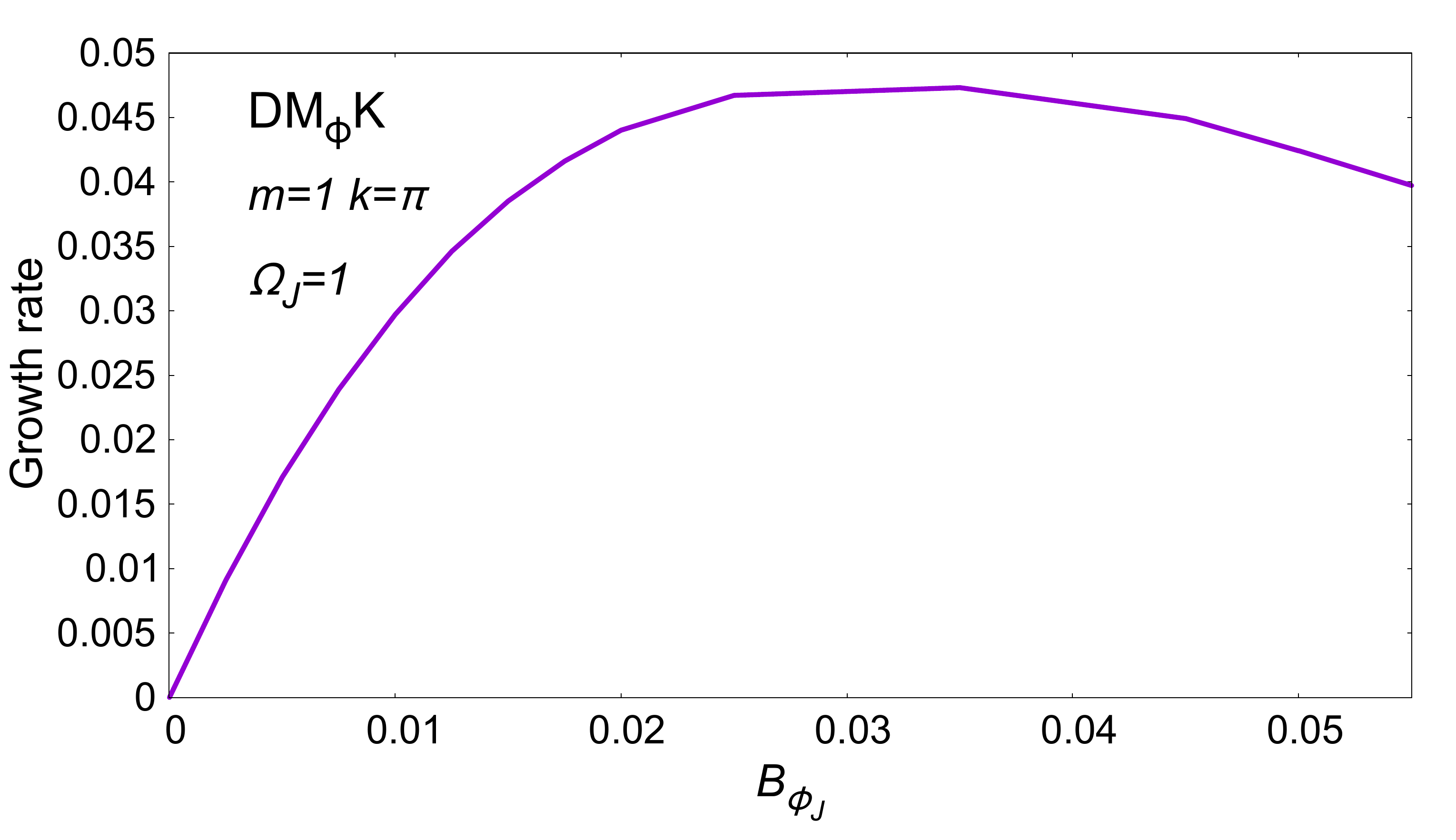}\label{fig:MRIBphiBphi}}
\subfigure[]{\includegraphics[scale=0.16]{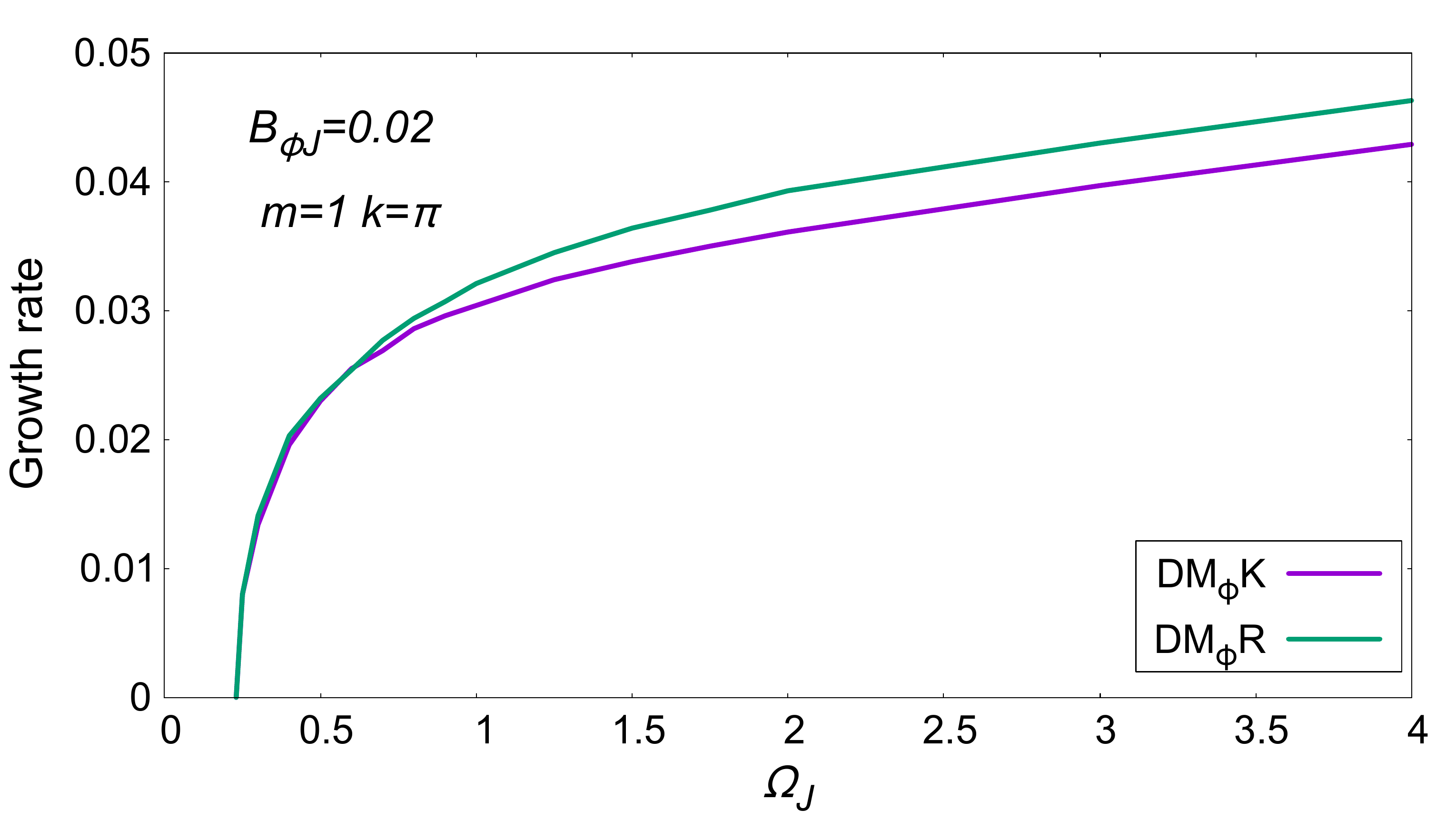}\label{fig:MRIBphiomega}}
    \caption{(a) The growth rates of $B_\phi$-induced MRI's in model $\rm{DM_\phi K}$ as functions of the wave number $k$. The purple, green, and blue curves correspond to the fundamental MRI modes with $m = 1$, $2$, and $3$, respectively. (b) The same as (a) but with the horizontal axis $B_{\phi J}$. The wave numbers are $m=1$ and $k=\pi$. (c) The growth rates as functions of the angular velocity $\Omega_J$ for the Keplerian-type rotation (model $\rm{DM_\phi K}$, purple) and the Rayleigh-merginal rotation (model $\rm{DM_\phi R}$, green). The wave numbers are set to $m=1$ and $k=\pi$.}
\end{figure}

Figure \ref{fig:MRIBphimk} shows the growth rates of $B_\phi$-induced MRI's as functions of the translational wave number $k$ for three azimuthal wave numbers: $m = 1$, $2$, and $3$ in model $\rm{DM_\phi K}$ as in Figure \ref{fig:MRIBzmk}. In this model, only the toroidal magnetic field is present with $B_{\phi J}=0.02$; the jet is rotating with the Keplerian-type rotation with $\kappa = 1$; $\Omega_J = 1$. At long wavelengths or small values of $k$, the modes with small $m$ tend to have larger growth rates, while the situation is reversed at short wavelengths, with smaller $m$ modes having smaller growth rates. Since $k$ does not affect the local magnetic tension in the azimuthal direction, $B_\phi$-induced MRI's are not stabilized by increasing $k$ as confirmed for all three cases {of $m$}. This is in sharp contrast to the $B_z$-induced MRI's. In this model, it seems that there are infinitely many unstable overtones although the fundamental mode has the greatest growth rates.

Figure \ref{fig:MRIBphiBphi} plots the growth rate as a function of $B_{\phi J}$ for the $B_\phi$-induced MRI with $k = \pi$ and $m = 1$ in the same model $\rm{DM_{\phi}K}$. As expected, the mode turns unstable once $B_{\phi}$ becomes nonvanishing. The growth rate increases with $B_{\phi}$ initially, since this mode is $B_{\phi}$-induced. As $B_{\phi}$ rises further, the growth rate {reaches} the peak and then starts to decrease. This is again just as expected. The $B_{\phi}$-induced MRI accompanies the bending of toroidal magnetic field lines, which {is} difficult when the magnetic tension becomes stronger. As we mentioned earlier, we do not find mode change/exchange between MRI and KKI. This may be understood from the fact that the KK mode is enhanced as $B_{\phi}$ increases whereas MRI is suppressed when $B_{\phi}$ exceeds a certain threshold.

Figure \ref{fig:MRIBphiomega} demonstrates how the growth rates of $B_\phi$-induced MRI's change with the angular velocity for the Keplerian-type rotation (model $\rm{DM_{\phi}K}$, purple) and the Rayleigh-marginal rotation (model $\rm{DM_{\phi}R}$, green). In these models, we set $B_{\phi J}=0.02$, $k=\pi$ and $m=1$. The $B_\phi$-induced MRI's appear when $\Omega_J$ exceeds a threshold value $\Omega_J \sim 0.22$, which is nearly the same for the two models. The angular-velocity dependence is also similar for the two models. The growth rate increases quickly right after $\Omega_J$ exceeds the threshold. As the rotation becomes even faster, the growth rate tends to increase with a constant rate, the behavior common to the two models. In this phase, the growth rate for the same $\Omega_J$ is larger for the Rayleigh-marginal rotation. This is because the differential rotation is stronger for this rotation profile.

\section{Discussions}

\subsection{Limitations}

There are admittedly many limitations in the current study. The jet configurations in our models are highly simplified: the jet is assumed to be an infinitely long cylinder that is axisymmetric and translationally invariant. We consider only ideal MHD and ignore non-ideal effects such as resistivity. Our models take no account of radiative processes, either, although the possible ramifications of the jet instability for radiations have been drawing much attention lately (\citealt{Nikhil_2021}; \citealt{BodoG_2021}). Our analysis is also limited to non-relativistic jets. The Hamiltonian formulation we employ in this study can be extended to relativistic jets in principle (\citealt{Holm_1984}; \citealt{Holm_1985}). However, some details are remaining to be worked out yet. The greatest limitation of all is that this is a linear analysis. It is applicable only to the early phase of instability that will last for a short period of the order of the growth time and may be even skipped in some cases with large initial perturbations.

The advantage of linear analysis is its mathematical rigor, to which we give the utmost importance in the first step of our project. The next step is to study the nonlinear evolutions of the unstable modes that we identify in this paper. Numerical simulations will be performed for this purpose. One of the merits of the current investigation is that we obtain the eigenfunctions in addition to the eigenvalues (or the growth rates and oscillation frequencies) of various unstable modes. We will use them for setting initial conditions of the simulations. Comparing the linear growth of the individual unstable modes between the simulations and the linear analysis, we will be also able to gauge accuracy of the numerical computations quantitatively and, moreover, to identify clearly when the nonlinear phase starts and how various modes are mixed with one another to reach saturation. The asymptotic states so obtained should be compared with observations or realistic simulations in detail. In so doing, we will incorporate neglected elements such as non-ideal MHD effects and radiations as well as more realistic jet configurations. We may be able to repeat the entire program for relativistic jets.

It is another interesting issue to investigate what consequences the mode transition observed in this paper may have in more realistic, possibly time-dependent, setting. The possibility of the mode change/exchange between MRI and KKI should be further pursued in broader contexts. The nonlinear evolution and saturation of such pair of modes should be also studied quantitatively by numerical simulations. These are all issues for forthcoming papers.

\subsection{Comparison with previous works}

In this paper, we employ the Laplace-Fourier transforms for linear stability analysis. In this method, instability modes manifest themselves as singularities (mostly poles) located to the right of the imaginary axis in the complex $s$ plane, where $s$ is the conjugate variable of time $t$ in the Laplace transform. Recently, \cite{Vlahakis_2024} reported a different but closely related method for finding unstable modes in the complex plane. The method is based on Fourier transform. The author essentially looks for $s$ that fails to satisfy Eq. (54) in the complex plane. Hence the method coincides with ours at the pole positions. At other general points in the complex $s$ plane, we are dealing with different (complex) functions. Note that the analytic nature of our function is guaranteed by the theory of Laplace transform.

In the Laplace-transform analysis, an instability may emerge as a branching singularity in the complex plane although it may not be a dominant instability (\cite{Laplace_Tran}). In fact, we observe some discontinuities associated with such branching points for some of our models such as $\rm{DC\Omega}$. Although we ignore them in this paper for simplicity, the contributions from those branching points should be also considered appropriately. Incidentally, we also observe some poles moving into or coming out of another Riemann sheet as a model parameter varies. This may or may not be related somehow to the mode changes we observe in this paper. These issues will be a future work, though.

\cite{Bodo_2016} utilized another scheme based on the equation of motion for the Lagrangian displacement vector (\cite{Frieman_1960}) in order to identify mode natures. While our Hamiltonian analysis employs energies, their method uses forces. As such, the latter is easier to understand the physical meaning of each term whereas the former has a merit as a direct extension of the conventional criteria for MHD instabilities (\cite{Awad_2022}). 

The relation between the two methods is not clear for the moment. In fact, the modes distinguished by the two schemes are a bit different from each other; the mode pairs involved in the observed mode transitions are not the same although this is mostly owing to the differences in the background models employed. Indeed, despite our results are broadly aligned with theirs that MRI never dominates over other instabilities and occurs only in rather restricted conditions, we find more: MRI's with $m = 1, 2$ and $3$ emerge in a much wider range of $k$ than in \cite{Bodo_2016}. Furthermore, we find unstable MRI's with $m=0$ and $m > 3$, which were lacking in their paper. On the other hand, \cite{Bodo_2016} considered the case, in which $B_z$ and $B_{\phi}$ are both nonvanishing, whereas we investigate the models with either $B_z$ or $B_{\phi}$ nonzero. They were mainly interested in the effect of pitch angle. The MRI's with $m=0$ are $B_z$-induced while those with $m>3$ that we find in our models are both $B_{z}$- and $B_\phi$-induced. It may indicate that MRI occurs more easily when one of the magnetic field components is dominant over the others. Since the growth rate of MRI becomes larger for faster rotations and lower jet velocities, it may be important in the jet formation region.

In the models considered so far no unstable sausage mode, i.e., the current-driven instability with $m=0$, is found. Different magnetic field configurations such as $B_{\phi0}\propto r$ up to a larger radius could have induced sausage instability (see Section 4.3). Further investigations on the dependence of various unstable modes on the magnetic field configuration are certainly needed anyway.

One of the main findings in this paper is the continuous change of the mode nature. It is true that the classification of instability modes depends, more or less, on the diagnostic scheme we choose but we do not think that our results are its artifact. In fact, the mixed nature of the mode during the transition is evident not only in the Hamiltonian analysis but also in its eigenfunctions (see Figs. \ref{F4} and \ref{F6}). Bodo et al. (2016) reported a similar continuous transition of mode nature in their highly restricted models: the gas pressure and the jet velocity are neglected; the jet and ambient matter have the same density. More importantly, their diagnostic is different, being based not on the energies (terms in the Hamiltonian) but on forces (terms in the equation of motion), as mentioned earlier.

It is not surprising that the growth rate of an unstable mode changes continuously with the parameter values that specify the physical condition of the jet and ambient matter. We naively expected that an unstable mode would be either suppressed (or just remain unstable) and another unstable mode would occur anew as one of the parameter values is varied continuously and the driving force changes from one to another as a result. That is not the case. A single mode changes its nature continuously from one to another. Indeed, the pole in the complex s-plane that corresponds to the initial instability neither disappears nor moves to the stability region ($\rm{Re(s)} \leq 0$). It moves continuously inside the instability region ($\rm{Re(s)} >0$), changing its nature along it. Potential consequences that the mode transition may have in realistic settings are an interesting issue. We will postpone its study to forthcoming papers that will investigate the nonlinear evolutions of various unstable modes by numerical simulations.

\subsection{Applications}

The limitations of the current formulation we discussed in Section 4.1 hamper our attempts to make quantitative comparisons of our results with observations or realistic simulations. Bearing the caveat in mind, we consider two applications below.

The protostellar jets are well-collimated outflows with velocities higher than those of winds. They are supposed to be driven by the magnetic field (\citealt{Lee_2018}, \citealt{Kölligan_2018}) and are prone to instabilities like KHI and CDI. We pay attention here to the motions of knots in the protostellar jet in Herbig-Haro 211, one of the best observed objects of the kind (\citealt{Lee_2018}, \citealt{Caratti_2024}). We consider a possibility that those knots are generated by one of the MHD instabilities and moving at the group velocity of the unstable mode. It is well-known that a wave packet formed by an unstable mode travels at the velocity given by ${d\operatorname{Im(s)}/dk}$ along the jet (\citealt{LIFSHITZ1981265}).

We test this conjecture, building a new model referred to as model $\rm{HH211}$ (see Table \ref{tablemodel}). We set $k=\pi$ , based on the observation (\citealt{Lee_2018}; \citealt{Caratti_2024}). We also modify the $B_{\phi}$ profile so that $B_{\phi0}\propto r$ in the entire jet. We find that the dominant instability is KHI with $m = 0$. Its growth rate is $\sim0.33 \rm{yr^{-1}}$ and the group velocity is $\sim 108\rm{km/s}$. The latter is close to the knot velocity observed in HH 211 $\sim 100\ \rm{km/s}$. The KHI with $m=1$ has a slightly lower growth rate $\sim0.32 \rm{yr^{-1}}$ and group velocity $\sim 107\rm{km/s}$. In this model, a sausage mode is unstable in addition to KKI (see the comment in the third last paragraph in Section 4.1). They are subdominant, though, with the growth rate $\sim 0.1 \rm{yr^{-1}}$ and $\sim 0.07 \rm{yr^{-1}}$, respectively. The existence of KKI, though subdominant, could have a potential to disrupt the jet, which would contradict the observation that the jet of HH 211 is well collimated. We surmise that $B_z$ is non-vanishing in HH 211 and stabilizes the KKI, since a weak poloidal magnetic field is sufficient to suppress KKI. In this model, the plasma beta, or the ratio of gas pressure to magnetic pressure, is $\beta\approx 0.01$ near the jet boundary, which is larger than the observed value $\beta \sim 0.001$. There are also uncertainties in the observational estimates of the density and jet velocity profiles.

The second target is the CCNSe jets. We take two snapshots at 10ms after bounce from realistic 3D MHD simulations of CCSNe by \citet{Bugli_2021}. In all models they observed a jet formation and a subsequent growth of KKI at different postbounce times. We use the (azimuthally averaged) distributions of matter and magnetic field at $z = 50 {\rm km}$ as a function of the distance from the $z$-axis. This corresponds to the region where the linear growth of KKI was seen most clearly (see Fig. 8 in \citealt{Bugli_2021} for details). We approximate the profiles with the analytic ones we employ in this paper (with a slight modifiation for $B_{\phi}$ so that the peak should be located at $r=0.3$ instead of $r=1$, i.e., at the jet boundary). This is admittedly a very crude approximation, since the numerical results are much more complex. We also adjust matter pressure to ensure force balance, since the original configurations are not steady completely.

The two models so obtained are referred to as models L1-0 and L1-90 after the original model names in \cite{Bugli_2021} and given in the last two rows in Table 1. At $10\rm{ms}$ postbounce, KKI has already set in for the original model L1-0 of \cite{Bugli_2021} whereas it has not yet started in the original model L1-90. In fact, matter is still accreting, i.e., the jet has a negative velocity in the latter model. Regardless, we apply the same linear analysis to these models.

In model L1-0, we find that there is an unstable KK mode, which has the highest growth rate $\sim0.5\rm{ms^{-1}}$ at $k=0.9\pi$. All other types of instabilities are stable. This is consistent with the numerical result and confirms authors' claim that the unstable mode is a kink mode. The linear growth rate inferred from the numerical data is $\sim0.57\rm{ms^{-1}}$. This is also in good agreement with the result of the linear analysis. This may be accidental, though. The growth rate is highly sensitive to the value of $B_z$: the growth rate rises to $\sim2.35\rm{ms^{-1}}$ if we reduce the value of $B_z$ by $5\%$ while the mode becomes stable if we raise the value of $B_z$ by $5\%$. In this model KHI and MRI are stable because of the strong poloidal magnetic field again.

In model L1-90 we find no unstable mode. This time the poloidal magnetic field is strong enough to suppress all instabilties. This is again consistent with the result of the simulation, where the KK mode did not show a clear growth around this point of time (see the solid black line in Figure 8 of \citealt{Bugli_2021}).

A word of caution is necessary here. The actual configurations of matter and magnetic-field obtained in the simulations are far more complex than our models and they are not steady, either. We do not expect a perfect agreement between the simulations and the linear analysis from the beginning. Nevertheless, the above results are encouraging for further investigations of the instabilities of MHD jets with our models.

\section{Conclusions}

In this paper, we have presented the results of the linear stability analysis for idealized models of non-relativistic (rotational) MHD jets that are surrounded by rather dense ambient matter, which are also rotating and magnetized in general. We have in mind applications to the jets propagating in the stellar cores and envelopes such as those produced in core-collapse supernovae. We have employed not the ordinary Fourier mode analysis but used the Laplace transform in time and Fourier transform in space. Although we have been focused on unstable eigenmodes in this paper, the method can be used to analyze which modes are excited more preferentially from a particular initial perturbation. We have utilized the Hamiltonian analysis to identify the driving mechanism of individual unstable modes unequivocally. We have demonstrated that this is particularly useful in distinguishing Kelvin-Helmholtz instability (KHI) from magnetorotational instability (MRI), since they do not have remarkable features in their eigenfunctions to distinguish one from the other.

We have found that the mode nature can change continuously from one to another as a model parameter such as the jet velocity varies. In fact, we have observed in one of the non-rotating jet models with only toroidal magnetic fields that an unstable kink (KK) mode turned into a Kelven-Helmholtz (KH) mode continuously as the jet velocity increases. We have also witnessed the opposite transition in another mode. They are most clearly demonstrated by the Hamiltonian analysis, in which the dominant negative contributor changes from one to another indeed. The transitions are also evident in the behavior of the growth rates as well as in the eigenfunctions in this case. The simultaneous mode transitions from KKI to KHI and vice versa happen for two coexisting modes in some cases. The poles corresponding to these unstable modes come close to each other in the complex plane during the transitions, suggesting that this close encounter may be a trigger of the mode exchanges observed. However, there is another case, in which a single mode changes its identity without an encounter with another mode. For the moment, we have no idea what triggers the mode change in general. The mode transition is also observed between MRI and KHI in some of our rotational jet models. The Hamiltonian analysis turns out to be indispensable in this case, since the growth rates and eigenfunctions are not so clear-cut in distinguishing them. We have also studied how the results depend on the jet profiles and other model parameters. We have shown that the main results mentioned above  not change qualitatively if the jet boundary is not sharp but rather continuous.

There are a couple of directions of extension conceivable. The same Laplace-Fourier formulation can be applied immediately to relativistic MHD jets, which will be relevant for gamma-ray bursts. We need to extend the Hamiltonian analysis to the relativistic MHD, though. Non-linear evolutions of the (linearly) unstable modes we have found in this paper should be studied quantitatively, probably numerically. The linear evolutions we have obtained in this paper may be used for the code check. Nonlinear couplings of different modes and the subsequent saturation will be investigated systematically. We may also discuss the relation of the asymptotic state with the perturbation imposed initially.

As we mentioned in Introduction of this paper, we are particularly interested in the stability of the jet observed in some MHD simulations of CCSNe. \cite{Mösta_2014} was the first to demonstrate in 3D simulations that the jet generated by the strong magnetic fields that are wound up by rotation and amplified by MRI may be unstable probably as KKI. Such jets had been commonly observed in 2D simulations with axisymmetry imposed. They are well-collimated in the direction of the rotation axis. In the 3D simulations, they were found to become helical and eventually disrupted. Their results suggested that kink instability plays a role here. A similar conclusion was obtained later by \cite{Bugli_2021}, in which they claimed to have observed a clear linear growth phase. We have extracted two snapshots from their results, approximated them with our own models and applied the linear analysis. We have found that our results are consistent with the numerical results, confirming the claim that the unstable mode is a KK mode and deriving the growth rate that is close to the value deduced from the numerical data. However, further elaborations in modeling are certainly needed, since the matter distribution and magnetic field configurations in the numerical simulations are much more complex than our simplified models. Numerical investigations of both the linear and nonlinear evolutions of each unstable mode we identified in this paper will be also conducted in the forthcoming papers.

\begin{acknowledgments}

We appreciate Matteo Bugli of I.A.P. and Martin Obergaulinger of the Univeristy of Valencia who kindly provided us with their simulation data.

Y.S. is partially supported by the Grants-in-Aid for Scientific Research (21H01083) and the Waseda University Grant for Special Research Projects (project No. 2024C-56, 2024Q-014). He is also supported by the Institute for Advanced Theoretical and Experimental Physics, Waseda University.
\end{acknowledgments}


\bibliography{sample631}{}

\begin{thebibliography}{}
\expandafter\ifx\csname natexlab\endcsname\relax\def\natexlab#1{#1}\fi
\providecommand{\url}[1]{\href{#1}{#1}}
\providecommand{\dodoi}[1]{doi:~\href{http://doi.org/#1}{\nolinkurl{#1}}}
\providecommand{\doeprint}[1]{\href{http://ascl.net/#1}{\nolinkurl{http://ascl.net/#1}}}
\providecommand{\doarXiv}[1]{\href{https://arxiv.org/abs/#1}{\nolinkurl{https://arxiv.org/abs/#1}}}

\bibitem[{{Akiyama} {et~al.}(2003){Akiyama}, {Wheeler}, {Meier}, \& {Lichtenstadt}}]{Akiyama_2003}
{Akiyama}, S., {Wheeler}, J.~C., {Meier}, D.~L., \& {Lichtenstadt}, I. 2003, \apj, 584, 954, \dodoi{10.1086/344135}

\bibitem[{Alhasi \& Elmabrok(2022)}]{Awad_2022}
Alhasi, A., \& Elmabrok, A. 2022, AlQalam Journal of Medical and Applied Sciences, 5, 314–320

\bibitem[{{Appl}(1996)}]{Appl_1996}
{Appl}, S. 1996, \aap, 314, 995

\bibitem[{{Appl} \& {Camenzind}(1992)}]{Appl_1992}
{Appl}, S., \& {Camenzind}, M. 1992, \aap, 256, 354

\bibitem[{{Appl} {et~al.}(2000){Appl}, {Lery}, \& {Baty}}]{Appl_2000}
{Appl}, S., {Lery}, T., \& {Baty}, H. 2000, \aap, 355, 818

\bibitem[{{Balbus} \& {Hawley}(1991)}]{Balbus_1991}
{Balbus}, S.~A., \& {Hawley}, J.~F. 1991, \apj, 376, 214, \dodoi{10.1086/170270}

\bibitem[{{Balbus} \& {Hawley}(1998)}]{Balbus_1998}
---. 1998, Reviews of Modern Physics, 70, 1, \dodoi{10.1103/RevModPhys.70.1}

\bibitem[{{Bateman}(1978)}]{Bateman_1978}
{Bateman}, G. 1978, {MHD instabilities}

\bibitem[{Begelman(1998)}]{Begelman_1998}
Begelman, M.~C. 1998, The Astrophysical Journal, 493, 291, \dodoi{10.1086/305119}

\bibitem[{Bellman \& Roth(1984)}]{Laplace_Tran}
Bellman, R., \& Roth, R. 1984, The Laplace transform, 1st edn., Series in modern applied mathematics ; v. 3 (Singapore: World Scientific)

\bibitem[{{Berlok} \& {Pfrommer}(2019)}]{Berlok_2019}
{Berlok}, T., \& {Pfrommer}, C. 2019, \mnras, 485, 908, \dodoi{10.1093/mnras/stz379}

\bibitem[{Blandford \& Pringle(1976)}]{Blandford_1976}
Blandford, R.~D., \& Pringle, J.~E. 1976, Monthly Notices of the Royal Astronomical Society, 176, 443, \dodoi{10.1093/mnras/176.2.443}

\bibitem[{{Bodo} {et~al.}(2013){Bodo}, {Mamatsashvili}, {Rossi}, \& {Mignone}}]{Bodo_2013}
{Bodo}, G., {Mamatsashvili}, G., {Rossi}, P., \& {Mignone}, A. 2013, \mnras, 434, 3030, \dodoi{10.1093/mnras/stt1225}

\bibitem[{{Bodo} {et~al.}(2016){Bodo}, {Mamatsashvili}, {Rossi}, \& {Mignone}}]{Bodo_2016}
---. 2016, \mnras, 462, 3031, \dodoi{10.1093/mnras/stw1650}

\bibitem[{{Bodo} {et~al.}(2019){Bodo}, {Mamatsashvili}, {Rossi}, \& {Mignone}}]{Bodo_2019}
---. 2019, \mnras, 485, 2909, \dodoi{10.1093/mnras/stz591}

\bibitem[{{Bodo} {et~al.}(2022){Bodo}, {Mamatsashvili}, {Rossi}, \& {Mignone}}]{Bodo_2022}
---. 2022, \mnras, 510, 2391, \dodoi{10.1093/mnras/stab3492}

\bibitem[{{Bodo} {et~al.}(1989){Bodo}, {Rosner}, {Ferrari}, \& {Knobloch}}]{Bodo_1989}
{Bodo}, G., {Rosner}, R., {Ferrari}, A., \& {Knobloch}, E. 1989, \apj, 341, 631, \dodoi{10.1086/167522}

\bibitem[{{Bodo} {et~al.}(1996){Bodo}, {Rosner}, {Ferrari}, \& {Knobloch}}]{Bodo_1996}
---. 1996, \apj, 470, 797, \dodoi{10.1086/177910}

\bibitem[{{Bodo} {et~al.}(2021){Bodo}, {Tavecchio}, \& {Sironi}}]{BodoG_2021}
{Bodo}, G., {Tavecchio}, F., \& {Sironi}, L. 2021, \mnras, 501, 2836, \dodoi{10.1093/mnras/staa3620}

\bibitem[{{Bonanno} \& {Urpin}(2011)}]{Bonanno_2011}
{Bonanno}, A., \& {Urpin}, V. 2011, \aap, 525, A100, \dodoi{10.1051/0004-6361/200913836}

\bibitem[{{Borse} {et~al.}(2021){Borse}, {Acharya, Sriyasriti}, {Vaidya, Bhargav}, {Mukherjee, Dipanjan}, {Bodo, Gianluigi}, {Rossi, Paola}, \& {Mignone, Andrea}}]{Nikhil_2021}
{Borse}, N., {Acharya, Sriyasriti}, {Vaidya, Bhargav}, {et~al.} 2021, A\&A, 649, A150, \dodoi{10.1051/0004-6361/202140440}

\bibitem[{{Bromberg} {et~al.}(2019){Bromberg}, {Singh}, {Davelaar}, \& {Philippov}}]{Bromberg_2019}
{Bromberg}, O., {Singh}, C.~B., {Davelaar}, J., \& {Philippov}, A.~A. 2019, \apj, 884, 39, \dodoi{10.3847/1538-4357/ab3fa5}

\bibitem[{{Bromberg} \& {Tchekhovskoy}(2016)}]{Bromberg_2016}
{Bromberg}, O., \& {Tchekhovskoy}, A. 2016, \mnras, 456, 1739, \dodoi{10.1093/mnras/stv2591}

\bibitem[{{Bugli} {et~al.}(2021){Bugli}, {Guilet}, \& {Obergaulinger}}]{Bugli_2021}
{Bugli}, M., {Guilet}, J., \& {Obergaulinger}, M. 2021, \mnras, 507, 443, \dodoi{10.1093/mnras/stab2161}

\bibitem[{{Caratti o Garatti, A.} {et~al.}(2024){Caratti o Garatti, A.}, {Ray, T. P.}, {Kavanagh, P. J.}, {McCaughrean, M. J.}, {Gieser, C.}, {Giannini, T.}, {van Dishoeck, E. F.}, {Justtanont, K.}, {van Gelder, M. L.}, {Francis, L.}, {Beuther, H.}, {Tychoniec, Ł.}, {Nisini, B.}, {Navarro, M. G.}, {Devaraj, R.}, {Reyes, S.}, {Nazari, P.}, {Klaassen, P.}, {Güdel, M.}, {Henning, Th.}, {Lagage, P. O.}, {Östlin, G.}, {Vandenbussche, B.}, {Waelkens, C.}, \& {Wright, G.}}]{Caratti_2024}
{Caratti o Garatti, A.}, {Ray, T. P.}, {Kavanagh, P. J.}, {et~al.} 2024, A\&A, 691, A134, \dodoi{10.1051/0004-6361/202451350}

\bibitem[{{Carey} \& {Sovinec}(2009)}]{Carey_2009}
{Carey}, C.~S., \& {Sovinec}, C.~R. 2009, \apj, 699, 362, \dodoi{10.1088/0004-637X/699/1/362}

\bibitem[{{Das} \& {Begelman}(2019)}]{Das_2019}
{Das}, U., \& {Begelman}, M.~C. 2019, \mnras, 482, 2107, \dodoi{10.1093/mnras/sty2675}

\bibitem[{Donati \& Landstreet(2009)}]{Donati_2009}
Donati, J.-F., \& Landstreet, J. 2009, Annual Review of Astronomy and Astrophysics, 47, 333, \dodoi{10.1146/annurev-astro-082708-101833}

\bibitem[{{Ferrari} {et~al.}(1978){Ferrari}, {Trussoni}, \& {Zaninetti}}]{Ferrari_1978}
{Ferrari}, A., {Trussoni}, E., \& {Zaninetti}, L. 1978, \aap, 64, 43

\bibitem[{Frieman \& Rotenberg(1960)}]{Frieman_1960}
Frieman, E., \& Rotenberg, M. 1960, Revs. Modern Phys., Vol:32, \dodoi{10.1103/RevModPhys.32.898}

\bibitem[{{Gourgouliatos} \& {Komissarov}(2018)}]{Gourgouliatos_2018}
{Gourgouliatos}, K.~N., \& {Komissarov}, S.~S. 2018, \mnras, 475, L125, \dodoi{10.1093/mnrasl/sly016}

\bibitem[{{Hardee}(2007)}]{Hardee_2007}
{Hardee}, P.~E. 2007, \apj, 664, 26, \dodoi{10.1086/518409}

\bibitem[{{Hardee} \& {Stone}(1997)}]{Hardee_1997}
{Hardee}, P.~E., \& {Stone}, J.~M. 1997, \apj, 483, 121, \dodoi{10.1086/304208}

\bibitem[{{Holm}(1985)}]{Holm_1985}
{Holm}, D.~D. 1985, Physica D Nonlinear Phenomena, 17, 1, \dodoi{10.1016/0167-2789(85)90131-9}

\bibitem[{{Holm} \& {Kupershmidt}(1984)}]{Holm_1984}
{Holm}, D.~D., \& {Kupershmidt}, B.~A. 1984, Physics Letters A, 101, 23, \dodoi{10.1016/0375-9601(84)90083-5}

\bibitem[{Igoshev {et~al.}(2021)Igoshev, Popov, \& Hollerbach}]{Igoshev_2021}
Igoshev, A.~P., Popov, S.~B., \& Hollerbach, R. 2021, Universe, 7, \dodoi{10.3390/universe7090351}

\bibitem[{{Janka}(2012)}]{Janka_2012}
{Janka}, H.-T. 2012, Annual Review of Nuclear and Particle Science, 62, 407, \dodoi{10.1146/annurev-nucl-102711-094901}

\bibitem[{Keszthelyi(2023)}]{Keszthelyi_2023}
Keszthelyi, Z. 2023, Galaxies, 11, \dodoi{10.3390/galaxies11020040}

\bibitem[{{Kim} {et~al.}(2017){Kim}, {Balsara}, {Lyutikov}, \& {Komissarov}}]{Kim_2017}
{Kim}, J., {Balsara}, D.~S., {Lyutikov}, M., \& {Komissarov}, S.~S. 2017, \mnras, 467, 4647, \dodoi{10.1093/mnras/stx409}

\bibitem[{{Kim} {et~al.}(2018){Kim}, {Balsara}, {Lyutikov}, \& {Komissarov}}]{Kim_2018}
---. 2018, \mnras, 474, 3954, \dodoi{10.1093/mnras/stx3065}

\bibitem[{{Kim} {et~al.}(2015){Kim}, {Balsara}, {Lyutikov}, {Komissarov}, {George}, \& {Siddireddy}}]{Kim_2015}
{Kim}, J., {Balsara}, D.~S., {Lyutikov}, M., {et~al.} 2015, \mnras, 450, 982, \dodoi{10.1093/mnras/stv606}

\bibitem[{{Kim} \& {Ostriker}(2000)}]{Kim_2000}
{Kim}, W.-T., \& {Ostriker}, E.~C. 2000, \apj, 540, 372, \dodoi{10.1086/309293}

\bibitem[{{Komissarov} {et~al.}(2019){Komissarov}, {Gourgouliatos}, \& {Matsumoto}}]{Komissarov_2019}
{Komissarov}, S.~S., {Gourgouliatos}, K.~N., \& {Matsumoto}, J. 2019, \mnras, 488, 4061, \dodoi{10.1093/mnras/stz1973}

\bibitem[{{Konar}(2017)}]{Konar_2017}
{Konar}, S. 2017, Journal of Astrophysics and Astronomy, 38, 47, \dodoi{10.1007/s12036-017-9467-4}

\bibitem[{{Kumar} \& {Zhang}(2015)}]{Kumar_2015}
{Kumar}, P., \& {Zhang}, B. 2015, \physrep, 561, 1, \dodoi{10.1016/j.physrep.2014.09.008}

\bibitem[{{Kuroda} {et~al.}(2020){Kuroda}, {Arcones}, {Takiwaki}, \& {Kotake}}]{Kuroda_2020}
{Kuroda}, T., {Arcones}, A., {Takiwaki}, T., \& {Kotake}, K. 2020, \apj, 896, 102, \dodoi{10.3847/1538-4357/ab9308}

\bibitem[{{Kölligan, A.} \& {Kuiper, R.}(2018)}]{Kölligan_2018}
{Kölligan, A.}, \& {Kuiper, R.} 2018, A\&A, 620, A182, \dodoi{10.1051/0004-6361/201833686}

\bibitem[{{Lee} {et~al.}(2018){Lee}, {Hwang}, {Ching}, {Hirano}, {Lai}, {Rao}, \& {Ho}}]{Lee_2018}
{Lee}, C.-F., {Hwang}, H.-C., {Ching}, T.-C., {et~al.} 2018, Nature Communications, 9, 4636, \dodoi{10.1038/s41467-018-07143-8}

\bibitem[{Lifshitz \& Pitaevski(1981)}]{LIFSHITZ1981265}
Lifshitz, E., \& Pitaevski, L. 1981, in Course of Theoretical Physics, Vol.~10, Physical Kinetics, ed. E.~LIFSHITZ \& L.~PITAEVSKI (Amsterdam: Pergamon), 265--283, \dodoi{https://doi.org/10.1016/B978-0-08-026480-6.50011-4}

\bibitem[{{Lyubarskii}(1999)}]{Lyubarskii_1999}
{Lyubarskii}, Y.~E. 1999, \mnras, 308, 1006, \dodoi{10.1046/j.1365-8711.1999.02763.x}

\bibitem[{MacFadyen \& Woosley(1999)}]{MacFadyen_1999}
MacFadyen, A.~I., \& Woosley, S.~E. 1999, The Astrophysical Journal, 524, 262, \dodoi{10.1086/307790}

\bibitem[{Matsumoto {et~al.}(2022)Matsumoto, Asahina, Takiwaki, Kotake, \& Takahashi}]{Matsumoto_2022}
Matsumoto, J., Asahina, Y., Takiwaki, T., Kotake, K., \& Takahashi, H.~R. 2022, Monthly Notices of the Royal Astronomical Society, 516, 1752, \dodoi{10.1093/mnras/stac2335}

\bibitem[{{Matsumoto} {et~al.}(2021){Matsumoto}, {Komissarov}, \& {Gourgouliatos}}]{Matsumoto_2021}
{Matsumoto}, J., {Komissarov}, S.~S., \& {Gourgouliatos}, K.~N. 2021, \mnras, 503, 4918, \dodoi{10.1093/mnras/stab828}

\bibitem[{Mattia {et~al.}(2023)Mattia, Del~Zanna, Bugli, Pavan, Ciolfi, Bodo, \& Mignone}]{Mattia_2023}
Mattia, G., Del~Zanna, L., Bugli, M., {et~al.} 2023, Astronomy \&amp; Astrophysics, 679, A49, \dodoi{10.1051/0004-6361/202347126}

\bibitem[{{Mizuno} {et~al.}(2009){Mizuno}, {Lyubarsky}, {Nishikawa}, \& {Hardee}}]{Mizuno_2009}
{Mizuno}, Y., {Lyubarsky}, Y., {Nishikawa}, K.-I., \& {Hardee}, P.~E. 2009, \apj, 700, 684, \dodoi{10.1088/0004-637X/700/1/684}

\bibitem[{{Musso} {et~al.}(2024){Musso}, {Bodo}, {Mamatsashvili}, {Rossi}, \& {Mignone}}]{Musso_2024}
{Musso}, M., {Bodo}, G., {Mamatsashvili}, G., {Rossi}, P., \& {Mignone}, A. 2024, \mnras, 532, 4810, \dodoi{10.1093/mnras/stae1788}

\bibitem[{Mösta {et~al.}(2014)Mösta, Richers, Ott, Haas, Piro, Boydstun, Abdikamalov, Reisswig, \& Schnetter}]{Mösta_2014}
Mösta, P., Richers, S., Ott, C.~D., {et~al.} 2014, The Astrophysical Journal Letters, 785, L29, \dodoi{10.1088/2041-8205/785/2/L29}

\bibitem[{{Nakamura} \& {Meier}(2004)}]{Nakamura_2004}
{Nakamura}, M., \& {Meier}, D.~L. 2004, \apj, 617, 123, \dodoi{10.1086/425337}

\bibitem[{{Nalewajko} \& {Begelman}(2012)}]{Nalewajko_2012}
{Nalewajko}, K., \& {Begelman}, M.~C. 2012, \mnras, 427, 2480, \dodoi{10.1111/j.1365-2966.2012.22117.x}

\bibitem[{{Narayan} {et~al.}(2009){Narayan}, {Li}, \& {Tchekhovskoy}}]{Narayan_2009}
{Narayan}, R., {Li}, J., \& {Tchekhovskoy}, A. 2009, \apj, 697, 1681, \dodoi{10.1088/0004-637X/697/2/1681}

\bibitem[{{Obergaulinger} \& {Aloy}(2017)}]{Obergaulinger_2017}
{Obergaulinger}, M., \& {Aloy}, M.~{\'A}. 2017, \mnras, 469, L43, \dodoi{10.1093/mnrasl/slx046}

\bibitem[{{Obergaulinger} \& {Aloy}(2020)}]{Obergaulinger_2020}
---. 2020, \mnras, 492, 4613, \dodoi{10.1093/mnras/staa096}

\bibitem[{{Obergaulinger} {et~al.}(2009){Obergaulinger}, {Cerd{\'a}-Dur{\'a}n}, {M{\"u}ller}, \& {Aloy}}]{Obergaulinger_2009}
{Obergaulinger}, M., {Cerd{\'a}-Dur{\'a}n}, P., {M{\"u}ller}, E., \& {Aloy}, M.~A. 2009, \aap, 498, 241, \dodoi{10.1051/0004-6361/200811323}

\bibitem[{{O'Neill} {et~al.}(2012){O'Neill}, {Beckwith}, \& {Begelman}}]{O'Neill_2012}
{O'Neill}, S.~M., {Beckwith}, K., \& {Begelman}, M.~C. 2012, \mnras, 422, 1436, \dodoi{10.1111/j.1365-2966.2012.20721.x}

\bibitem[{{Perucho} {et~al.}(2004){Perucho}, {Hanasz}, {Mart{\'\i}}, \& {Sol}}]{Perucho_2004}
{Perucho}, M., {Hanasz}, M., {Mart{\'\i}}, J.~M., \& {Sol}, H. 2004, \aap, 427, 415, \dodoi{10.1051/0004-6361:20040349}

\bibitem[{{Pessah} \& {Psaltis}(2005)}]{Pessah_2005}
{Pessah}, M.~E., \& {Psaltis}, D. 2005, \apj, 628, 879, \dodoi{10.1086/430940}

\bibitem[{{Porth} \& {Komissarov}(2015)}]{Porth_2015}
{Porth}, O., \& {Komissarov}, S.~S. 2015, \mnras, 452, 1089, \dodoi{10.1093/mnras/stv1295}

\bibitem[{Rembiasz {et~al.}(2016)Rembiasz, Guilet, Obergaulinger, Cerdá-Durán, Aloy, \& Müller}]{Rembiasz_2016}
Rembiasz, T., Guilet, J., Obergaulinger, M., {et~al.} 2016, Monthly Notices of the Royal Astronomical Society, 460, 3316, \dodoi{10.1093/mnras/stw1201}

\bibitem[{{Singh} {et~al.}(2016){Singh}, {Mizuno}, \& {de Gouveia Dal Pino}}]{Singh_2016}
{Singh}, C.~B., {Mizuno}, Y., \& {de Gouveia Dal Pino}, E.~M. 2016, \apj, 824, 48, \dodoi{10.3847/0004-637X/824/1/48}

\bibitem[{{Sobacchi} \& {Lyubarsky}(2018)}]{Sobacchi_2017}
{Sobacchi}, E., \& {Lyubarsky}, Y.~E. 2018, \mnras, 473, 2813, \dodoi{10.1093/mnras/stx2592}

\bibitem[{{Spruit} {et~al.}(1997){Spruit}, {Foglizzo}, \& {Stehle}}]{Spruit_1997}
{Spruit}, H.~C., {Foglizzo}, T., \& {Stehle}, R. 1997, \mnras, 288, 333, \dodoi{10.1093/mnras/288.2.333}

\bibitem[{{Takahashi} {et~al.}(2016){Takahashi}, {Iwakami}, {Yamamoto}, \& {Yamada}}]{Takahashi_2016}
{Takahashi}, K., {Iwakami}, W., {Yamamoto}, Y., \& {Yamada}, S. 2016, \apj, 831, 75, \dodoi{10.3847/0004-637X/831/1/75}

\bibitem[{Turland \& Scheuer(1976)}]{Turland_1976}
Turland, B.~D., \& Scheuer, P. A.~G. 1976, Monthly Notices of the Royal Astronomical Society, 176, 421, \dodoi{10.1093/mnras/176.2.421}

\bibitem[{Varma {et~al.}(2022)Varma, Müller, \& Schneider}]{Varma_2023}
Varma, V., Müller, B., \& Schneider, F. R.~N. 2022, Monthly Notices of the Royal Astronomical Society, 518, 3622, \dodoi{10.1093/mnras/stac3247}

\bibitem[{{Vlahakis}(2024)}]{Vlahakis_2024}
{Vlahakis}, N. 2024, Universe, 10, 183, \dodoi{10.3390/universe10040183}

\bibitem[{Zhou \& Qin(2013)}]{Qin_2013}
Zhou, Y., \& Qin, H. 2013, \dodoi{10.2172/1072366}

\end{thebibliography}
\bibliographystyle{aasjournal}

\end{document}